\documentclass[a4paper, 11pt]{article}
\pdfoutput=1
\usepackage{jheppub}
\usepackage{graphicx}
\usepackage{subcaption}

\usepackage[T1]{fontenc}

\usepackage{csquotes}
\usepackage[normalem]{ulem}
\usepackage{soul}

\newcommand{\ket}[1]{|#1\rangle}

\newcommand{\indicatrixwidth}{0.47\textwidth}

\newcommand{\tc}{C_t}

\addtocontents{toc}{\protect\setcounter{tocdepth}{2}}

\title{The Geometry of Quantum Complexity in Open Systems}

\author[a]{Ezra Acalapati,}
\author[b]{Kausik Ghosh,}
\author[a]{Giuseppe Policastro}
\affiliation[a]{Laboratoire de Physique de l'\'{E}cole Normale Supérieure, ENS, Universit\'{e} PSL, CNRS, Sorbonne Universit\'{e}, Universit\'{e} de Paris, F-75005 Paris, France}
\affiliation[b]{Department of Mathematics, King's College London, Strand, London, WC2R 2LS, United Kingdom}

\emailAdd{ezra.acalapati@phys.ens.psl.eu}
\emailAdd{kau.rock91@gmail.com}
\emailAdd{giuseppe.policastro@phys.ens.psl.eu}

\abstract{We extend Nielsen's geometric approach for quantum complexity from closed to open quantum systems, whose dynamics is governed by Lindbladian evolution.  In this framework, complexity is defined through an optimal-control problem on the space of mixed states, with a cost assigned to both unitary and non-unitary generators. We show that the resulting geometric structure differs fundamentally from the Riemannian geometry that emerges in the case of unitary evolution. In the open-system setting, the natural geometry is typically sub-Finslerian. Dissipation makes the geodesics non-reversible, while the admissible tangent directions are restricted by the physically allowed controls.  We analyze several physically motivated examples, including a single qubit subject to depolarizing and amplitude-damping channels, as well as the damped harmonic oscillator. We show that, similarly to the unitary case, varying the penalty factors in the cost functional modifies the geometric properties through changes in the flag curvature, the Finslerian analog of sectional curvature. Our results provide a geometric framework for quantifying the abstract notion of complexity in dissipative quantum systems, with potential connections to experimentally realizable setups.
}

\begin{document}
\maketitle
\flushbottom

\section{Introduction}

The notion of quantum complexity has emerged as a unifying theme across different fields, such as quantum computing, quantum information, high-energy theory, and statistical physics, each field having its own motivations and approaches. A physical motivation is to differentiate the behavior of physical systems in terms of the delocalization (or {\it scrambling}) of the information contained in a subsystem along the time evolution and possibly defining a notion of quantum chaos. In this context, many quantities have been proposed and studied (circuit complexity, Krylov complexity, spread complexity etc., see the review \cite{Baiguera:2025dkc}). Among these, the notion of circuit complexity is closest to the origin of the concept in quantum computation, as a measure of the algorithmic difficulty to solve a problem with a quantum computer. An algorithm is implemented as a {\it circuit}, that is a succession of {\it gates} (unitary operations) acting on a string of qubits. The simplest measure is to just count the number of required gates, and a natural refinement is to attribute different weights (or penalty factors) to different gates. A slight abstraction leads us from circuits to continuous paths in the space of unitary operators, and from the penalty factors to a suitable {\it cost functional} defined on the space of paths. This viewpoint, pioneered by Nielsen \cite{Nielsen:2006cea,Dowling:2006tnk}, allows one to interpret the complexity of a unitary operator\footnote{The complexity of a state is defined as the minimum complexity of an operator that creates that state from a given reference state.} as the length of an optimal trajectory in a space of unitary transformations equipped with a certain metric structure. While the original circuit complexity is hard to compute, Nielsen's version is more amenable to computation, and can be expected to provide a good approximation for circuits with a large number of gates. It was also  shown by Nielsen to provide rigorous bounds on the circuit complexity defined by gate-counting.

This geometric formulation has proven fruitful in diverse contexts ranging from quantum computation to holography \cite{Chapman:2021jbh}.  Indeed, Susskind and collaborators proposed that the quantum complexity of a state has a direct interpretation in the holographic correspondence in terms of geometric quantities in the bulk AdS spacetime \cite{Susskind:2014rva,Susskind:2014moa,Brown:2015bva} (see again \cite{Baiguera:2025dkc} for subsequent developments). In particular, they suggested that  the growth of the volume of the region inside the black hole horizon (which is invisible from the exterior) can be explained by equating it with the growth of the complexity of the state dual to the black hole, which is the thermofield double state of the dual theory.  The complexity of a generic state is expected to grow linearly for a time proportional to the dimension of the Hilbert space of the system, and then to saturate to a constant value. In the holographic correspondence, the saturation time is infinite in the classical approximation, and should be made finite by quantum gravity corrections in the bulk. Through this proposal, quantum complexity becomes an essential ingredient in our understanding of the emergence of spacetime in holography.

As emphasized in \cite{Brown:2016wib,Brown:2019whu} the {\it complexity geometry} defined by Nielsen's approach has a rich structure, which appears even for the simplest conceivable system composed of a single qubit. The main qualitative features are: the geometry on the  group of unitary transformations of the system is not left- and right-invariant, and depending on the penalty factors, it can develop regions of negative sectional curvature. Heuristically, this means that nearby geodesics diverge, a behavior that resembles that of a classically chaotic system, and could point out to a more precise connection between quantum complexity and quantum chaos.

The picture presented so far applies to {\it closed} systems, whose time evolution is a unitary transformation. However, there are many reasons to consider also \emph{open} systems. The most obvious reason is that realistic quantum systems — such as those encountered in quantum simulation platforms, cold atom experiments, and noisy intermediate-scale quantum (NISQ) devices — are inevitably subject to dissipation, decoherence, and environmental interactions. These effects have to be taken into account for the algorithm optimization. From the holographic point of view, if we believe that complexity is related to the emergence of spacetime behind the horizon, we would also like to understand what happens when the black hole evaporates, which in the dual boundary theory requires adding a coupling to an external bath. The introduction of noise can also be a useful mathematical device to extract information about the closed system by taking the limit of weak dissipation \cite{Mori:2023qbd}. 
Another motivation comes from considering systems that are invariant under a symmetry group (for instance, in holographic applications one would consider the conformal group). Then one can use the group structure to obtain results for the relative complexity between states that live in the same representation of the group (as for instance in \cite{Flory:2020eot,Chagnet:2021uvi}). However if one wants to relate states in different representations, one has to use gates that break the symmetry. The representations will then all mix with each other and it will be very difficult to relate two states with definite but different representations. If however one introduces noise, that can be used to ``dissipate'' away the symmetry charge, and one can naturally define complexity with respect to the singlet state. 

These reasons motivate us to extend the notion of quantum complexity to open systems. In this case, the dynamics is governed not by unitary evolution but in general by completely positive trace-preserving (CPTP) maps, often described as quantum channels. For a continuous evolution, under certain conditions, the most general evolution can be put in the form of the Lindblad equation. In this paper we will consider only this case\footnote{
See also \cite{Lloyd:2000dwz,Carlini:2007fqn,PhysRevA.110.042601} for previous work on quantum optimal control for open systems and noisy gates, including approaches based on embedding the open-system dynamics into an enlarged unitary evolution, but with a different focus.}.

The geometric framework of Nielsen can be adapted easily to the open system case. The Lindblad equation defines a motion in the space of mixed states of a system; the motion can be controlled using unitary gates and non-unitary operations that encode the effects of noise or of monitoring the system with measurements. The cost functional can be extended to the non-unitary gates, and the minimization of the cost can be translated in the language of {\it optimal control}. However, the resulting geometry is very different from the unitary case: instead of a Riemannian geometry, the construction yields in general a {\it sub-Finslerian geometry} on the space of states.\footnote{A Finslerian geometry can appear already in the unitary case for some choices of cost functionals \cite{Jefferson:2017sdb}, however in that case it is not of the most general type as the one we describe here. In particular, there is no irreversibility, which is the hallmark of a non-unitary evolution.} We will describe in detail its definition and  properties in the main text; here we just anticipate the main differences from the Riemannian case. While in Finslerian geometry one can define distance and therefore geodesics, they are usually non-reversible, as an effect of the dissipation. Moreover, the tangent vector of the geodesic is restricted to lie in a subspace of the tangent space (this motivates the {\it sub-} part of the name). This has the effect that the geodesics are generally only piecewise-smooth. Since the Lindblad equation does not correspond to a group (the Lindbladian only generates a semigroup), the geometry is never homogeneous, and the Finslerian function, which plays the role of the metric, is a non-trivial function of both positions and velocities.

In this paper we apply the construction found in \cite{Lopez:2000} to describe how the sub-Finslerian geometry arises from the non-unitary version of Nielsen's complexity, and we analyze its properties in some simple and physically motivated examples: a single qubit driven by {\it depolarizing} and {\it amplitude-damping} dynamics, and the damped harmonic oscillator. We also allow for the presence of a drift, i.e. an operation that acts on the system but is not controlled; we study its effect when it is either in the unitary or in the non-unitary part of the Lindbladian. We derive general formulas for the Finslerian function that applies to all the cases we consider, and elucidate in which cases we find a genuine Finslerian structure and in which cases we can recover a more standard Riemannian structure. We also study the dependence of the geometry on the choice of penalty factors,  and show that they affect the flag curvature (the Finslerian analogue of the sectional curvature), in particular its sign which controls the behavior of neighboring geodesics.

Our results represent a step toward establishing a concrete connection between abstract theoretical notions of complexity and physical systems that can be implemented and probed experimentally. The extension of Nielsen's approach to open systems provides an operational framework for quantifying and analyzing complexity in realistic settings.  Our results may be used to study problems of optimization and the associated trade-offs between different physical and computational resources, across a broad range of platforms that include noisy intermediate-scale quantum devices and driven-dissipative nonequilibrium systems.

The rest of the paper is organized as follows.  In section \ref{sec:DynamicsOpen} we recall the basics of open systems and the Lindblad equation that describe their continuous-time dynamics. In section \ref{sec:ComplexityGeometry}, we review Nielsen's approach and the emergent complexity geometry. In section \ref{sec:FinslerGeometry}, we review the definition and some basic properties of Finsler geometry. In section \ref{sec:OprimalControl}, we describe how the definition of complexity in the open case turns into an optimal control problem, and how this in turn gives rise to a Finsler geometry. In section \ref{sec:SingleQubit}, we apply the construction of the previous section to derive explicitly the Finsler geometry in the case of a system composed of one single qubit, with generic choices for the Lindbladian and the cost functional. In section \ref{sec:Examples}, we illustrate our results on some examples where the Lindbladian describes depolarizing and amplitude damping processes without drift, a depolarizing process with a unitary or non-unitary drift, and finally the damped harmonic oscillator restricted to Gaussian states, which we show is equivalent to the depolarizing process of a qubit. In section \ref{sec:Discussion}, we conclude with a discussion of the results and possible future directions. One Appendix contains some details on the computation of the geodesics in the geometry.

\section{Dynamics of open systems}\label{sec:DynamicsOpen}

We recall here some basic facts about open systems and their dynamics. For a more complete treatment the reader can consult the textbooks \cite{Breuer:2002pc,Rivas:2012ugu}.

As an open system interacts with the environment, the evolution does not preserve the purity of the state, therefore we have to consider the system as described by a density matrix. The evolution of the total system, including the environment, is unitary, and it induces on the reduced density matrix of the system an evolution described by a CPTP map (completely positive and trace preserving) $\rho \to L_t \rho$.  Completely positive means that the associated map on the system coupled to an ancilla system of dimension $n$
\begin{eqnarray}
    \rho_A \otimes \rho \to \rho_A \otimes L_t \rho 
\end{eqnarray}
where $\rho_A$ is any state on the ancilla, is positive for any $n$. 
Conversely, according to the Stinespring dilation theorem \cite{Stinespring1955,NielsenChuang2010,Watrous2018}, every CPTP map acting on a system can always be understood as the result of tracing out some ancilla system.

The most general form of a completely positive map is given by the Kraus representation,
\begin{equation} \label{eq:Kraus}
    \rho \rightarrow \rho' = \sum_a K_a \rho K_a^\dagger,
\end{equation}
where the operators $K_a$ capture the effect of the environment on the system. Physically, each Kraus operator can be interpreted as corresponding to a particular channel through which the system can evolve, for example due to different environmental interactions or  measurement operators. 

It can easily be seen that trace preservation requires that
\begin{equation}
    \sum_a K_a^\dagger K_a = I \,.
\end{equation}


The Kraus representation describes a finite-time evolution, or the result of a measurement (without post-selection of the state based on the outcome of the measurement). In order to describe continuous-time dynamics we need the notion of a {\it quantum dynamical semigroup} (QDS), also called a quantum Markov semigroup (QMS). 
This is a one-parameter family $\{\phi_t\}_{t\ge 0}$ of CPTP linear maps acting on the algebra of bounded operators $\mathcal{B}(\mathcal{H})$ over a Hilbert space $\mathcal{H}$,
\begin{equation}
    \phi_t: \mathcal{B}(\mathcal{H}) \to \mathcal{B}(\mathcal{H}), \qquad t\ge 0.
\end{equation}
The family $\{\phi_t\}_{t\ge 0}$ of CPTP maps must satisfy the following properties:

\begin{enumerate}
    \item \textbf{Identity (monoid) property:}
    \begin{equation}
        \phi_0 = \mathrm{id}, \qquad \phi_0(\rho)=\rho \quad \forall\,\rho\in\mathcal{B}(\mathcal{H}).
    \end{equation}

    \item \textbf{Semigroup (Markov) property:}
    \begin{equation}
        \phi_{t+s} = \phi_t \circ \phi_s, \qquad \phi_{t+s}(\rho)=\phi_t\!\big(\phi_s(\rho)\big), \quad \forall\,s,t\ge 0.
    \end{equation}
    This expresses the memoryless (Markovian) character of the evolution: the future depends only on the present state.


    \item \textbf{(Strong) continuity:}
    \begin{equation}
        \lim_{t\to 0^+}\|\phi_t(\rho)-\rho\|_1 = 0 \,.
    \end{equation}
\end{enumerate}

Unlike a group, a semigroup does \emph{not} require invertibility. This reflects the intrinsic irreversibility of open quantum dynamics: information is lost to the environment and cannot, in general, be recovered by a CPTP inverse~\cite{Rivas:2012ugu}. In finite dimension, a CPTP map admits a CPTP inverse if and only if it is unitary; hence generic dissipative channels are irreversible.

The condition 3 ensures the existence of a generator, namely a (bounded) superoperator  $\mathcal{L}$, called the Lindbladian, 
 such that
\begin{equation}
    \frac{d}{dt}\rho_t \,=\, \mathcal{L}(\rho_t), 
    \qquad \rho_t := \phi_t(\rho_0), 
    \qquad \phi_t \,=\, e^{t\mathcal{L}}.
\end{equation}

The general form of the (time-homogeneous) generator of a QDS is given by the Lindblad (GKSL) master equation, named after Gorini, Kossakowski, Sudarshan, and Lindblad, where the former three and the latter derived said equation independently~\cite{Gorini:1975nb,Lindblad:1975ef,Chruscinski:2017}.
In its canonical GKSL form, the master equation reads 
\begin{equation}
    \label{eq: LinMas}
    \Dot{\rho} = -i\big[ H,\rho \big] + \sum_{i} \Big( L_i\rho L_i^\dagger - \frac{1}{2} \big\{ L_i^\dagger L_i,\rho \big\} \Big),
\end{equation}
with $H=H^\dagger$ describing the unitary part of the evolution, and Lindblad operators $\{L_i\}$ describing the dissipative channels.

The trace-preserving property is easily checked to follow from the structure of the master equation. The complete positivity is less apparent, but it can be shown by relating the evolution after a small time interval to the Kraus representation \eqref{eq:Kraus}, with the identifications 
    \begin{eqnarray}
        \label{eq: KraHLin}
        K_0 &\approx& I - i H dt - \frac{1}{2} \sum_{i} L_i^\dagger L_i dt,\\
        K_i &\approx& L_i \sqrt{dt}.
    \end{eqnarray}

 An important property of the QDS evolution is that it is contractive for distinguishability measures such as the trace distance, and suitable entropic quantities decrease monotonically, reflecting the thermodynamic arrow of time and non-negative entropy production~\cite{Breuer:2002pc,Rivas:2012ugu}.

\paragraph{Examples}
\begin{itemize}
    \item \emph{Amplitude damping:} this describes the process of spontaneous emission in a two-level system.  
    For small $\Delta t$, a Kraus representation for a single qubit is
    \[
        K_0 = |0\rangle\!\langle 0| + \sqrt{1-\gamma \Delta t}~\,|1\rangle\!\langle 1|,
        \qquad 
        K_1 = \sqrt{\gamma \Delta t}~\,|0\rangle\!\langle 1|,
    \]
    describing the irreversible decay $\ket{1}\to\ket{0}$. With more degrees of freedom, we can have similar amplitude-damping processes such as in $T_1$ relaxation in superconducting qubits~\cite{Krantz:2019jkw, Kubica:2022fqg}, photon loss~\cite{Fisher:2011hvf, Albert:2018reb}, exciton recombination~\cite{PhysRevB.80.155307} and decay of an excited NV-center level~\cite{DOHERTY20131, Goldman:2015ohd}.
    
    \item \emph{Depolarizing channel:}  this models the loss of information for a system. The Kraus representation for a single qubit is
     \[
        K_0 = \sqrt{1-3\gamma\Delta t} \, I,
        \qquad 
        K_1 = \sqrt{\gamma\Delta t} \, \sigma_x,
        \qquad
        K_2 = \sqrt{\gamma\Delta t} \, \sigma_y,
        \qquad
        K_3 = \sqrt{\gamma\Delta t} \, \sigma_z.
    \]
     Analogous higher-dimensional examples, though not identical to the single-qubit process considered here, arise as effective descriptions in randomized benchmarking and twirling protocols~\cite{Zyczkowski:2005jvo,Dankert:2009yux,Magesan:2012mfg}, are related to noise-tailoring ideas in randomized compiling~\cite{Wallman:2015uzh,Hashim:2020cop}, and can be engineered in photonic polarization systems~\cite{Karpinski:2008ehv,Shaham:2010iju,Jeong:2012jbn} or generalized to higher-spin systems~\cite{Denis:2022ckr}.

\end{itemize}

\section{Complexity geometry} \label{sec:ComplexityGeometry}

We review here the notions of Nielsen's continuous circuit complexity, first applied to unitary circuit and then to the Lindblad equation. 

We start by constructing a path in the space of unitary operators $U(N)$ by means of a time-dependent Hamiltonian: 
\begin{equation}\label{controlprob1}
U(t) = \overleftarrow{\mathcal{P}} \exp \left(i  \int_0^{t} H (s) ds \right) . 
\end{equation}
The Hamiltonian can be expanded in a basis of hermitian operators, with coefficients that serve as control functions:
\begin{equation}\label{controlprob2}
 H (t) = \sum_I \alpha^I (t)  \mathcal{O}_I .
\end{equation}
The complexity of a given unitary $U_T$ is defined by the minimization, over all paths \eqref{controlprob1} such that $U(t_f)=U_T$, of a {\it cost
functional} $F [U,\dot U]$
\begin{equation}
    \mathcal{C }_F [U_T] = \min_{\{ U(t) \}} \int_0^{t_f} dt \, F [U(t), \dot U(t)] \,.
\end{equation}
Notice that the final time $t_f$ is in general part of the minimization problem. Under certain conditions that we will spell out shortly, the complexity defines a distance on the space of the unitaries. 

If we see the operators ${\cal O}_I$ as infinitesimal gates, it is natural to choose a cost functional that depends only on the control functions $\alpha^I = \gamma^{IJ} {\rm Tr} (\dot U \, U^{-1}{\cal O_J})$, where $\gamma_{IJ} = {\rm Tr} ({\cal O_I}{\cal O_J})$. If moreover the cost function has the form $F = \sqrt{p_{IJ} \alpha^I \alpha^J}$, then the distance is associated to a Riemannian metric on $U(N)$ which is homogeneous and right-invariant but in general not left-invariant. Notice that in this case, and more generally when $F$ is a homogeneous function of the velocities of degree 1, one can always reparametrize the time coordinate in order to fix $t_f=1$.

As an illustrative example, it is good to keep in mind the case of the single qubit \cite{Brown:2019whu}, with operators given by the Pauli matrices $\{ \sigma_x,\sigma_y,\sigma_z\}$, and cost function $F^2=\alpha_x^2+\alpha_y^2 + p \, \alpha_z^2$. The corresponding Riemannian metric on $SU(2) \approx S^3$ in Euler angles is 
\begin{eqnarray}
    ds^2 = cos^2(2 \theta_y) d\theta_x^2 + p (d\theta_z+sin 2\theta_y d\theta_x)^2 \,.
\end{eqnarray}
For $p=1$ this reduces to the round metric on the 3-sphere. At the identity, the sectional curvature has components $K_{xy}^{~~yx} = 4-3p$, so it can become negative for sufficiently large penalty factor. As mentioned in the introduction, this behavior is associated to the divergence of nearby geodesics. 

The cost function however can have a more general form, and it will correspond not to a Riemannian metric but to a Finsler structure. We will review this notion in the next section. 

The non-unitary complexity is defined in a completely analogous way. Even though we consider state and not operator complexity, the state is given by a density matrix, so the formalism actually is close to the unitary operator complexity. The evolution equation \eqref{controlprob1}, or $\dot U = i U H$, is replaced by the equation \eqref{eq: LinMas}, and we expand also the Lindblad operators in a basis:
\begin{eqnarray}\label{controlprob3}
    L_i(t) = \sum_J \beta_i^J {\cal O}_J \,. 
\end{eqnarray}
We can use without loss of generality the same basis as in \eqref{controlprob2}, however the $L_i$ do not have to be hermitian. The cost function will now be a function of all the control parameters $\alpha_i, \beta_i^J$. 

The crucial difference with the unitary case is that, for solutions of the Lindblad equation, the derivative $\dot \rho$ does not span the full tangent space of the state manifold. For instance, we can compute the variation of the purity of the state ${\rm Tr} \rho^2$ along the evolution \cite{Lidar:2005gzv}
\begin{equation}
    \begin{split}
    \dot p \equiv  \frac12   \frac{d}{dt} {\rm Tr} \rho^2 & = \sum_i {\rm Tr} (\rho L_i\rho L_i^\dagger - \rho L_i^\dagger L_i \rho) \\
     & = -\frac{1}{2} \sum_i {\rm Tr}([\rho,L_i]^\dagger [\rho,L_i]) + \frac{1}{2} \sum_i {\rm Tr} (\rho^2 [L_i,L_i^\dagger] )\,.
    \end{split}
\end{equation}
From this we see that a sufficient condition for the purity to decrease is \[{\cal L}(I) = \sum_i [L_i,L_i^\dagger] \leq 0 \,.\]
Since ${\cal L}$ is a positive map, this actually implies ${\cal L}(I)=0$. The condition is also necessary, as can be seen by taking $\rho = I + \epsilon A$. To linear order in $\epsilon$, $\dot p \sim \epsilon \, {\rm Tr}(A {\cal L}(I))$. This cannot have a definite sign for all $A$ unless it is zero.\footnote{The derivation presented here assumes finite dimension;  see \cite{Lidar:2005gzv} for the infinite-dimensional case. A simple counter-example to the necessary condition is to take a single operator  $L=a^\dagger$, the bosonic creation operator. The evolution is purity-decreasing but ${\cal L}(I) = -I$.}

Under the condition of a unital Lindbladian, then, 
the derivative $\dot \rho$ has to lie in the half-space of the directions of decreasing purity. This is a general constraint that does not depend on the specific form of the Lindblad operators. We will see in examples that the full constraints can be much more intricate. The control functions in the Lindbladian determine a non-linear subset of allowed velocities, and  the minimization of the cost function has to be addressed as an  {\it optimal control} problem. We will see shortly how this gives rise to a sub-Finslerian structure. 

\section{Finsler geometry} \label{sec:FinslerGeometry}

At this point we have to give an account of Finsler geometry, that we have already mentioned several times. Finsler geometry provides a natural generalization of Riemannian geometry in which the notion of length depends not only on position but also on direction in the tangent space. For more details the reader can consult e.g. \cite{Bao:2000}


Let $M$ be a smooth manifold. A Finsler structure on $M$ is defined by a function on its tangent bundle 
\begin{equation}
    F : TM \to [0,\infty],
\end{equation}
(where $\infty$ is an allowed value) 
which assigns a norm to each tangent vector $y \in T_x M$, and satisfies:
\begin{enumerate}
    \item \textbf{Regularity:} $F$ is smooth on $TM \setminus \{0\}$,
    \item \textbf{Positive homogeneity:} $F$ is a function positively homogeneous of order one in the velocities, i.e.
    \begin{equation}
        F(x,\lambda y) = \lambda F(x,y), \qquad \lambda > 0,
    \end{equation}
    \item \textbf{Strong convexity:} the Hessian of $F^2$ with respect to $y$,
    \begin{equation}
        g_{ij}(x,y) = \frac{1}{2} \frac{\partial^2 F^2}{\partial y^i \partial y^j},
    \end{equation}
    is positive definite.
\end{enumerate}

The tensor $g_{ij}(x,y)$ is known as the \emph{fundamental tensor} and plays the role of a direction-dependent metric. Unlike the Riemannian case, it depends explicitly on the tangent vector $y$, reflecting the anisotropic nature of the geometry.

\subsection{Geodesics and spray}

Geodesics in Finsler geometry are curves that locally extremize the length functional
\begin{equation}
    \int F(x,\dot{x})\, dt.
\end{equation}
Comparing with the previous section, we see that the Finsler function is the same as the cost function. 
The geodesics are determined locally by the Euler–Lagrange equations, which can be written in the form
\begin{equation}
    \ddot{x}^i + 2 G^i(x,\dot{x}) = 0,
\end{equation}
where $G^i(x,y)$ are the spray coefficients, defined in terms of $F$ by
\begin{equation}
    G^i = \frac{1}{4} g^{ij}\left( \frac{\partial^2 F^2}{\partial x^k \partial y^j} y^k - \frac{\partial F^2}{\partial x^j} \right).
\end{equation}
The spray encodes the intrinsic geometry and generalizes the Levi-Civita connection of Riemannian geometry: 

\begin{equation}
R^{i}{}_{k}(y)
=
2\,\frac{\partial G^{i}}{\partial x^{k}}
-
y^{j}\frac{\partial^{2}G^{i}}{\partial x^{j}\partial y^{k}}
+
2\,G^{j}\frac{\partial^{2}G^{i}}{\partial y^{j}\partial y^{k}}
-
\frac{\partial G^{i}}{\partial y^{j}}
\frac{\partial G^{j}}{\partial y^{k}} \, .
\end{equation}

\subsection{Curvature and flag curvature}

The intrinsic curvature of a Finsler manifold is described by the \emph{flag curvature}, which generalizes the sectional curvature of Riemannian geometry. Since the geometry depends on direction, curvature must be defined with respect to both a point and a chosen tangent vector.

A flag is given by a pair $(y, v)$, where $y \in T_x M$ is a nonzero tangent vector (the \emph{flagpole}) and $v \in T_x M$ spans a two-dimensional plane together with $y$. The flag curvature is defined as
\begin{equation}
    K(x,y,v)
    =
    \frac{g_y\big(R_y(v), v\big)}
    {g_y(y,y)\,g_y(v,v) - g_y(y,v)^2},
\end{equation}
where $R_y(v)$ denotes the curvature operator derived from the spray, and $g_y$ is the fundamental tensor evaluated at $y$.
In contrast to the Riemannian case, where sectional curvature depends only on the plane, the flag curvature depends explicitly on the choice of direction $y$ within the plane. 

When the Finsler function takes the quadratic form
\begin{equation}\label{eq:quadratic-finsler}
    F(x,y) = \sqrt{g_{ij}(x) y^i y^j},
\end{equation}
the dependence on $y$ disappears from the fundamental tensor, and the geometry reduces to the Riemannian case. In this limit, the spray coincides with the Levi-Civita connection and the flag curvature reduces to the sectional curvature.
An important class of examples is given by Minkowski norms, where $F$ depends only on the tangent vector $y$ and not on the base point $x$. In this case, the spray coefficients vanish,
\begin{equation}
    G^i = 0,
\end{equation}
and the curvature tensor is identically zero. Thus, despite possible anisotropy or non-Euclidean shape of the unit ball, the intrinsic geometry is flat.

Finsler geometry can be viewed as equipping each tangent space with a possibly asymmetric and direction-dependent norm, whose unit level set (the indicatrix) determines the local geometry. The curvature measures how this norm changes as one moves along the manifold.
This perspective is particularly useful in contexts where evolution is constrained or anisotropic, as it allows one to encode dynamical properties directly into the geometry of the state space.

\section{Optimal control problem for autonomous systems} \label{sec:OprimalControl}

In this section we describe how a general optimal control problem with a cost functional can be reformulated in terms of Pontryagin's Maximum Principle (PMP), and how this naturally induces a (sub-)Finsler geometric structure. We restrict to the autonomous case, where the dynamics and cost do not depend explicitly on time. The presentation and notations in this section follow closely the reference \cite{Lopez:2000} (see also \cite{Boscain:2021jlj} for a recent pedagogical exposition).

\subsection{Control system and cost functional}

Consider an autonomous control system on a manifold $M$ with local coordinates $x^i$,
\begin{equation}
    \dot{x}^i = f^i(x,u), \qquad u \in U_x,
\end{equation}
where $u$ denotes the control variables (collectively denoting the $\alpha_i$ of \eqref{controlprob2} and $\beta_i^J$ of \eqref{controlprob3}), and $U_x$ is the admissible control set that can generally depend on the point $x$, although in the examples we will consider it will be the same at all points.  

We associate to each trajectory a cost functional of the form
\begin{equation}
    \mathcal{C}[\gamma] = \int_0^T L(x(t),u(t))\, dt,
\end{equation}
where $L(x,u)$ is a positive function representing the instantaneous cost. We assume that $L$ is continuous and positively homogeneous in the control variables when appropriate.

The optimal control problem consists in minimizing $\mathcal{C}$ among all admissible trajectories connecting given endpoints.

\subsection{Pontryagin's Maximum Principle}

A systematic method to find the solutions of the optimal control problem is Pontryagin’s Maximum Principle (PMP) \cite{PMP}. 
One introduces the costate variables $p_i \in T_x^\ast M$ and define the Pontryagin Hamiltonian\footnote{In general one has to introduce an additional costate variable $p_0$ conjugate to the cost functional, and distinguish the normal case $p_0\neq 0$ from the abnormal case $p_0=0$. We will only consider the normal case, in which one can always use the normalization $p_0=-1$.}
\begin{equation}\label{Pontryagin}
    H(x,p,u) = p_i f^i(x,u) - L(x,u).
\end{equation}

Pontryagin's Maximum Principle states that an optimal trajectory $(x(t),u(t))$ admits a costate $p(t)$ such that
\begin{align} \label{eq:HamiltonPMP}
    \dot{x}^i &= \frac{\partial H}{\partial p_i} = f^i(x,u),\\
    \dot{p}_i &= -\frac{\partial H}{\partial x^i},
\end{align}
and the optimal control $u(t)$ maximizes the Hamiltonian:
\begin{equation}
    H(x(t),p(t),u(t)) = \max_{u \in U_x} H(x(t),p(t),u).
\end{equation}
In the autonomous case, the Hamiltonian is conserved along optimal trajectories. When the final time is a free parameter, one has the  transversality condition that the optimal trajectories lie on the $H=0$ surface.

The PMP reduces the optimal control problem to a pointwise maximization over controls, which
under suitable convexity assumptions on the control set and cost, defines an effective Hamiltonian
\begin{equation}
    \mathcal{H}(x,p) := \max_{u \in U_x} \left( p_i f^i(x,u) - L(x,u) \right) 
\end{equation}
The effective Hamiltonian encodes the optimal dynamics entirely in terms of $(x,p)$. 

It is convenient to introduce an auxiliary system with coordinates $x^i,t$ and a new time parameter $s$ such that 
\begin{equation}
    \frac{ds}{dt} = L \,.
\end{equation}
In other words, the new time tracks the evolution of the cost function along the trajectory. It can be seen that the equations \eqref{eq:HamiltonPMP} for the optimal trajectory can be obtained from the auxiliary Hamiltonian 
\begin{equation}\label{eq:auxHam}
    \widetilde H = p_i \frac{f^i(x,u)}{L(x,u)} + p_t \frac{1}{L(x,u)} -1 
\end{equation}
with the condition that $p_t = -E$ where $E$ is the constant value of $H$ along the trajectory.
The new Hamiltonian corresponds to a time-optimal problem; indeed comparing with \eqref{Pontryagin} we can see that it is the Pontryagin Hamiltonian for a cost function identically equal to 1 and so the cost is just the total time taken by the trajectory. Since the new final time is now also a free parameter, one should consistently take solutions with the two transversality conditions, which read
\begin{equation}\label{eq:transCond}
    E = 0, \qquad \widetilde H =0.
\end{equation}


\subsection{Induced geometric structure}

We can see now how the Finslerian structure arises. Let us start from the case of the time-optimal problem, i.e. $L=1$. 

The control system defines, at each point $x$, a subset of admissible velocities
\begin{equation}
    S_x = \{ f(x,u) \mid u \in U_x \} \subset T_x M.
\end{equation}
We define the set of possible optimal controls $U_x^*$ as the set of $u \in U_x$ that are maxima  of $H(x,p,u)$ for some $p \in T_x^*M$. Since maximizing the Hamiltonian means maximizing the projection of the velocity along $p$, the corresponding subset of possible optimal velocities $S_x^*$ is contained in the set of longest admissible velocities 
\begin{equation}
      S_x^0 = \{(x,v) \in TM | v \in S_x, \lambda v \notin S_x \forall \lambda>1 \}
\end{equation}
Consider the cone $D_x$ generated by elements of $S_x^0$, and the function 
\begin{equation}
    F: D_x \to \mathbb{R}^+ \,, \qquad F(v) = \lambda \,\, {\rm for} \, v = \lambda v_0, \, v_0 \in S_x^0 \,.
\end{equation}
This is the Finslerian function; it is positively homogeneous of degree 1 by construction. The geometry is sub-Finslerian, because $F$ is defined only on the cone of admissible directions and not on the whole tangent bundle. 

The unit level set
\begin{equation}
    \{ v \in T_x M \mid F(x,v) = 1 \}
\end{equation}
is called the {\it indicatrix} and it contains the set of optimal velocities. Contrary to the case of Riemannian geometry, the indicatrix does not have to be topologically a sphere around the origin, it does not even need to be closed or connected. 

Going back now to the general case, the optimal control has to maximize the auxiliary Hamiltonian \eqref{eq:auxHam}. The condition $\partial \widetilde H/\partial u =0$, together with $\widetilde H=0$ yields
\begin{equation}\label{eq:optimalc}
    \frac{\partial}{\partial u} (p_i f^i(x,u)-L(x,u)) = 0\,.
\end{equation}
The condition is now no longer homogeneous in $p$, so it does not depend only on the ray.
It is convenient to work with the rescaled velocities which appear in the auxiliary Hamiltonian \eqref{eq:auxHam}
\begin{equation}\label{eq:rescaled-v}
    v^i = \frac{f^i}{L} \,.
\end{equation}
Since the admissible velocities are constrained, we can solve for them in terms of a subset of independent ones $v^\alpha$. 

\subsection{Final time and terminal  cost}
The formulas of the previous subsection are of course reminiscent of classical mechanics. If we interpret the cost function as a Lagrangian, then the complexity of a state is the minimal action of a trajectory that arrives at that state at the time $t_f$. However we have considered the case where the final time is free, and so it is one of the parameters over which we minimize. If we consider the case of a free particle moving on a manifold, with $L_{free} = \frac12 g_{ij} \dot x^i \dot x^j$, since the energy is conserved, the final time of a trajectory with energy $E$ will be $t_f \sim 1/\sqrt{E}$ and the action is $S = t_f E \sim \sqrt{E}$. This is minimized by taking $E=0$ and it does not give a useful notion of complexity. One solution is to take the geodesic length, which is finite and independent of the energy, however the price is that the control problem becomes more complicated. Alternatively, we can introduce an additional {\it terminal cost} \cite{Liberzon:2012} which could be in general an arbitrary function of the final time, but we will consider the case where it is just a constant shift $(\tc)$ of the cost function: $L = L_{free}+ \tc $. The equations of motion are unaffected, so the action becomes $S = t_f (\tc+E)$ and its minimization with respect to $t_f$ gives $E = \tc$, and $S\sim \sqrt{\tc}$. The action is then equal to the geodesic length up to a constant multiplicative factor. We will use the terminal cost in the following, but as we will see, when the geometry is not Riemannian, the dependence on $\tc$ does not have to be simply an overall factor. 

\section{Finslerian geometry for a single qubit dynamics} \label{sec:SingleQubit}

In this section we give a concrete illustration of the procedure explained in the previous section, in the case of the Lindbladian evolution of a single qubit.  We parametrize the state of the qubit by the coordinates $x^i = {\rm tr}(\rho \sigma_i)$, with $\sigma_i$ the Pauli matrices. 

The Lindbladian equation will take the following general form:
\begin{equation}\label{eq:EoM}
  \dot x^i =   f^i(x,u) = A^i_\alpha(x) u^\alpha + \Delta^i(x) \,. \qquad i=1,2,\ldots, n, \, \alpha=1,2, \ldots, n-1 \,. \qquad n =3.
\end{equation}
We distinguish between the controls $u^\alpha$ and the drift terms $\Delta^i$ that cannot be controlled. 
In a three-dimensional space we need to consider two controls, as with three controls we can span the full tangent space of velocities, and with one control the trajectories are determined and we cannot find a non-trivial geometry. 
For the cost function we take a quadratic function of the controls, plus the terminal cost, identified with the energy as explained above: 
\begin{equation}\label{eq:CostFunction}
    L(u) = \tc+ \frac12 g_{\alpha\beta} u^\alpha u^\beta \,.
\end{equation}
Since there are two controls for three variables, the velocities obey one constraint, i.e. there is one disallowed direction. Given a covector $y_i$ such that $y_i A^i_\alpha=0$, the constraint on the admissible velocities is 
\begin{equation}
    y_i (\dot x^i - \Delta^i) =0 \,.
\end{equation}
The solution of the optimal control equation \eqref{eq:optimalc} is 
\begin{equation}
    u_*^\alpha = g^{\alpha\beta} A_\beta^i \, p_i \,,
\end{equation}
Using $\widetilde H=0$, for the free final-time problem, we find that on the optimal solution locus we have 
\begin{equation}\label{eq:Lopt}
    L(u_*)  = 2 \tc - p_i \Delta^i \,,
\end{equation}
and the momenta have to obey  the following quadratic equation:
\begin{equation}\label{eq:momentum-const}
    \frac12 g^{\alpha\beta} A_\alpha^i A_\beta^j \, p_i p_j + p_i \Delta^i = \tc \,.
\end{equation}

Let us consider first the case without drift: $\Delta^i=0$. In this case, if we decompose $p_i = \alpha y_i + \widetilde p_i$, we see that the component along $y$ decouples from eq. \eqref{eq:momentum-const}. The momentum is then effectively bidimensional, and the quadratic equation determines a line in the space of $\widetilde p$. This in term determines a line in the space of controls, and a corresponding line in the space of allowed velocities, which is the indicatrix. 

We can also determine the indicatrix by first solving for the controls in terms of the velocities, by finding a left-inverse of $A^i_\alpha$: $B^\alpha_{~i} A^{i}_{\beta} = \delta^\alpha_\beta$. Then $u^\alpha= L B^\alpha_{~i} v^i$, and 
\begin{equation}
    L = \tc+ \frac12 L^2  \Gamma_{ij} \, v^i v^j \,,
\end{equation}
where
\[\Gamma_{ij}=g_{\alpha\beta} B^\alpha_{~i} B^\beta_{~j} \,.\] 

Since we have found in \eqref{eq:Lopt} that on the indicatrix 
$L_* = 2\tc$, we have the equation for the indicatrix 
\begin{equation}
     \Gamma_{ij} v_*^i v_*^j = \frac{1}{2 \tc} \,.
\end{equation}
The Finslerian function is determined for each allowed vector $v$ by finding the rescaling that brings it to the optimal line, $v = \lambda v_*$, and equating $F$ with $\lambda$. This implies that 
\begin{equation}\label{eq:FInsler-Riemann}
    F(v) = \sqrt{
    2\tc~ \Gamma_{ij} v^i v^j }.
\end{equation}
In this case the Finslerian function takes the quadratic form \eqref{eq:quadratic-finsler}, so the geometry is in fact sub-Riemannian. 

Notice that the left-inverse is not unique, as it can be shifted by $y_i$, but the Finslerian function is well defined on the admissible velocities. In the case without drift the admissibility constraint is linear, so all the velocities in the cone generated by the indicatrix are admissible.

Let us now consider the case with non-zero drift, $\Delta^i \neq 0$. In this case the component of $p$ along $y$ does not decouple, and eq. \eqref{eq:momentum-const} can be used to determine it. Therefore the space of momenta is bidimensional, and for every $\widetilde p$ there is a corresponding optimal velocity. Every ray emanating from the origin in velocity space will contain a single admissible velocity, so in this case the indicatrix is just given by the admissible velocities, which form an affine plane (actually a half-plane). 
Consider the admissibility constraint for the rescaled velocities: $L y_i v^i = y_i \Delta^i$. From this we read that on the indicatrix
\begin{equation}\label{eq:lambdat}
L_* = \frac{y \cdot \Delta}{y \cdot v} \equiv \frac{1}{\lambda_t(v)} \,.    
\end{equation}
If we now again solve for the controls in terms of the velocities, $u^\alpha=B^\alpha_{~i} (L v^i - \Delta^i)$, and plugging this in the cost function, we have  
\begin{equation}
    L(v) = \tc+  \frac12 \Gamma_{ij} (L^2 v^i v^j -2 L \Delta^i v^j+ \Delta^i \Delta^j). 
\end{equation}
This equation is satisfied on the indicatrix. As before, we write a generic vector $v=\lambda v_*$, with $v_*$ in the indicatrix; this determines the Finslerian function as
\begin{equation}\label{eq:Finsler-with-drift}
    F(v) = \tc \lambda_t(v) + \frac12 \frac{\Gamma_{ij}}{\lambda_t(v)} (v^i - \lambda_t(v) \Delta^i)(v^j - \lambda_t(v) \Delta^j) \,.
\end{equation}

The geometry is now genuinely sub-Finslerian. Notice that 
the dependence on 
$C_t$ is not simply as a prefactor. 
One can also rewrite \eqref{eq:Finsler-with-drift} as
\begin{equation}\label{eq:Finsler-with-drift-v2}
    F(v) = \tc \lambda_t(v) + \frac12 \frac{\widetilde{\Gamma}_{ij}}{\lambda_t(v)} v^i v^j \,,
\end{equation}
where
\begin{equation}
    \widetilde{\Gamma}_{ij} = \Gamma_{kl} \left( \delta^k_i - \frac{y_i \Delta^k}{y \cdot \Delta} \right) \left( \delta^l_j - \frac{y_j \Delta^l}{y \cdot \Delta} \right).
\end{equation}


We still need to consider a special case: there can be a singular locus $X_s$ in the space of states, where $y(x)\cdot \Delta(x)=0, x\in X_s$. The admissibility condition on $X_s$ degenerates to $y \cdot v=0$, so the admissible velocities form a plane orthogonal to $y$. The length of the optimal velocities is not fixed and has to be determined by minimizing the cost. Equivalently, $\lambda_t$ is not determined by \eqref{eq:lambdat}, instead it can be considered as a free parameter over which to minimize the cost \eqref{eq:Finsler-with-drift}. After the minimization we find 
\begin{equation}
\label{eq: genFins}
    F(v) = -\Gamma_{ij}\Delta^i v^j + \sqrt{ 2\Gamma_{ij}v^i v^j \left( \tc +\frac{1}{2}\Gamma_{ij}\Delta^i\Delta^j \right)}.
\end{equation}


Equation \eqref{eq: genFins} describes the cost on the singular locus. Away from this locus, the admissibility constraint fixes the reparametrization factor, so the speed associated with a projected velocity direction is fixed. On the singular locus, the constraint becomes homogeneous: the admissible velocities are restricted to the projected plane, while the reparametrization factor becomes a free lifting variable. The singular Finslerian function is therefore obtained by optimizing over this remaining freedom. It need not agree with the ordinary limit of the regular Finslerian function. Dynamically, the singular locus behaves as a switching surface: the trajectory in state space can remain continuous, while the optimal speed or lifted control branch changes discontinuously. However, the effect is only practically relevant if the trajectory reaches the singular locus in finite time and then continues to move within it, rather than merely passing through it instantaneously.

Equation \eqref{eq: genFins} is also valid in the fully actuated case i.e., the number of independent controls equals the dimension of the state space. In the underactuated system above, where the independent controls are fewer than the state space dimension, it appears only on the singular stratum, where the admissibility constraint degenerates. When the number of controls equals the number of state variables and the control matrix is invertible, no such constraint fixes the reparametrization factor. Every velocity direction can be generated by an appropriate choice of controls, so this factor remains free everywhere. Optimizing over it gives the same functional form as Eq. \eqref{eq: genFins}. Thus, in the fully actuated case, Eq. \eqref{eq: genFins} is not a singular correction, but the general Finslerian function induced by a quadratic control cost in the presence of drift.

In conclusion, eqs. \eqref{eq:FInsler-Riemann}, \eqref{eq:Finsler-with-drift},\eqref{eq: genFins} are our main results for the general form of the Finslerian function in the case of optimal control problems described by the equations of motion \eqref{eq:EoM} and the cost functional \eqref{eq:CostFunction}.

\section{Examples} \label{sec:Examples}

We now illustrate the general formalism developed in the previous section by applying to the examples of the processes described at the end of section \ref{sec:DynamicsOpen}. In all those cases, there is one non-unitary control parameter denoted by $\gamma$. We also include a unitary part in the evolution, of the form $H = \frac{\omega}{2} \sigma_z$, which is a controllable rotation around the z-axis. The control parameters are then $u^\alpha=\{\gamma,\omega\}$. As in the general discussion, we can also add an uncontrolled drift, which can come either from the Hamiltonian or from the Lindbladian operator. 

For all the examples, we take the total cost to be a quadratic function of the controls:
\begin{equation}
L
=
C_t
+
p_\gamma\gamma^2
+
p_{\gamma\omega}\gamma\omega
+
p_\omega\omega^2 .
\label{eq:cost-function}
\end{equation}
For positivity of the control cost, we assume
\begin{equation}
p_\gamma>0,
\qquad
p_\omega>0,
\qquad
p_{\gamma\omega}^2<4p_\gamma p_\omega .
\end{equation}
Once the controls are reconstructed from the rescaled velocity \(v^i=\dot x^i/L\), the control-dependent part of the cost always defines a quadratic form
\begin{equation}
G(x,v)
:=
p_\gamma A(x,v)^2
+p_{\gamma\omega}A(x,v)B(x,v)
+p_\omega B(x,v)^2
=
\frac12 \widetilde{\Gamma}_{ij}(x)v^iv^j ,
\label{eq:quadratic-cost-function}
\end{equation}
where \(A\) and \(B\) denote the reconstructed depolarizing and rotational controls, up to the common scaling factor. Therefore, in the driftless cases the reduced cost is simply
\begin{equation}
\widetilde L(v)=\tc+G(x,v).
\label{eq:cost-function-form}
\end{equation}
When a drift is present, the admissibility condition instead fixes a scaling function \(\lambda_t(v)\). The corresponding conic Finsler function takes the universal form as in \eqref{eq:Finsler-with-drift-v2} 
%
%
Thus, in each example it is enough to specify the admissibility condition, the functions \(A,B\) or \(G\), and \(\lambda_t\). 

In the examples, we compute the indicatrices, curvatures and, for the depolarizing case with unitary drift, the optimal trajectories. It is straightforward to get the trajectories for all the other cases numerically, following similar steps.


\subsection{Example 1: depolarizing dynamics without drift}
\label{sec:depol_nodrift}
For the depolarizing channel, we take the Lindblad operators to be
\begin{equation}
    L_1=\sqrt{\frac{\gamma}{2}}\,\sigma_x,\qquad
    L_2=\sqrt{\frac{\gamma}{2}}\,\sigma_y,\qquad
    L_3=\sqrt{\frac{\gamma}{2}}\,\sigma_z .
\end{equation}
In addition, we allow for a controllable unitary rotation around the $z$-axis. At the level of the Lindblad equation this is implemented by the Hamiltonian, $ H=\frac{\omega}{2}\sigma_z $.
The factor of $1/2$ is the usual spin-$1/2$ normalization. Indeed, the corresponding unitary evolution is
\begin{equation}
    U(t)=e^{-iHt}
    =
    e^{-i\frac{\omega t}{2}\sigma_z},
\end{equation}
which induces on the Bloch vector the ordinary rotation
\begin{equation}
\begin{pmatrix}
x(t)\\
y(t)\\
z(t)
\end{pmatrix}
=
\begin{pmatrix}
\cos(\omega t)&-\sin(\omega t)&0\\
\sin(\omega t)&\cos(\omega t)&0\\
0&0&1
\end{pmatrix}
\begin{pmatrix}
x(0)\\
y(0)\\
z(0)
\end{pmatrix}.
\end{equation}
Thus the usual three-dimensional rotation matrix is recovered after exponentiating the Hamiltonian flow. In the master equation, however, we work directly with the infinitesimal generator.

Writing the density matrix as
\begin{equation}
    \rho=\frac{1}{2}\left(I+x\sigma_x+y\sigma_y+z\sigma_z\right),
\end{equation}
the Hamiltonian part $-i[H,\rho]$ gives
\begin{equation}
    \dot{x}=-\omega y,\qquad
    \dot{y}=\omega x,\qquad
    \dot{z}=0 .
\end{equation}
On the other hand, the depolarizing Lindblad operators shrink the Bloch vector isotropically and give
\begin{equation}
    \dot{x}=-2\gamma x,\qquad
    \dot{y}=-2\gamma y,\qquad
    \dot{z}=-2\gamma z .
\end{equation}
Combining the coherent rotation with the depolarizing noise, we obtain
\begin{equation}
\begin{split}
    \dot{x} &=-2\gamma x-\omega y,\\
    \dot{y} &=\omega x-2\gamma y,\\
    \dot{z} &=-2\gamma z .
\end{split}
\label{eq:depol-nodrift-eom}
\end{equation}
Therefore the dynamics consists of a rotation in the $xy$-plane, generated by $H=\omega\sigma_z/2$, together with an isotropic contraction of the Bloch vector due to depolarization.
It is convenient to introduce cylindrical coordinates $(r_\perp,\theta,z)$ where 
\begin{equation}
r_\perp^2:=x^2+y^2.
\end{equation}
Using also \(\gamma=-\dot z/(2z)\), we arrive at the admissibility constraint
\begin{equation}
z \dot r_\perp - r_\perp \dot z = 0 \,.
\label{eq:depol-nodrift-constraint}
\end{equation}
Thus, at each point \((x,y,z)\), the allowed velocities span a two-dimensional plane in the tangent space; imposing \(\gamma\ge0\) restricts this further to a half-plane.

\begin{figure}[htbp]
\centering

\begin{subfigure}[t]{\indicatrixwidth}
    \centering
    \includegraphics[width=\linewidth]{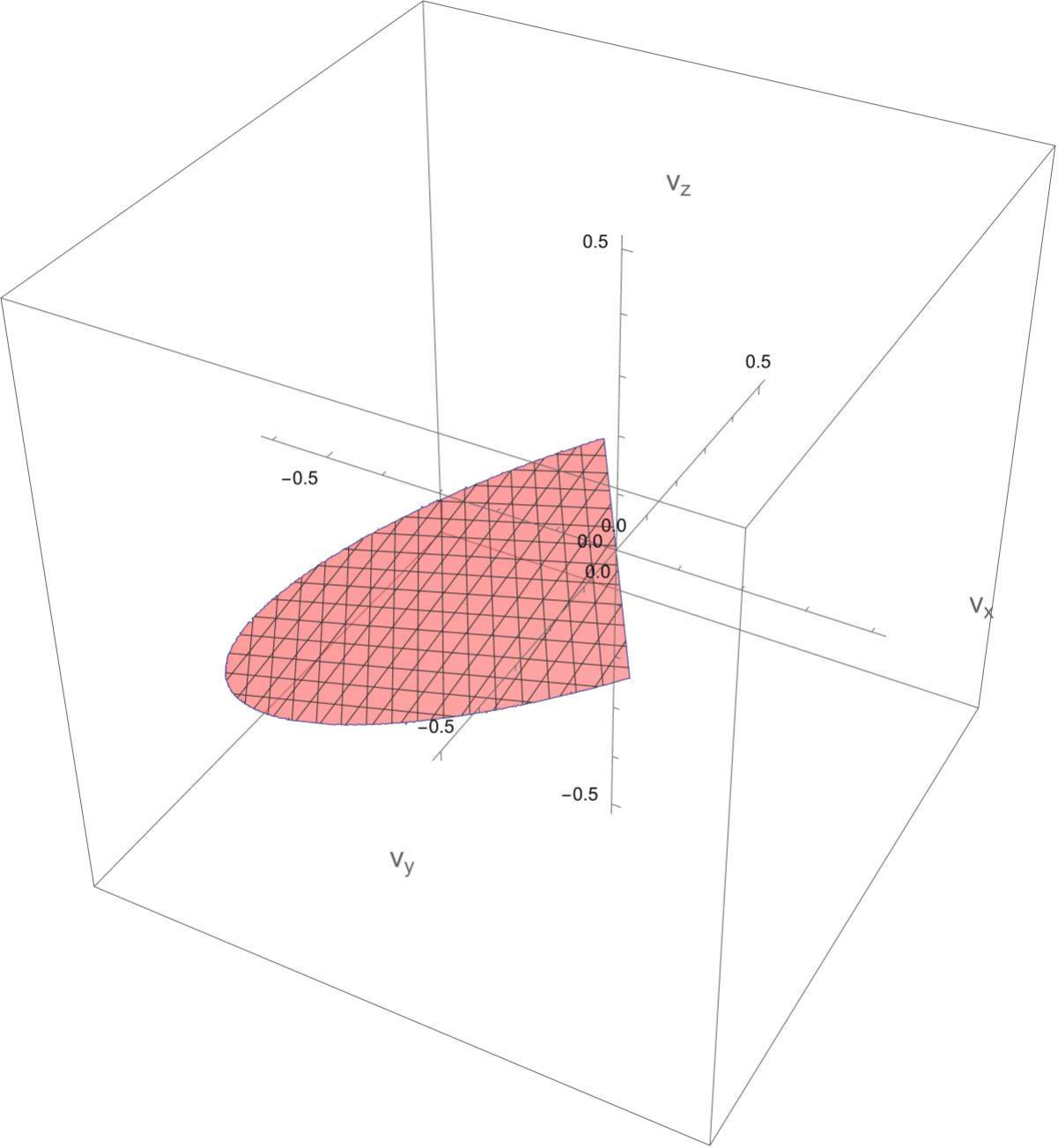}
    \caption{$p_\gamma = 1$, $p_\omega = 1$ and $p_{\gamma\omega} = 0$.}
    \label{fig:indicatrix-depnodrift-1}
\end{subfigure}
\hfill
\begin{subfigure}[t]{\indicatrixwidth}
    \centering
    \includegraphics[width=\linewidth]{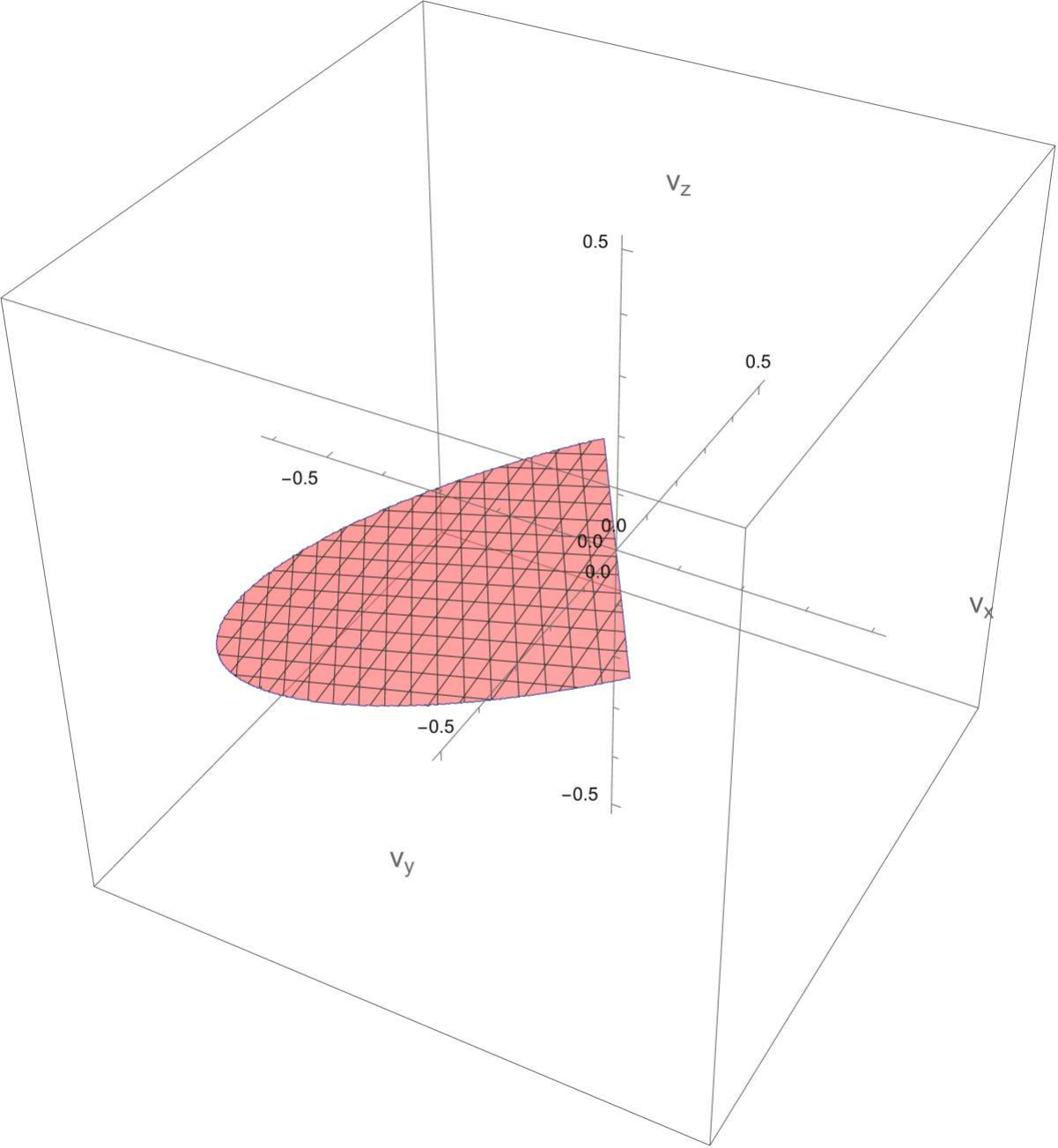}
    \caption{$p_\gamma = 1$, $p_\omega = 1$ and $p_{\gamma\omega} = \frac12 $.}
    \label{fig:indicatrix-depnodrift-2}
\end{subfigure}

\vspace{0.5cm}

\begin{subfigure}[t]{\indicatrixwidth}
    \centering
    \includegraphics[width=\linewidth]{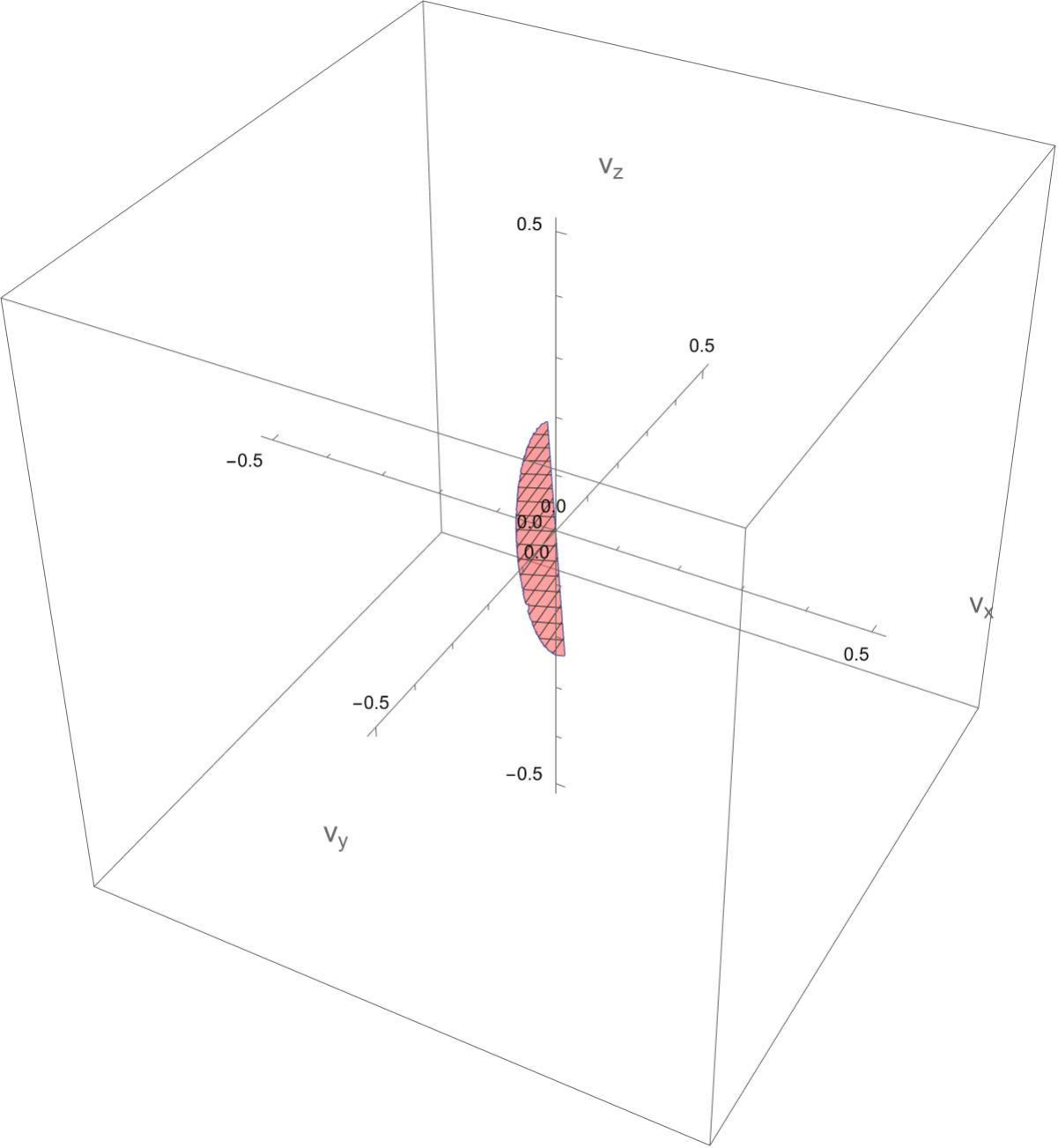}
    \caption{$p_\gamma = 100$, $p_\omega = 1$ and $p_{\gamma\omega} = 0$.}
    \label{fig:indicatrix-depnodrift-3}
\end{subfigure}
\hfill
\begin{subfigure}[t]{\indicatrixwidth}
    \centering
    \includegraphics[width=\linewidth]{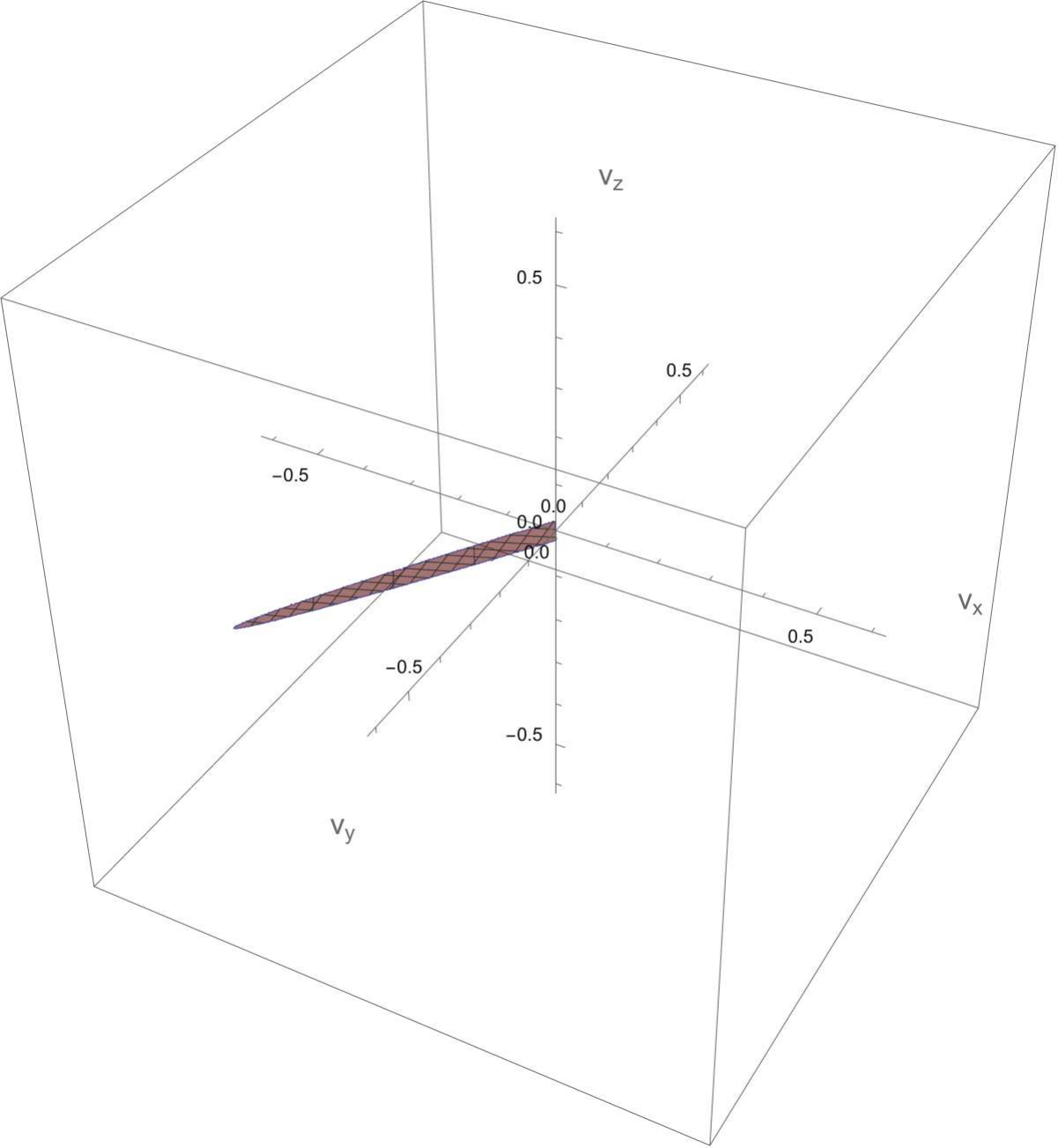}
    \caption{$p_\gamma = 1$, $p_\omega = 100$ and $p_{\gamma\omega} = 0$.}
    \label{fig:indicatrix-depnodrift-4}
\end{subfigure}

    \caption{
    The indicatrices for the depolarizing dynamics without drift at base point \(x=(0.5,0.3,0.2)\) with $C_t = 1$ as boundaries of the surfaces shown. The four plots show how the indicatrix changes with respect to the control-cost-weights \(p_\gamma\), \(p_{\gamma\omega}\), and \(p_\omega\).
    }
    \label{fig:indicatrices-depnodrift}
\end{figure}

To construct the reduced geometry, we introduce the rescaled velocity
\begin{equation}
v^i=\frac{\dot x^i}{L},
\qquad\Longleftrightarrow\qquad
\dot x^i=Lv^i,
\label{eq:depol-nodrift-rescaled-v}
\end{equation}
where \(L\) is the cost function in Eq.~\eqref{eq:cost-function}. In terms of \(v\), the admissibility constraint becomes
\begin{equation}
L \left[z(xv_x+yv_y)-r_\perp^2v_z \right]=0,
\label{eq:depol-nodrift-v-constraint}
\end{equation}
which gives the indicatrix of this dynamics as shown in figure \ref{fig:indicatrices-depnodrift}. 
This shows the indicatrices for the depolarizing dynamics without drift at a fixed base point. Since the system has two controls, $\gamma$ and $\omega$, the admissible velocities form a two-dimensional surface in the three-dimensional tangent space. Changing the cost weights deforms this surface: increasing a weight suppresses the corresponding control direction, while decreasing it makes that direction more favorable. Thus, in plot \ref{fig:indicatrix-depnodrift-3} the velocities are biased toward the rotational $\omega$ direction, while in plot \ref{fig:indicatrix-depnodrift-4} they are biased toward the dissipative $\gamma$ direction. The mixed term $p_{\gamma\omega}$ in plot \ref{fig:indicatrix-depnodrift-2} skews the indicatrix by coupling the two controls.

Substituting \eqref{eq:depol-nodrift-rescaled-v} into \eqref{eq:depol-nodrift-eom} gives
\begin{equation}
\gamma=-L\frac{xv_x+yv_y}{2r_\perp^2},
\qquad
\omega=L\frac{xv_y-yv_x}{r_\perp^2}.
\end{equation}
Hence the cost takes the form
\begin{equation}
L-\tc=L^2\,(\widetilde L(v) - \tc),
\qquad
\widetilde L(v) - \tc=\frac{L-\tc}{L^2},
\end{equation}
with
\begin{equation}
\widetilde L(v)
=
p_\gamma \frac{(xv_x+yv_y)^2}{4r_\perp^4}
-
p_{\gamma \omega} \frac{(xv_x+yv_y)(xv_y-yv_x)}{2r_\perp^4}
+
p_\omega \frac{(xv_y-yv_x)^2}{r_\perp^4}
+ \tc,
\label{eq:depol-nodrift-Ltilde}
\end{equation}
which is in the form as in \eqref{eq:cost-function-form}.
The components of the quadratic form are
\begin{equation}
    \begin{aligned}
        \widetilde{\Gamma}_{xx} &= \frac{ p_\gamma x^2 + 2 p_{\gamma \omega} xy + 4 p_\omega y^2}{2r_\perp^4},\quad
        \widetilde{\Gamma}_{yy}= \frac{p_\gamma y^2 - 2 p_{\gamma \omega} xy + 4 p_\omega x^2}{2r_\perp^4},\\
        \widetilde{\Gamma}_{xy} &=\widetilde{\Gamma}_{yx}=\frac{ p_\gamma xy  - p_{\gamma \omega} (x^2-y^2) - 4 p_\omega xy}{2r_\perp^4},\\
        \widetilde{\Gamma}_{xz}&=\widetilde{\Gamma}_{yz}=\widetilde{\Gamma}_{zz}=0.
    \end{aligned}
    \label{eq:depol-nodrift-metric-components}
\end{equation}
The vanishing of the \(z\)-components reflects the fact that \(v_z\) is not an independent direction, but is fixed by \eqref{eq:depol-nodrift-v-constraint}. The resulting structure is therefore defined on the admissible two-dimensional distribution:

\begin{equation}
F(v)
=
\sqrt{
\frac{\tc}{r_\perp^4}
\Big[p_{\gamma}(xv_x+yv_y)^2- 2p_{\gamma\omega} (xv_x+yv_y)(xv_y-yv_x)  +4p_{\omega}(xv_y-yv_x)^2\Big]
}.
\label{eq:depol-nodrift-finsler-explicit}
\end{equation}
Since the reduced cost is quadratic on the admissible distribution, the resulting geometry is in fact sub-Riemannian. A useful check is that the metric defined by \eqref{eq:depol-nodrift-finsler-explicit} is flat. 
To see this, introduce polar coordinates
\[
x=r_\perp\cos\phi,\qquad y=r_\perp\sin\phi .
\]
The corresponding components of the rescaled velocity \(v\) are
\[
v_{r_\perp}
=
\frac{xv_x+yv_y}{r_\perp},
\qquad
v_\phi
=
\frac{xv_y-yv_x}{r_\perp^2}.
\]
Equivalently,
\[
xv_x+yv_y=r_\perp v_{r_\perp},
\qquad
xv_y-yv_x=r_\perp^2 v_\phi .
\]
Substituting this in \eqref{eq:depol-nodrift-finsler-explicit}, we obtain
\begin{equation}
F^2
=
C_t\left[
\frac{p_\gamma}{r_\perp^2}v_{r_\perp}^2
-\frac{2p_{\gamma\omega}}{r_\perp}v_{r_\perp}v_\phi
+4p_\omega v_\phi^2
\right].
\end{equation}
Now define
\begin{equation}
u=\log r_\perp,
\qquad
v_u=\frac{v_{r_\perp}}{r_\perp}.
\end{equation}
Then
\begin{equation}
F^2
=
C_t\left[
p_\gamma v_u^2
-2p_{\gamma\omega}v_u v_\phi
+4p_\omega v_\phi^2
\right].
\end{equation}
The coefficients are constant in the coordinates $(u,\phi)$. Therefore the induced two-dimensional metric is flat, and the Gaussian curvature, equivalently the flag curvature in this two-dimensional Riemannian case, is
\[
K=0.
\]


\subsection{Example 2: amplitude damping of a single qubit without drift}

Consider the Lindbladian dynamics for an amplitude damping process toward $y = 1$ with unitary rotational control along the $z$-axis with the following Lindbladian operators:
\begin{equation}
    L_1 = \frac{\sqrt{\gamma}}{2} (\sigma_z + i \sigma_x),
\end{equation}
which result in the following Lindblad equations:
\begin{equation}
\begin{split}
    \dot{x} &= -\frac{\gamma}{2} x - \omega y,\\
    \dot{y} &= \omega x -\gamma y + \gamma,\\
    \dot{z} &= -\frac{\gamma}{2}z.
\end{split}
\label{eq:ampdamp-nodrift-eom}
\end{equation}
From the dynamics in \eqref{eq:ampdamp-nodrift-eom} one obtains

\begin{equation}
    \dot r_\perp = -\gamma \frac{ \frac{x^2}{2} + y^2 - y }{ r_\perp }, \qquad \dot \phi = \omega + \frac{\gamma}{r_\perp} \cos\phi - \frac{\gamma}{2} \sin\phi \cos\phi \,.
\end{equation}
We can also find the admissibility constraint as
\begin{equation}
    z \dot r_\perp - \frac{2 \left( \frac{x^2}{2} + y^2 - y \right)}{r_\perp} \dot z = 0 \,.
\label{eq:ampdamp-nodrift-constraint}
\end{equation}
As with the previous example, the allowed velocities span a two-dimensional plane in the tangent space at each point \((x,y,z)\) while the restriction \(\gamma\ge0\) gives us a half-plane.

\begin{figure}[htbp]
\centering

\begin{subfigure}[t]{\indicatrixwidth}
    \centering
    \includegraphics[width=\linewidth]{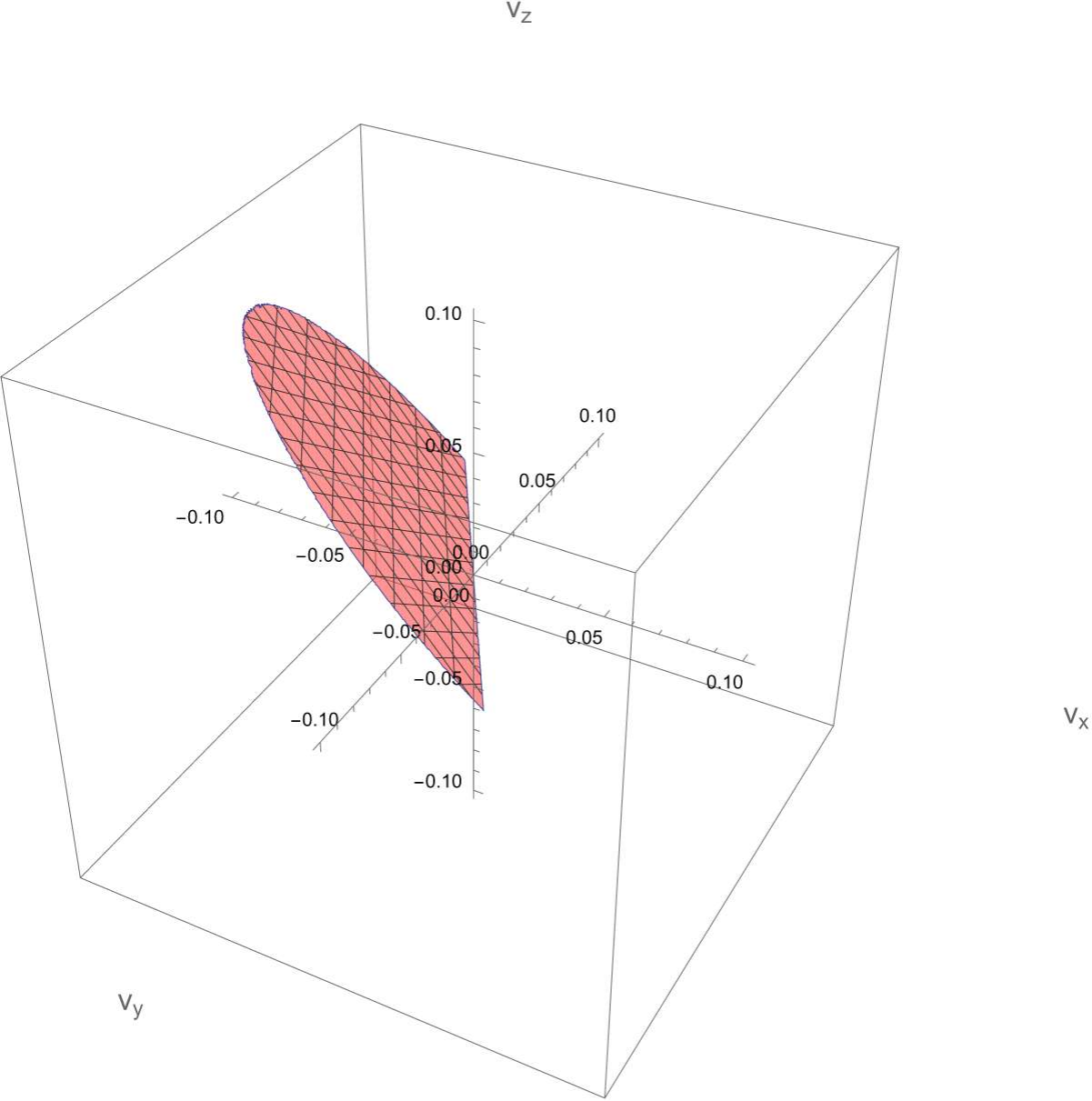}
    \caption{$p_\gamma = 1$, $p_\omega = 1$ and $p_{\gamma\omega} = 0$.}
    \label{fig:indicatrix-ampnodrift-1}
\end{subfigure}
\hfill
\begin{subfigure}[t]{\indicatrixwidth}
    \centering
    \includegraphics[width=\linewidth]{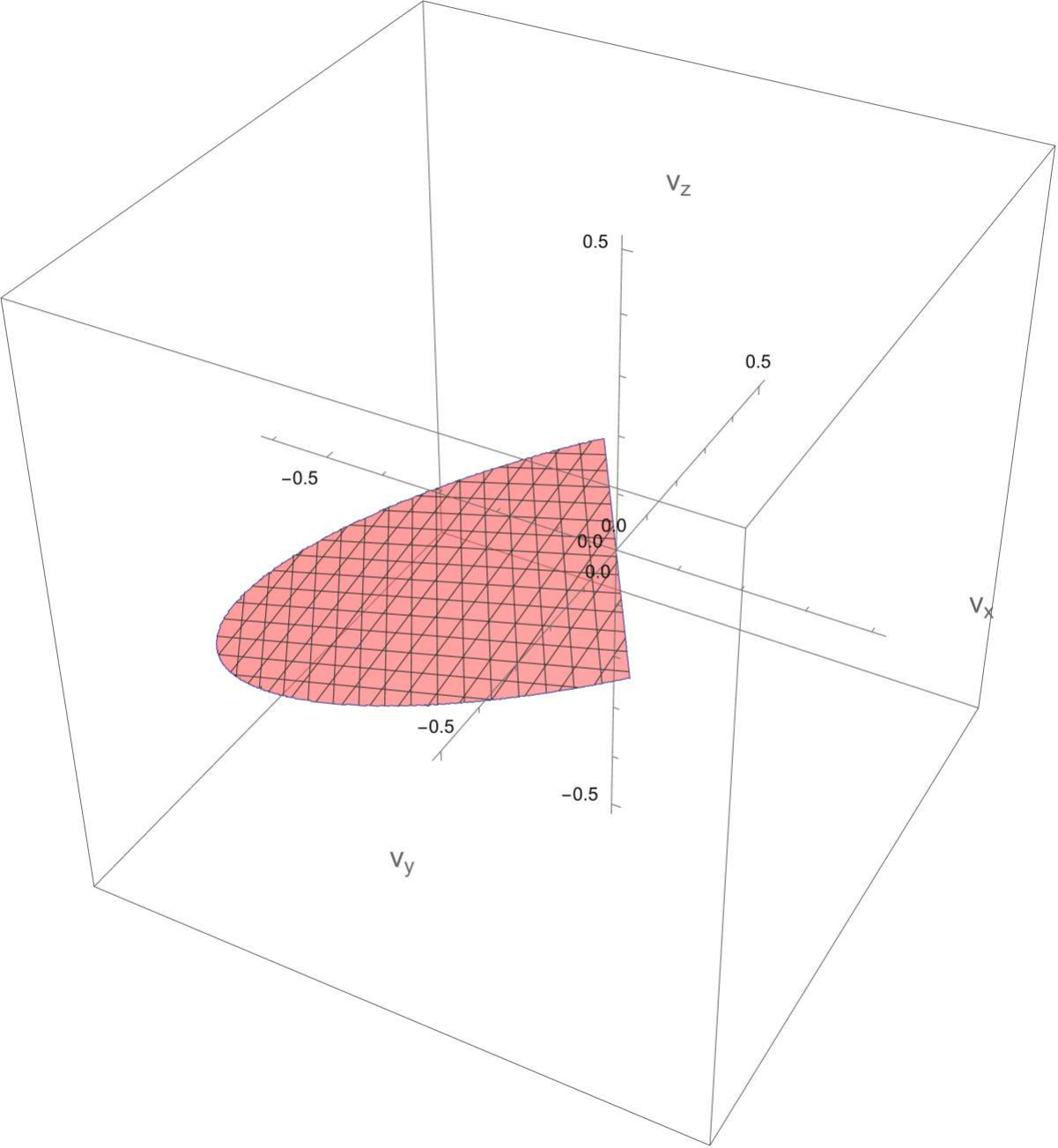}
    \caption{$p_\gamma = 1$, $p_\omega = 1$ and $p_{\gamma\omega} = \frac12 $.}
    \label{fig:indicatrix-ampnodrift-2}
\end{subfigure}

\vspace{0.5cm}

\begin{subfigure}[t]{\indicatrixwidth}
    \centering
    \includegraphics[width=\linewidth]{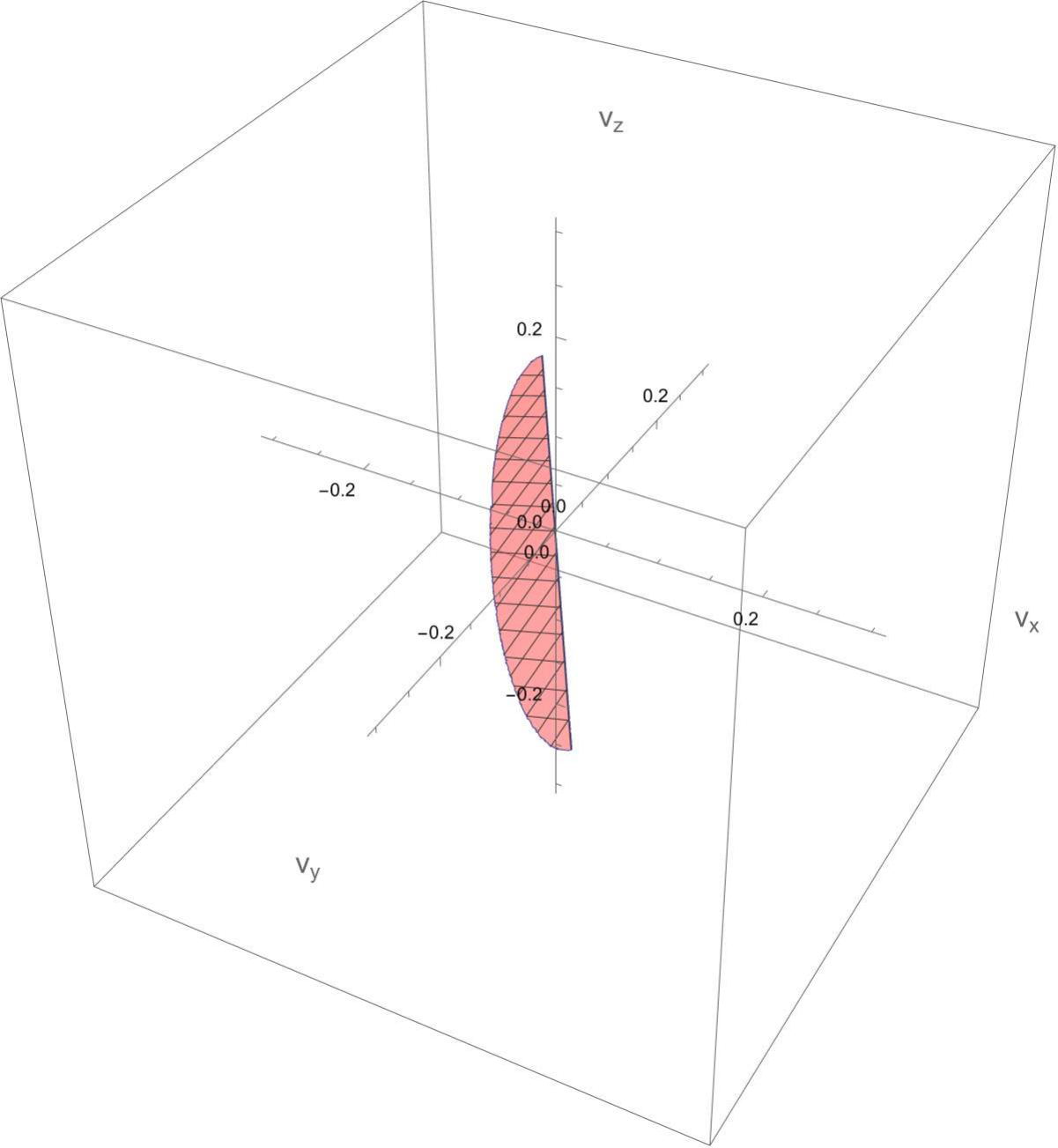}
    \caption{$p_\gamma = 100$, $p_\omega = 1$ and $p_{\gamma\omega} = 0$.}
    \label{fig:indicatrix-ampnodrift-3}
\end{subfigure}
\hfill
\begin{subfigure}[t]{\indicatrixwidth}
    \centering
    \includegraphics[width=\linewidth]{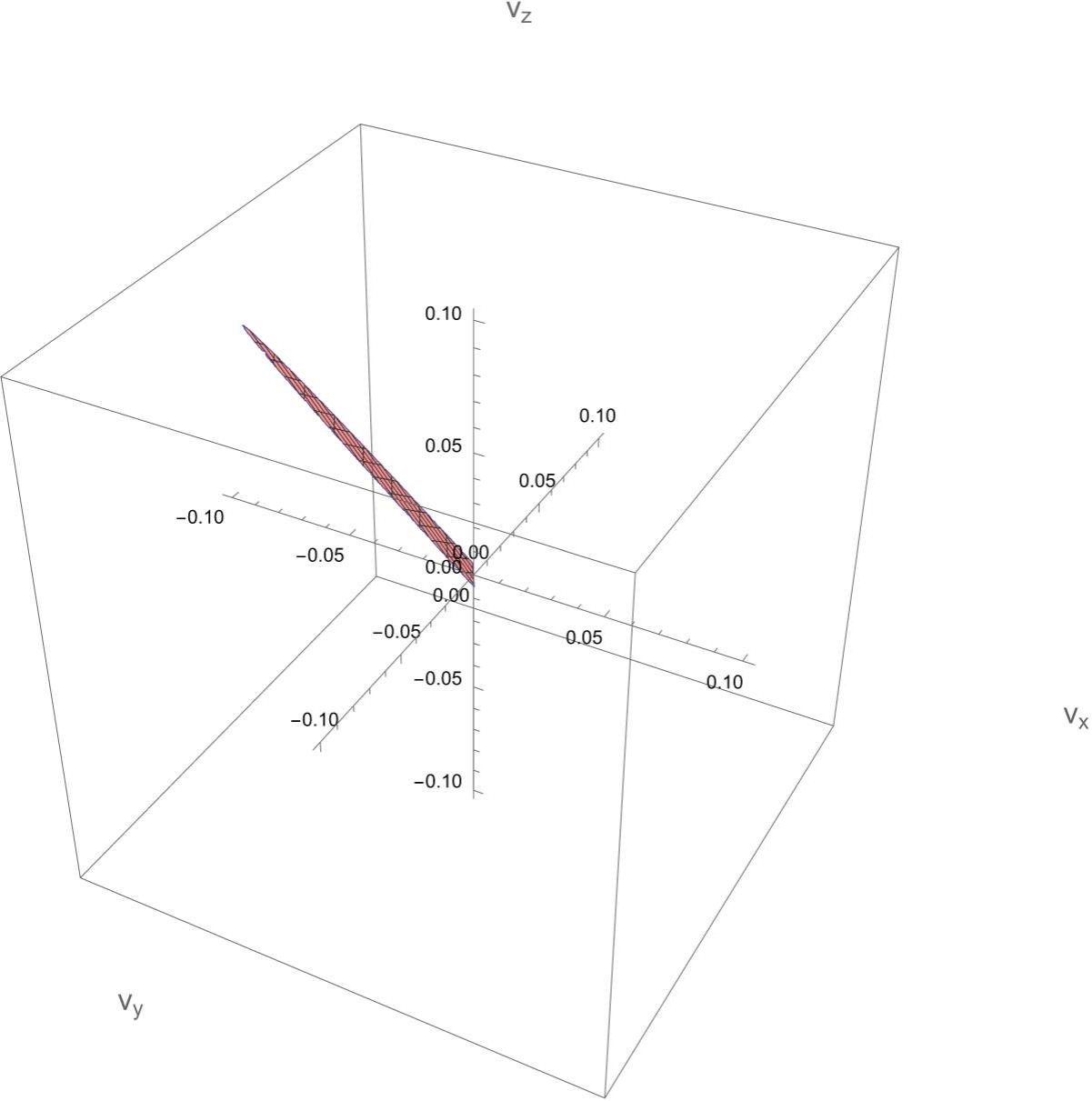}
    \caption{$p_\gamma = 1$, $p_\omega = 100$ and $p_{\gamma\omega} = 0$.}
    \label{fig:indicatrix-ampnodrift-4}
\end{subfigure}

    \caption{
    The indicatrices for the amplitude damping dynamics without drift at base point \(x=(0.5,0.3,0.2)\) with $\tc = 1$ as boundaries of the surfaces shown. The four plots show how the indicatrix changes with respect to the control-cost-weights \(p_\gamma\), \(p_{\gamma\omega}\), and \(p_\omega\).
    }
    
    \label{fig:indicatrices-ampnodrift}
\end{figure}

In terms of the rescaled velocity \(v\), the admissibility constraint becomes
\begin{equation}
    L \left[ z(xv_x+yv_y)-2 \left( \frac{x^2}{2} + y^2 - y \right)v_z \right]=0.
\label{eq:ampdamp-nodrift-v-constraint}
\end{equation}
This gives the indicatrix for the system as in figure \ref{fig:indicatrices-ampnodrift}. For the amplitude-damping case, the indicatrix has the same two-control origin as in the depolarizing example, but its shape is modified by the non-unitary damping flow. The $\gamma$ direction no longer gives a purely radial contraction toward the origin, so the admissible surface is tilted and less symmetric. Increasing $p_\gamma$ in plot \ref{fig:indicatrix-ampnodrift-3} suppresses the damping contribution and leaves a narrower surface dominated by rotational motion, while increasing $p_\omega$ in plot \ref{fig:indicatrix-ampnodrift-4} suppresses rotation and forces the velocities to align more strongly with the amplitude-damping direction. The mixed term in plot \ref{fig:indicatrix-ampnodrift-2} further skews the surface by correlating the two controls.

While the controls are expressed as
\begin{equation}
    \gamma=-L\frac{xv_x+yv_y}{ \frac{x^2}{2} + y^2 - y},
    \qquad
    \omega=L\frac{\frac{x}{2} v_y + (1-y) v_x}{\frac{x^2}{2} + y^2 - y}.
    \label{eq:ampdamp-nodrift-Ltilde}
\end{equation}
Thus, the components of the quadratic form of the parameterized cost function in \eqref{eq:cost-function-form} are
\begin{equation}
    \begin{aligned}
        \widetilde{\Gamma}_{xx} 
        &= 
        \frac{
        2p_\gamma x^2 
        - 2 p_{\gamma \omega } x (1-y) 
        + 2p_\omega (1-y)^2
        }{
        \left(\frac{x^2}{2}+y^2-y\right)^2
        },
        \\
        \widetilde{\Gamma}_{yy} 
        &= 
        \frac{
        2p_\gamma y^2 
        - p_{\gamma \omega} xy 
        + p_\omega \frac{x^2}{2}
        }{
        \left(\frac{x^2}{2}+y^2-y\right)^2
        },
        \\
        \widetilde{\Gamma}_{xy}
        &=
        \widetilde{\Gamma}_{yx}
        =
        \frac{
        2p_\gamma xy 
        - p_{\gamma \omega} \left( \frac{x^2}{2} + y-y^2 \right) 
        + p_\omega x (1-y)
        }{
        \left(\frac{x^2}{2}+y^2-y\right)^2
        },
        \\
        \widetilde{\Gamma}_{xz}
        &=
        \widetilde{\Gamma}_{yz}
        =
        \widetilde{\Gamma}_{zz}=0.
    \end{aligned}
    \label{eq:ampdamp-nodrift-metric-components}
\end{equation}
As in the previous example, the \(z\)-components vanish since \(v_z\) is not an independent direction. The structure is also defined on the admissible two-dimensional distribution.

Using \eqref{eq:ampdamp-nodrift-Ltilde}, this can be written explicitly as

\begin{equation}
F(v)
= 2
\sqrt{
\tc
\frac{
p_\gamma (xv_x+yv_y)^2
- p_{\gamma\omega} (xv_x+yv_y) 
\left[\frac{x}{2} v_y + (1-y) v_x\right]
+ p_\omega\left[\frac{x}{2} v_y + (1-y) v_x\right]^2
}{
\left(  \frac{x^2}{2} + y^2 - y \right)^2
}
}.
\label{eq:ampdampnodrift-finsler-explicit}
\end{equation}
As expected, the reduced cost is quadratic on the admissible distribution, the resulting geometry is in fact sub-Riemannian.

Compared with the depolarizing channel, amplitude damping introduces motion toward the ground state. This shifts the singular structure of the reduced geometry: the denominator appearing in the reconstructed controls now depends not only on the radial coordinate but also on the displacement from the damping fixed point. The admissible velocities still form a two-dimensional distribution, but the geometry is no longer simply radial. The amplitude-damping geometry has nonzero curvature in general, although the fully general expression is rather long and not very illuminating. To exhibit the effect explicitly, let us consider the simple choice
\[
\tc=1,\qquad p_\gamma=1,\qquad p_{\gamma\omega}=0,\qquad p_\omega=1 .
\]
For this choice, the Gaussian curvature, equivalently the flag curvature in this two-dimensional Riemannian case, is
\[
K(x,y)
=
\frac{
7x^4
-12x^2y^2
+28x^2y
-16x^2
-20y^4
+72y^3
-116y^2
+96y
-32
}{
16\left(x^2+2y^2-2y\right)^2
}.
\]
This expression is showing that the amplitude-damping cost geometry is genuinely curved. This should be contrasted with the depolarizing case, where a simple change of variables makes the metric components constant and hence gives vanishing curvature.

\subsection{Example 3: depolarizing dynamics with unitary drift}
\label{sec:depol_drift}

We now add an uncontrolled unitary drift. Geometrically, this means that the admissible velocity set is no longer centred around the origin in each tangent space and instead, the controllable directions are translated by the drift vector field.
This is done by including a Hamiltonian drift term to the depolarizing dynamics $H = H_0 + H_\Delta$, where
\begin{equation}
    H_\Delta = \frac{\Delta}{2} \sigma_y,
\end{equation}
which is a rotational drift in the y-axis.
The equations of motion are
\begin{eqnarray}
\dot{x}&=&-2\gamma x-\omega y+\Delta z,\nonumber\\
\dot{y}&=&\omega x-2\gamma y,\\
\dot{z}&=&-\Delta x-2\gamma z.\nonumber
\label{eq:depol-drift-eom}
\end{eqnarray}
Eliminating \(\gamma\) gives
\begin{equation}
\gamma
=
-\frac{x\dot{x}+y\dot{y}-\Delta xz}{2r_\perp^2}
=
-\frac{\dot z+\Delta x}{2z}.
\end{equation}
Therefore the admissibility condition is
\begin{equation}
z(x\dot{x}+y\dot{y})
-r_\perp^2\dot z
-\Delta x(r_\perp^2+z^2)
=0 .
\label{eq:depol-drift-constraint}
\end{equation}
For fixed \((x,y,z)\), this is an affine plane in velocity space. The condition
\(\gamma\ge 0\) further restricts this affine plane to a half-plane.

By rescaling the velocities, \eqref{eq:depol-drift-constraint} gives
\begin{equation}
L\Big[z(xv_x+yv_y)-r_\perp^2v_z\Big]
-\Delta x(r_\perp^2+z^2)=0 .
\end{equation}
Hence
\begin{equation}
L=\frac{1}{\lambda_t(v)},
\qquad
\lambda_t(v):=
\frac{
z(xv_x+yv_y)-r_\perp^2v_z
}
{
\Delta x(r_\perp^2+z^2)
}.
\label{eq:depol-drift-lambdat}
\end{equation}
The physical branch corresponds to \(\lambda_t(v)>0\), so that \(L>0\).
This expression is written in the local patch \(\Delta x\neq 0\).

\begin{figure}[htbp]
\centering

\begin{subfigure}[t]{\indicatrixwidth}
    \centering
    \includegraphics[width=\linewidth]{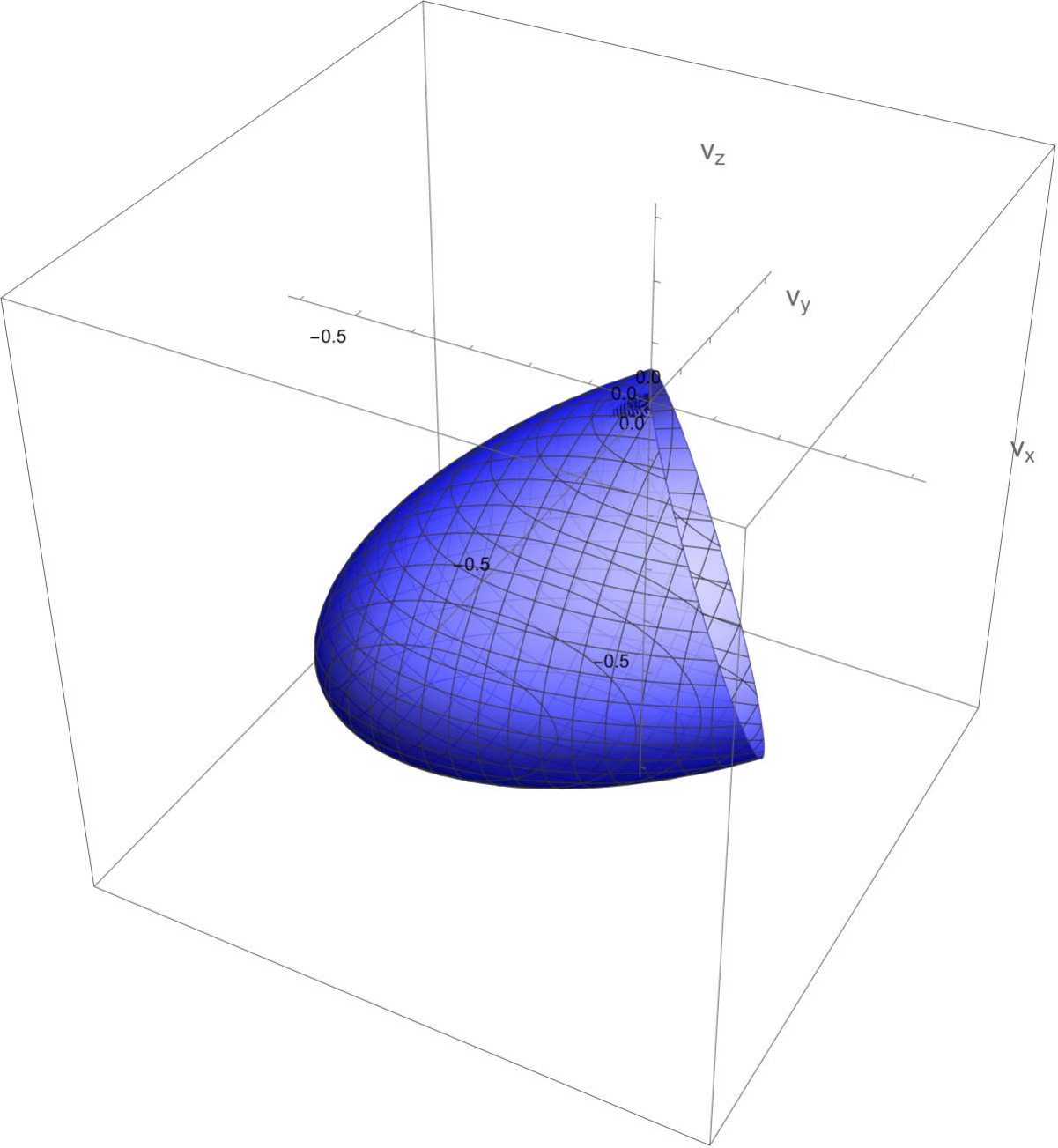}
    \caption{$p_\gamma = 1$, $p_\omega = 1$ and $p_{\gamma\omega} = 0$.}
    \label{fig:indicatrix-depuni-1}
\end{subfigure}
\hfill
\begin{subfigure}[t]{\indicatrixwidth}
    \centering
    \includegraphics[width=\linewidth]{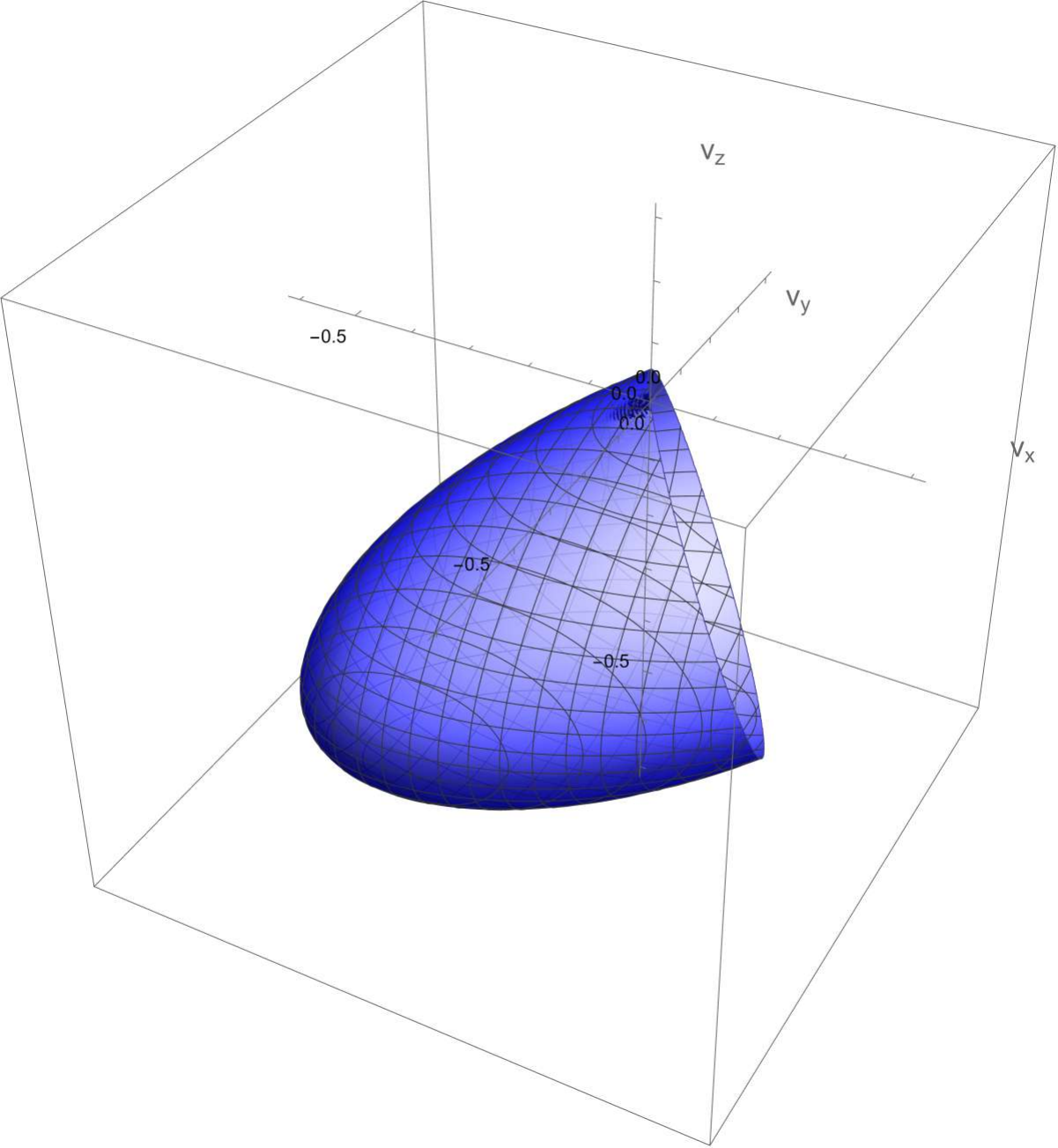}
    \caption{$p_\gamma = 1$, $p_\omega = 1$ and $p_{\gamma\omega} = \frac12 $.}
    \label{fig:indicatrix-depuni-2}
\end{subfigure}

\vspace{0.5cm}

\begin{subfigure}[t]{\indicatrixwidth}
    \centering
    \includegraphics[width=\linewidth]{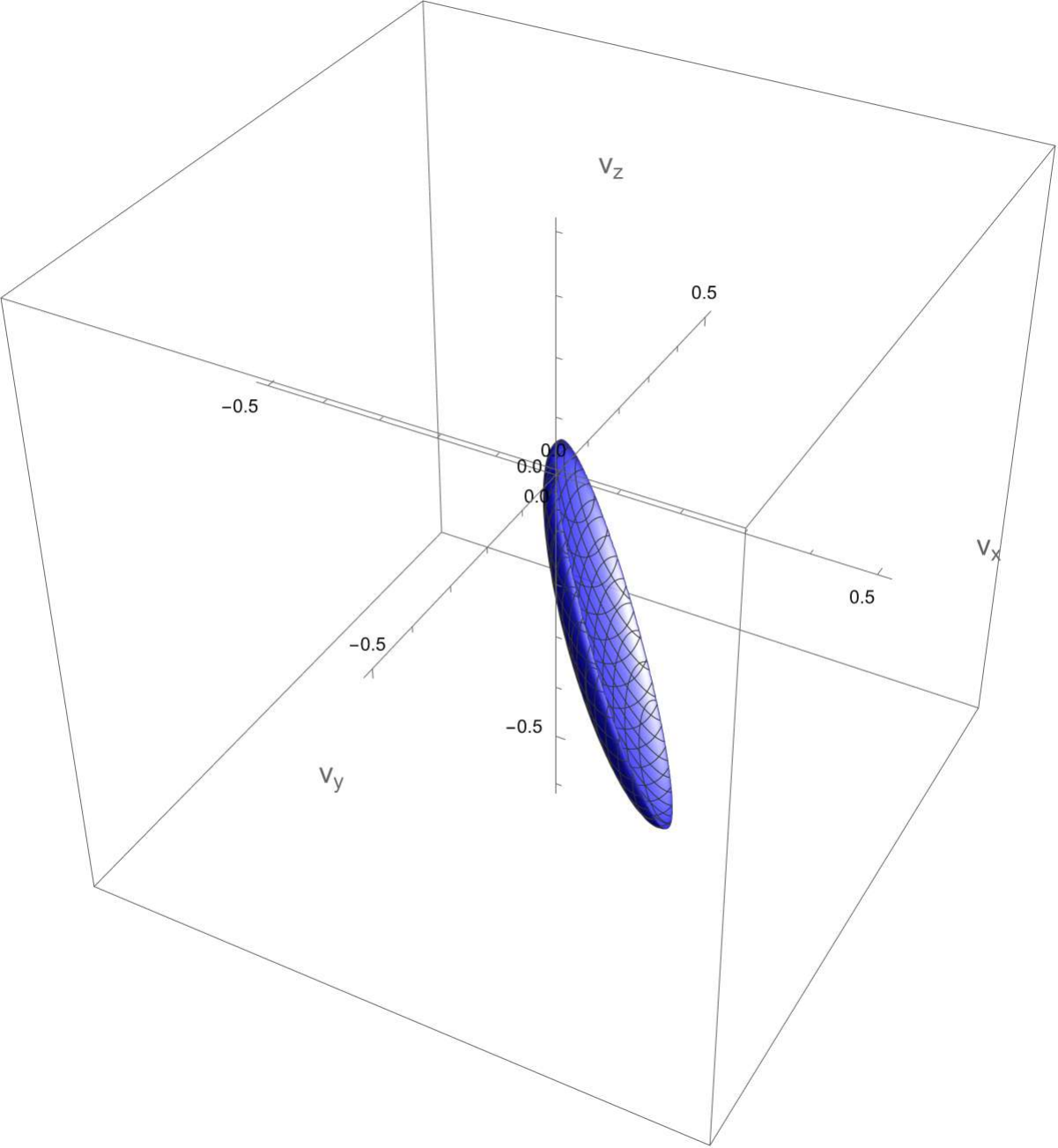}
    \caption{$p_\gamma = 100$, $p_\omega = 1$ and $p_{\gamma\omega} = 0$.}
    \label{fig:indicatrix-depuni-3}
\end{subfigure}
\hfill
\begin{subfigure}[t]{\indicatrixwidth}
    \centering
    \includegraphics[width=\linewidth]{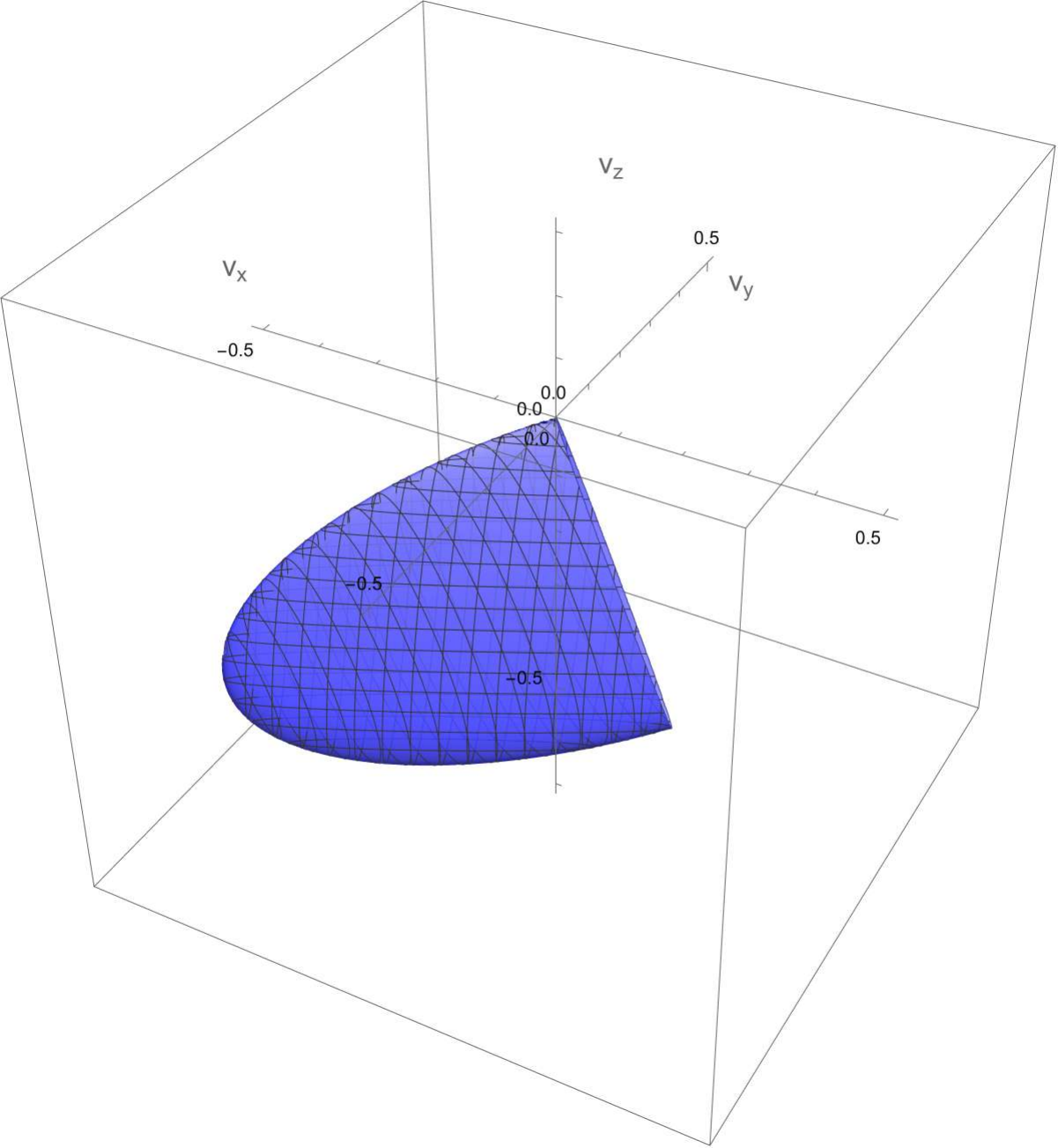}
    \caption{$p_\gamma = 1$, $p_\omega = 100$ and $p_{\gamma\omega} = 0$.}
    \label{fig:indicatrix-depuni-4}
\end{subfigure}

    \caption{
    The indicatrices for the depolarizing dynamics with unitary drift at base point \(x=(0.5,0.3,0.2)\) with $C_t = 1$ and $\Delta = 1$ as surfaces shown. The four plots show how the indicatrix changes with respect to the control-cost-weights \(p_\gamma\), \(p_{\gamma\omega}\), and \(p_\omega\).
    }
    \label{fig:indicatrices-depuni}
\end{figure}

The controls can be solved in terms of the physical velocity. From
\eqref{eq:depol-drift-eom}, one finds
\begin{equation}
\gamma
=
-\frac{x\dot x+y\dot y+z\dot z}
{2(r_\perp^2+z^2)}, \qquad
\omega
=
\frac{
(x^2+z^2)\dot y
-xy\dot x
-yz\dot z
}
{x(r_\perp^2+z^2)} .
\label{eq:depol-drift-velocity}
\end{equation}
%
Using rescaling \(\dot x^i=L v^i\), the cost in \eqref{eq:cost-function-form} becomes
\begin{equation}
L
=
\tc+L^2G(v).
\label{eq:depol-drift-cost-rescaled}
\end{equation}
Together with \(L=1/\lambda_t(v)\), this gives the indicatrix equation
\begin{equation}
\frac{1}{\lambda_t(v)}
=
\tc+\frac{G(v)}{\lambda_t(v)^2},
\label{eq:depol-drift-indicatrix}
\end{equation}
This gives the indicatrix for the system as in figure \ref{fig:indicatrices-depuni}. With the unitary drift, the indicatrix changes from the flat semi-elliptic surface of the driftless depolarizing case into a curved, hollow semi-ellipsoid-like surface. The fixed $\Delta$ term adds a background velocity, shifting the admissible unit-cost velocities out of the original control plane and making the surface genuinely three-dimensional. This also changes the extreme-weight behavior: unlike in the driftless case, where strongly penalizing one control makes the indicatrix collapse toward an almost one-dimensional curve, here the background drift keeps the admissible velocities effectively two-dimensional even when one controlled direction is suppressed. The weights still deform the surface in the expected way, with large $p_\gamma$ reducing the depolarizing contribution, large $p_\omega$ reducing the controlled rotation, and the mixed term $p_{\gamma\omega}$ skewing the indicatrix through the coupling of the two controls.

The Finsler scaling function in \eqref{eq:Finsler-with-drift-v2} defines a conic sub-Finsler norm on the admissible velocity cone. The
affine admissibility condition is encoded in \(\lambda_t\), while the
quadratic control cost is encoded in \(\widetilde{\Gamma}_{ij}\).

For compactness, the quadratic form in \eqref{eq:quadratic-cost-function} can be written directly in terms of
the rescaled controls
\begin{equation}
A(v)
=
-\frac{xv_x+yv_y+zv_z}
{2(r_\perp^2+z^2)},
\label{eq:depol-drift-Gamma}
\end{equation}
and
\begin{equation}
B(v)
=
\frac{
(x^2+z^2)v_y
-xyv_x
-yzv_z
}
{x(r_\perp^2+z^2)} .
\label{eq:depol-drift-Omega}
\end{equation}

Equivalently, the non-vanishing off-diagonal components of \(\widetilde{\Gamma}_{ij}\) are
\begin{equation}
\begin{split}
 \widetilde{\Gamma}_{xz}
&=
\frac{
z\bigl(p_\gamma x^2+2p_{\gamma\omega}xy+4p_\omega y^2\bigr)
}
{2x(r_\perp^2+z^2)^2},
\\
\widetilde{\Gamma}_{xy}
&=
\frac{
p_\gamma x^2y
-p_{\gamma\omega}x^3
+p_{\gamma\omega}xy^2
-p_{\gamma\omega}xz^2
-4p_\omega x^2y
-4p_\omega yz^2
}
{2x(r_\perp^2+z^2)^2},
\\[4pt]
\widetilde{\Gamma}_{yz}&=
\frac{
z\bigl(
p_\gamma x^2y
-p_{\gamma\omega}x^3
+p_{\gamma\omega}xy^2
-p_{\gamma\omega}xz^2
-4p_\omega x^2y
-4p_\omega yz^2
\bigr)
}
{2x^2(r_\perp^2+z^2)^2},
\end{split}
\label{eq:depol-drift-general-metric-components}
\end{equation}
and the diagonals take the following form,
\begin{equation}
\begin{split}
\widetilde{\Gamma}_{xx}
&=
\frac{
p_\gamma x^2+2p_{\gamma\omega}xy+4p_\omega y^2
}
{2(r_\perp^2+z^2)^2},\\
\widetilde{\Gamma}_{yy}
&=
\frac{
p_\gamma x^2y^2
-2p_{\gamma\omega}x^3y
-2p_{\gamma\omega}xyz^2
+4p_\omega x^4
+8p_\omega x^2z^2
+4p_\omega z^4
}
{2x^2(r_\perp^2+z^2)^2},
\\
\widetilde{\Gamma}_{zz}
&=
\frac{
z^2\bigl(p_\gamma x^2+2p_{\gamma\omega}xy+4p_\omega y^2\bigr)
}
{2x^2(r_\perp^2+z^2)^2}.
\end{split}
\end{equation}
Substituting \eqref{eq:depol-drift-lambdat} and
\eqref{eq:quadratic-cost-function} into
\eqref{eq:Finsler-with-drift-v2} gives the explicit Finsler scaling
for the drifted depolarizing process:
\begin{equation}
F(v)
=
C_t\,
\frac{
z(xv_x+yv_y)-r_\perp^2v_z
}
{
\Delta x(r_\perp^2+z^2)
}
+
\frac{
p_\gamma A(v)^2
+
p_{\gamma\omega}A(v)B(v)
+
p_\omega B(v)^2
}
{
\frac{
z(xv_x+yv_y)-r_\perp^2v_z
}
{
\Delta x(r_\perp^2+z^2)
}
}.
\label{eq:depol-drift-explicit-finsler}
\end{equation}

For the case $x = 0$, we encounter the singular locus, and the allowed velocity for a certain direction becomes multivalued. The Finsler scaling function can thus be found using \eqref{eq: genFins} as follows
\begin{equation}
    F_{x=0}(v) = -\Delta \frac{z \left( p_{\gamma \omega} v_y + 4 p_\omega v_x \right)}{2y^2} + 2 \sqrt{\left( \frac{p_\gamma v_y^2 + 2 p_{\gamma\omega} v_x v_y + 4 p_\omega v_x^2}{4y^2} \right) \left( \tc + \Delta^2\frac{p_\omega z^2}{y^2} \right)}.
    \label{eq:depol-drift-singular-finsler}
\end{equation}

\begin{figure}[htbp]
\centering

\begin{subfigure}[t]{0.48\linewidth}
    \centering
    \includegraphics[width=\linewidth]{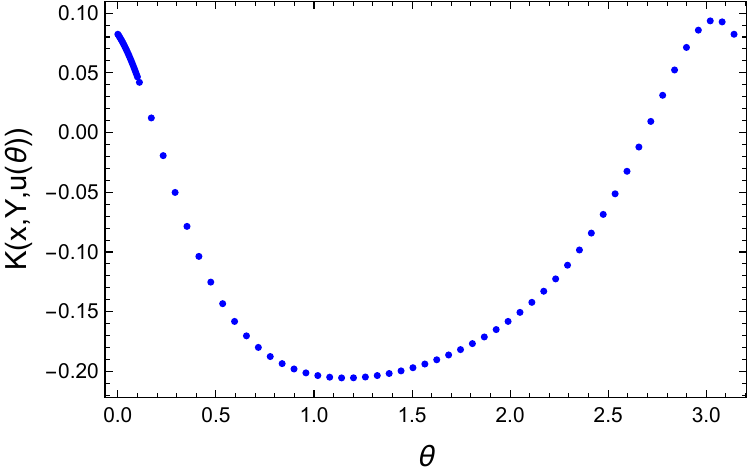}
    \caption{\(p_\gamma=1\), \(p_{\gamma\omega}=0\), \(p_\omega=1\).}
    \label{fig:unitary-drift-1-0-1}
\end{subfigure}
\hfill
\begin{subfigure}[t]{0.48\linewidth}
    \centering
    \includegraphics[width=\linewidth]{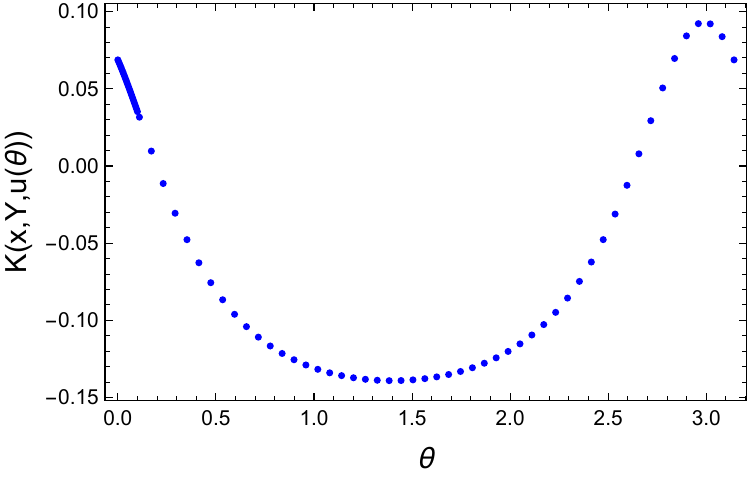}
    \caption{\(p_\gamma=1\), \(p_{\gamma\omega}=0.5\), \(p_\omega=1\).}
    \label{fig:unitary-drift-1-05-1}
\end{subfigure}

\vspace{0.4cm}

\begin{subfigure}[t]{0.48\linewidth}
    \centering
    \includegraphics[width=\linewidth]{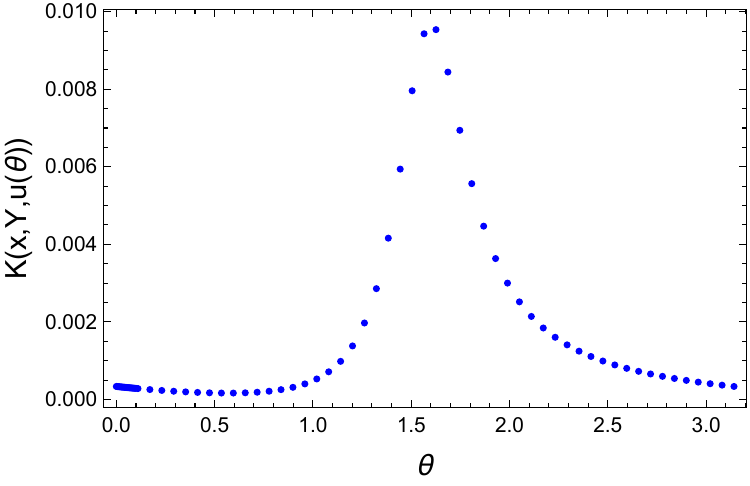}
    \caption{\(p_\gamma=100\), \(p_{\gamma\omega}=0\), \(p_\omega=1\).}
    \label{fig:unitary-drift-100-0-1}
\end{subfigure}
\hfill
\begin{subfigure}[t]{0.48\linewidth}
    \centering
    \includegraphics[width=\linewidth]{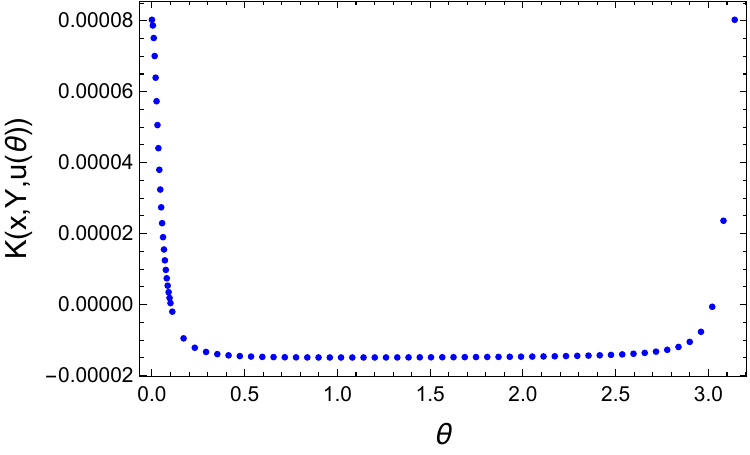}
    \caption{\(p_\gamma=1\), \(p_{\gamma\omega}=0\), \(p_\omega=100\).}
    \label{fig:unitary-drift-1-0-100}
\end{subfigure}

\caption{
Flag curvature for the depolarizing dynamics with unitary drift, evaluated at fixed base point \(x=(0.5,0.3,0.2)\) and fixed admissible flagpole \(Y\). The four panels show how \(K(x;Y,U(\theta))\) changes as the transverse flag direction \(U(\theta)\) is rotated in the physical control plane, for different choices of the quadratic control-cost weights \(p_\gamma\), \(p_{\gamma\omega}\), and \(p_\omega\). In all cases we take \(\Delta=1\) and \(\tc=1\).
}
\label{fig:unitary-drift-flag-curvature}
\end{figure}

In figure \ref{fig:unitary-drift-flag-curvature} the flag curvature \footnote{The intrinsic controlled distribution is two-dimensional and therefore has only one admissible flag plane, but in the full three-dimensional ambient space, the flag curvature can still vary nontrivially with the ambient flag plane containing the fixed flagpole.
} is evaluated at a fixed base point \(x=(0.5,0.3,0.2)\) and a fixed admissible flagpole \(Y\). To probe the dependence on the flag plane, we vary only the second flag vector while keeping \(x\) and \(Y\) fixed. More precisely, at the chosen base point we first determine the two-dimensional tangent plane to the admissible affine velocity space. Inside this plane we select a basis \(\{e_1,e_2\}\), with \(e_1\) chosen transverse to \(Y\), and then parametrize the second flag direction by
\begin{equation}
U(\theta)=\cos\theta\, e_1+\sin\theta\, e_2.
\end{equation}
In this way, changing \(\theta\) rotates the flag plane through \(Y\) while preserving the admissibility condition. Since \(U(\theta+\pi)=-U(\theta)\) spans the same plane together with \(Y\), the resulting flag curvature is naturally \(\pi\)-periodic. The plots therefore show how the local geometry depends on the choice of two-plane containing the fixed evolution direction \(Y\), rather than on the orientation of the transverse vector itself.

\begin{figure}[htbp]
\centering

\begin{subfigure}[t]{\indicatrixwidth}
    \centering
    \includegraphics[width=\linewidth]{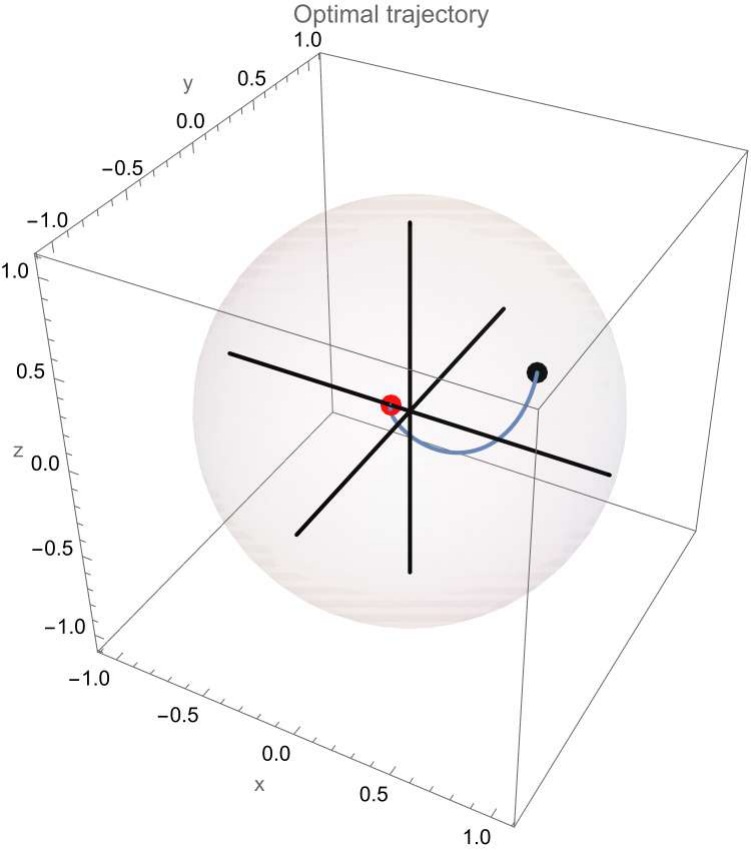}
    \caption{$p_\gamma = 1$, $p_\omega = 1$ and $p_{\gamma\omega} = 0$.}
    \label{fig:trajectory-depuni-1}
\end{subfigure}
\hfill
\begin{subfigure}[t]{\indicatrixwidth}
    \centering
    \includegraphics[width=\linewidth]{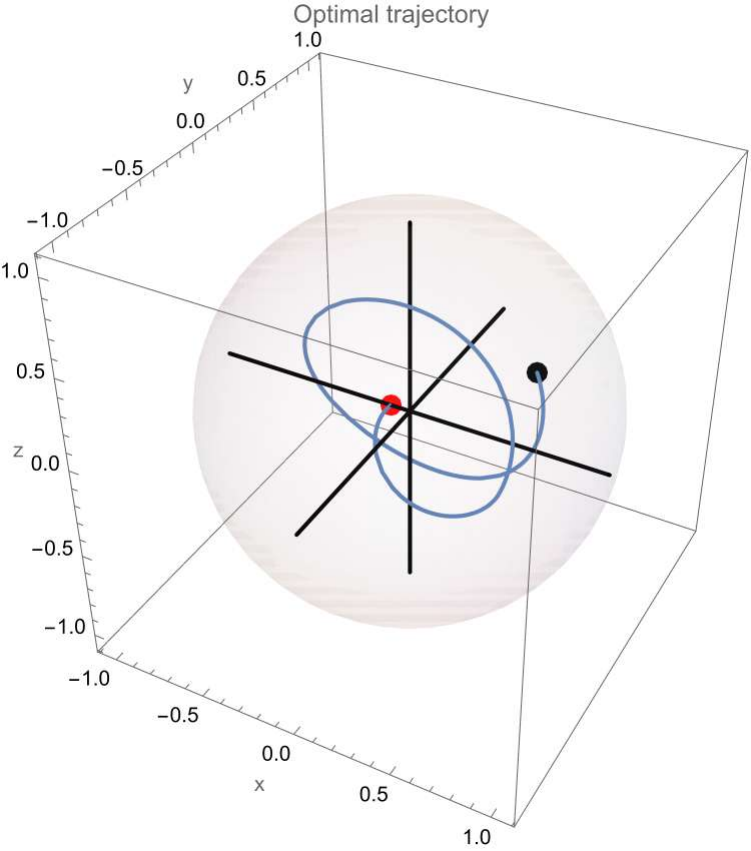}
    \caption{$p_\gamma = 1$, $p_\omega = 1$ and $p_{\gamma\omega} = 1.9 $.}
    \label{fig:trajectory-depuni-2}
\end{subfigure}

\vspace{0.5cm}

\begin{subfigure}[t]{\indicatrixwidth}
    \centering
    \includegraphics[width=\linewidth]{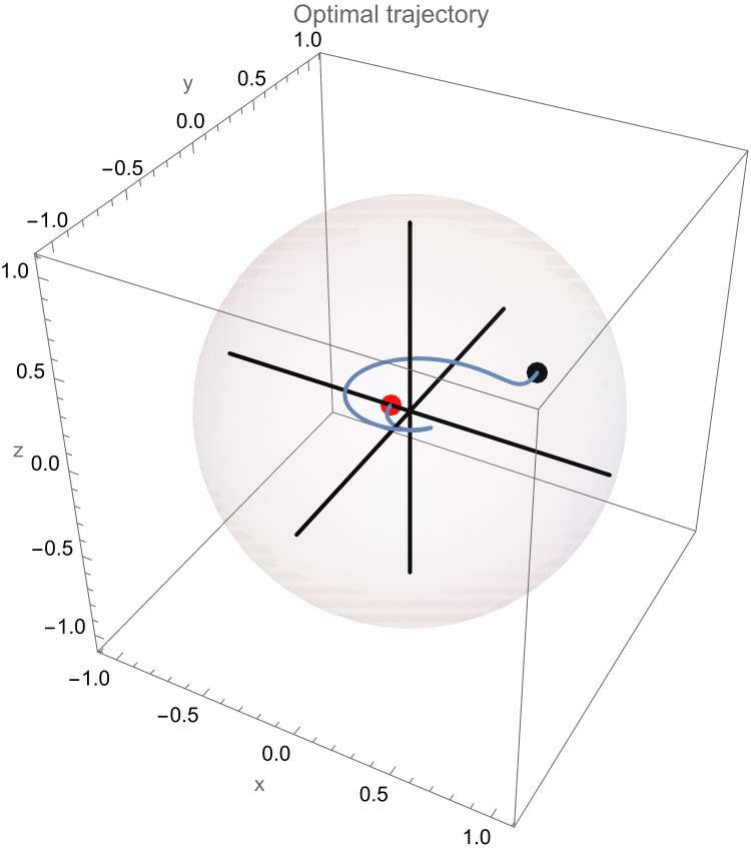}
    \caption{$p_\gamma = 1$, $p_\omega = 0.01$ and $p_{\gamma\omega} = 0$.}
    \label{fig:trajectory-depuni-3}
\end{subfigure}
\hfill
\begin{subfigure}[t]{\indicatrixwidth}
    \centering
    \includegraphics[width=\linewidth]{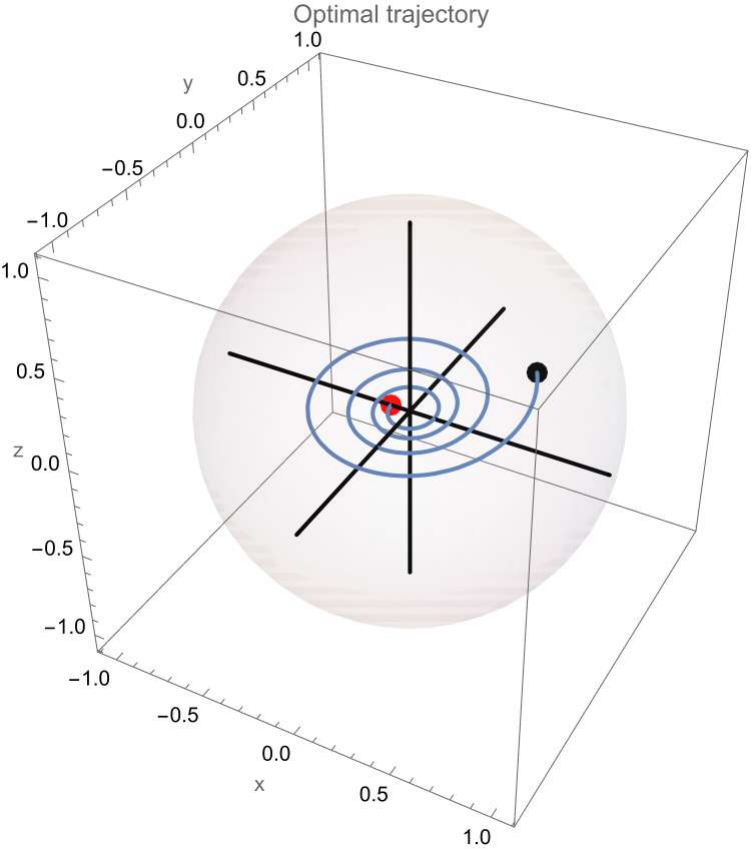}
    \caption{$p_\gamma = 10$, $p_\omega = 0.01$ and $p_{\gamma\omega} = 0$.}
    \label{fig:trajectory-depuni-4}
\end{subfigure}

    \caption{
    The optimal trajectories of the depolarizing with unitary drift computed numerically using the PMP approach with the initial point (black dot) at $x=(0.9, 0.0, 0.0)$ and the final point (red dot) at $x = (-0.1, 0.0, 0.0)$. The parameters are set to be $C_t = 1$ and $\Delta = 1$. The four plots show how the trajectory changes with respect to the control-cost weights \(p_\gamma\), \(p_{\gamma\omega}\), and \(p_\omega\).
    }
    \label{fig:trajectories-depuni}
\end{figure}

We compute the optimal trajectories for this case as shown in Fig. \ref{fig:trajectories-depuni}. 
The plots show how the optimal trajectory changes when the relative cost of the two controls is modified, while the fixed drift and endpoints are kept unchanged. In panel \ref{fig:trajectory-depuni-1}, where the controls are penalised equally and independently, the trajectory follows a relatively direct compromise between radial contraction, rotation, and the fixed drift. The main qualitative change occurs in panel \ref{fig:trajectory-depuni-2}: since $2p_{\gamma\omega}=1.9$, the cost strongly favours the combination $\omega\simeq-\gamma$ and suppresses $\omega\simeq\gamma$.
This explains the stronger bending and looping of the trajectory. In panel \ref{fig:trajectory-depuni-3}, $p_\omega$ is reduced while $p_\gamma$ is kept fixed, so rotations become cheaper relative to depolarization. Finally, in panel \ref{fig:trajectory-depuni-4}, increasing $p_\gamma$ makes radial contraction more expensive, so the trajectory relies more heavily on the cheap rotational direction, producing a tighter winding near the target.

\begin{figure}[htbp]
\centering

\begin{subfigure}[t]{\indicatrixwidth}
    \centering
    \includegraphics[width=\linewidth]{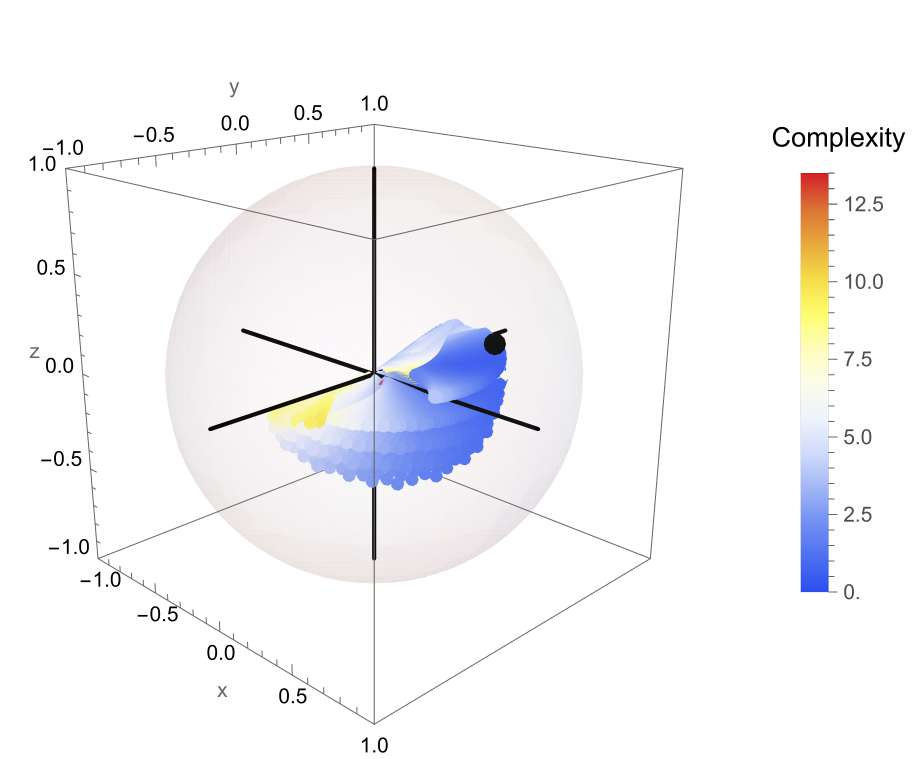}
    \caption{$p_\gamma = 1$, $p_\omega = 1$ and $p_{\gamma\omega} = 0$.}
    \label{fig:trajcost-depuni-1}
\end{subfigure}
\hfill
\begin{subfigure}[t]{\indicatrixwidth}
    \centering
    \includegraphics[width=\linewidth]{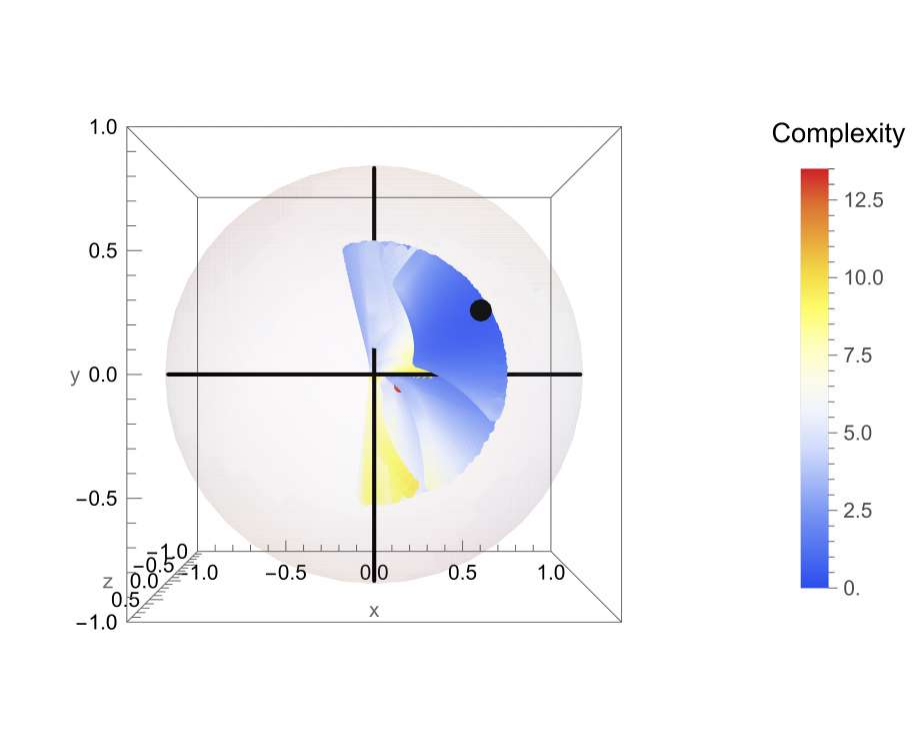}
    \caption{$p_\gamma = 1$, $p_\omega = 1$ and $p_{\gamma\omega} = 0 $.}
    \label{fig:trajcost-depuni-2}
\end{subfigure}

\vspace{0.5cm}

\begin{subfigure}[t]{\indicatrixwidth}
    \centering
    \includegraphics[width=\linewidth]{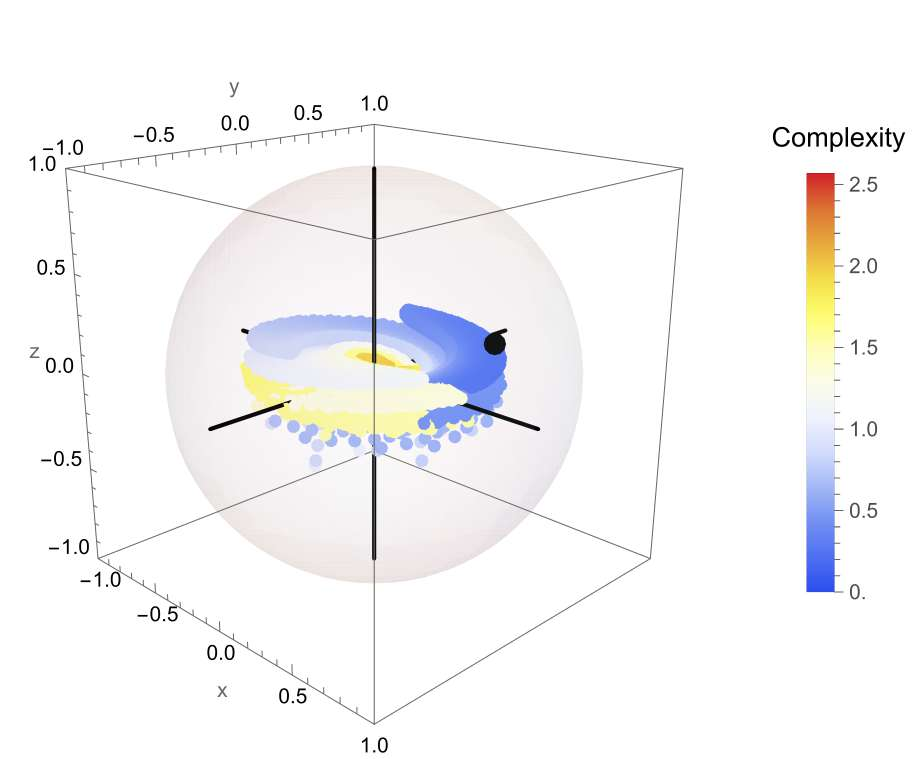}
    \caption{$p_\gamma = 1$, $p_\omega = 0.01$ and $p_{\gamma\omega} = 0$.}
    \label{fig:trajcost-depuni-3}
\end{subfigure}
\hfill
\begin{subfigure}[t]{\indicatrixwidth}
    \centering
    \includegraphics[width=\linewidth]{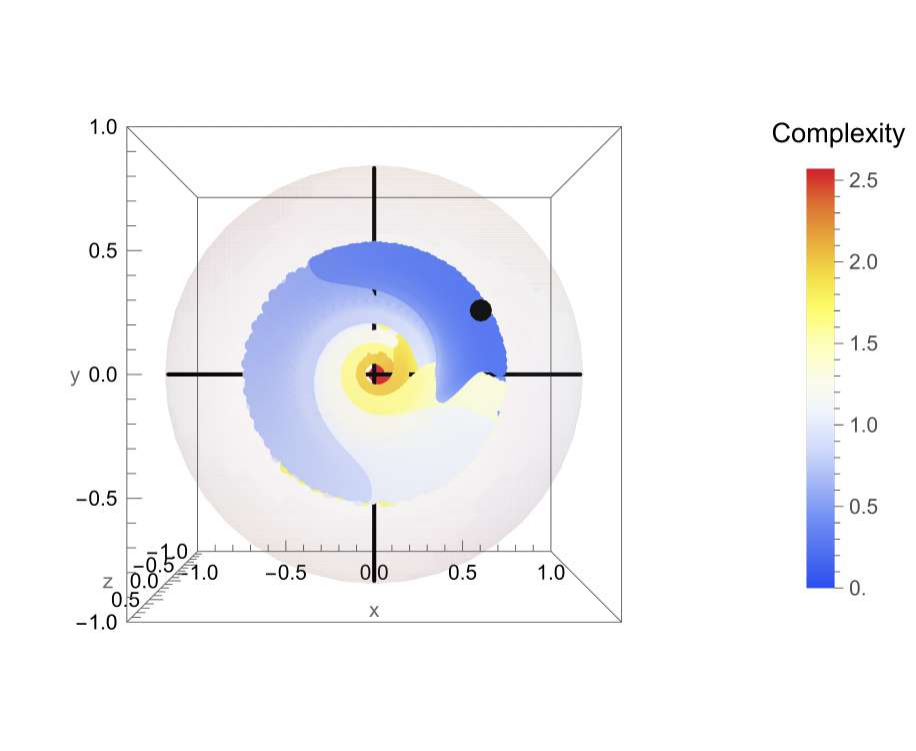}
    \caption{$p_\gamma = 1$, $p_\omega = 0.01$ and $p_{\gamma\omega} = 0$.}
    \label{fig:trajcost-depuni-4}
\end{subfigure}

    \caption{
    The trajectory costs gained from using the optimal trajectories for the depolarizing with unitary drift from the initial point (black dot) $x = (0.5, 0.3, 0.2)$ to final points reached by different initial velocities shown in color gradation. The parameters are set to be $C_t = 1$, $\Delta = 1$ and two sets of control-cost weights.
    }
    \label{fig:trajcost-depuni}
\end{figure}

We also compute the trajectory costs for this case with the cost function as in plot \ref{fig:trajectory-depuni-1}, which is shown by figure \ref{fig:trajcost-depuni}, by integrating the optimal trajectories from the initial state at
\((0.5,0.3,0.2)\) with different allowed initial velocities inside the Bloch ball. The color gradient reflects the cost to reach each final point: points closer to the initial state generally have smaller cost, while farther points require larger accumulated control cost. The remaining variation comes from the preferred directions induced by the depolarizing control and unitary drift. We performed the same computation for the other choices of cost functions used in the trajectory plots in Fig. \ref{fig:trajectories-depuni}.
%

\subsection{Example 4: depolarizing dynamics with non-unitary drift}

We consider depolarizing dynamics in the presence of a non-unitary amplitude-damping drift. The non-unitary drift is more restrictive: it adds an irreversible component to the dynamics that cannot be cancelled by the controls. As a result, the allowed cone of directions is shifted toward the dissipative flow.
The drift is proportional to \(\Delta\) which contributes to an additional Lindbladian operator:
\begin{equation}
    L_2 = \frac{\sqrt{\Delta}}{2} (\sigma_z + i \sigma_x)
\end{equation}
The equations of motion are
\begin{eqnarray}
\dot x&=&-2\gamma x-\omega y-\frac{\Delta}{2}x,\nonumber\\
\dot y&=&\omega x-2\gamma y-\Delta y+\Delta,\\
\dot z&=&-2\gamma z-\frac{\Delta}{2}z .\nonumber
\label{eq:nonunitary-drift-eom}
\end{eqnarray}
We introduce the rescaled velocity
\begin{equation}
\dot x^i=L v^i .
\end{equation}
The equations of motion then become
\begin{equation}
Lv_x=-2\gamma x-\omega y-\frac{\Delta}{2}x,
\qquad
Lv_y=\omega x-2\gamma y-\Delta y+\Delta,
\qquad
Lv_z=-2\gamma z-\frac{\Delta}{2}z .
\label{eq:nonunitary-drift-rescaled}
\end{equation}


Solving \eqref{eq:nonunitary-drift-rescaled} for \(L,\gamma,\omega\), one finds
\begin{equation}
L
=
\frac{
\Delta z\left(y-\frac{y^2}{2}\right)
}
{
z(xv_x+yv_y)-(x^2+y^2)v_z
}.
\label{eq:nonunitary-drift-L}
\end{equation}
It is therefore useful to define
\begin{equation}
D(v):=
z(xv_x+yv_y)-(x^2+y^2)v_z,
\qquad
A:=z\left(y-\frac{y^2}{2}\right).
\end{equation}
Then
\begin{equation}
L=\frac{\Delta A}{D(v)}
=
\frac{1}{\lambda_t(v)},
\qquad
\lambda_t(v):=
\frac{D(v)}{\Delta A}.
\label{eq:nonunitary-drift-lambdat}
\end{equation}
The physical branch corresponds to \(\lambda_t(v)>0\).

The corresponding controls are
\begin{equation}
\gamma
=
\frac{\Delta}{2}\,
\frac{
\left(y^2-y+\frac{x^2}{2}\right)v_z
-\frac{z}{2}(xv_x+yv_y)
}
{
z(xv_x+yv_y)-(x^2+y^2)v_z
}, \qquad
\omega
=
\Delta\,
\frac{
\left(\frac{y}{2}-1\right)(zv_x-xv_z)
}
{
z(xv_x+yv_y)-(x^2+y^2)v_z
}.
\label{eq:nonunitary-drift}
\end{equation}
The condition \(\gamma\ge0\) further restricts the admissible velocity space.

\begin{figure}[htbp]
\centering

\begin{subfigure}[t]{\indicatrixwidth}
    \centering
    \includegraphics[width=\linewidth]{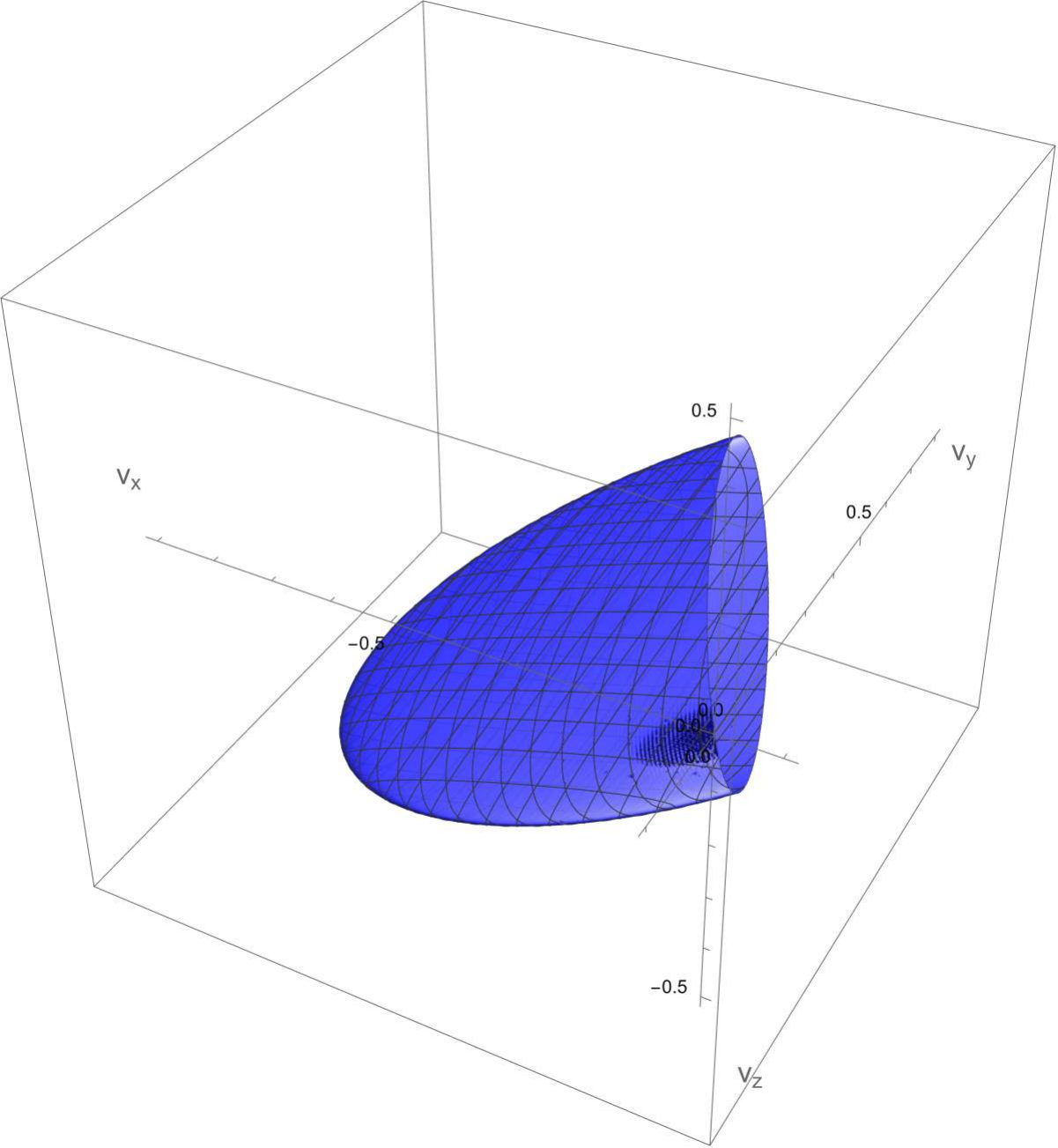}
    \caption{$p_\gamma = 1$, $p_\omega = 1$ and $p_{\gamma\omega} = 0$.}
    \label{fig:indicatrix-depnon-1}
\end{subfigure}
\hfill
\begin{subfigure}[t]{\indicatrixwidth}
    \centering
    \includegraphics[width=\linewidth]{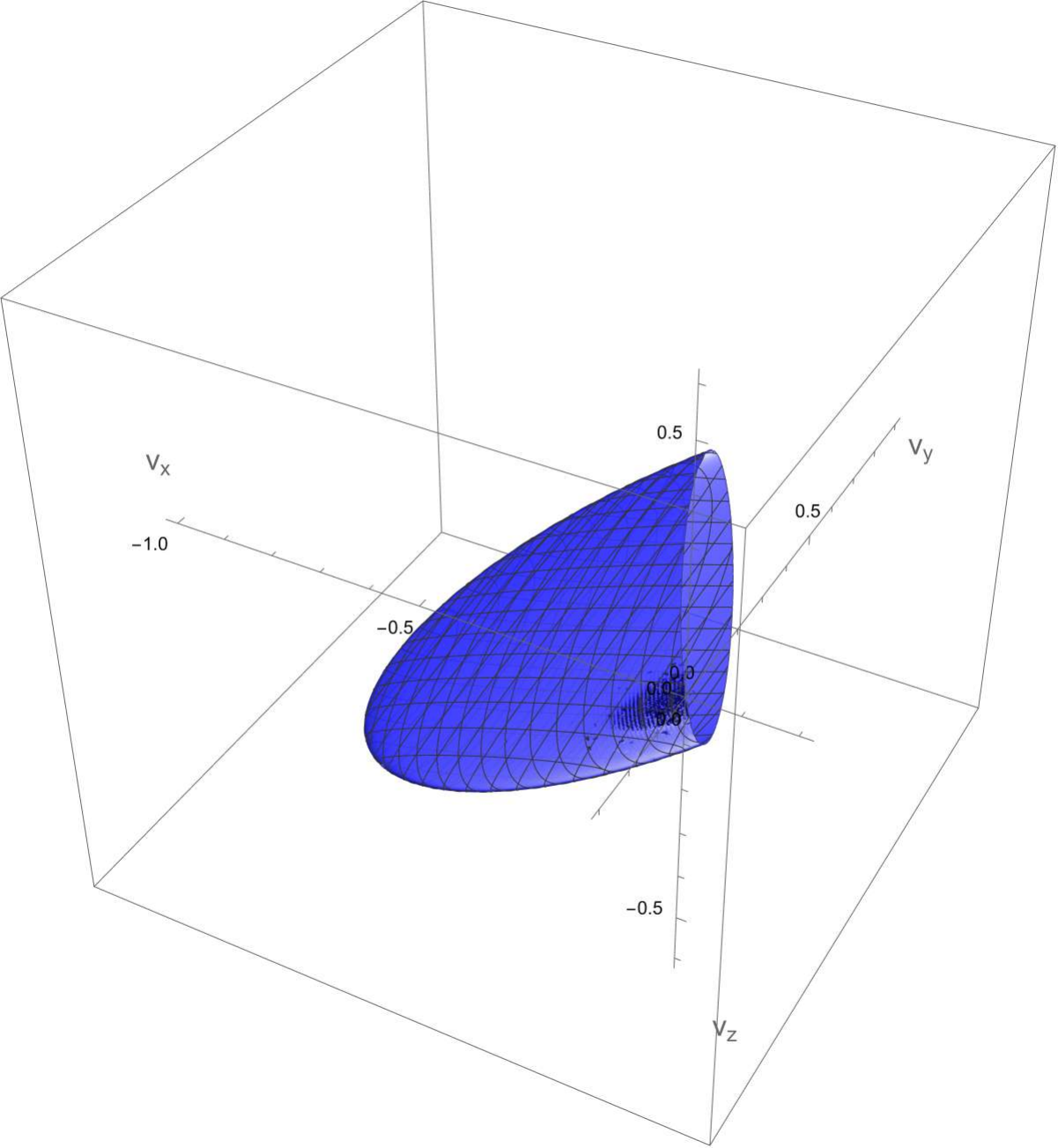}
    \caption{$p_\gamma = 1$, $p_\omega = 1$ and $p_{\gamma\omega} = \frac12 $.}
    \label{fig:indicatrix-depnon-2}
\end{subfigure}

\vspace{0.5cm}

\begin{subfigure}[t]{\indicatrixwidth}
    \centering
    \includegraphics[width=\linewidth]{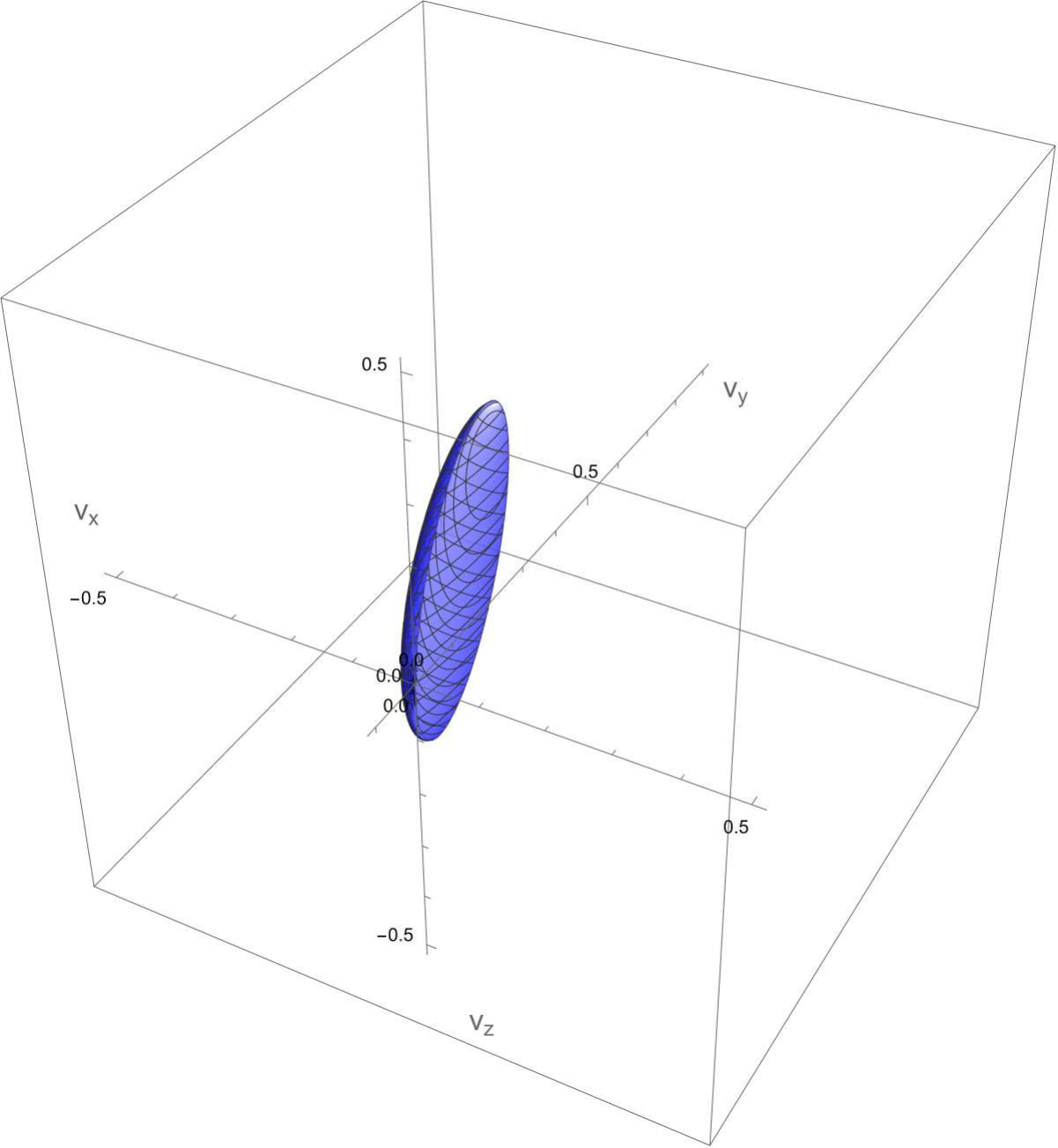}
    \caption{$p_\gamma = 100$, $p_\omega = 1$ and $p_{\gamma\omega} = 0$.}
    \label{fig:indicatrix-depnon-3}
\end{subfigure}
\hfill
\begin{subfigure}[t]{\indicatrixwidth}
    \centering
    \includegraphics[width=\linewidth]{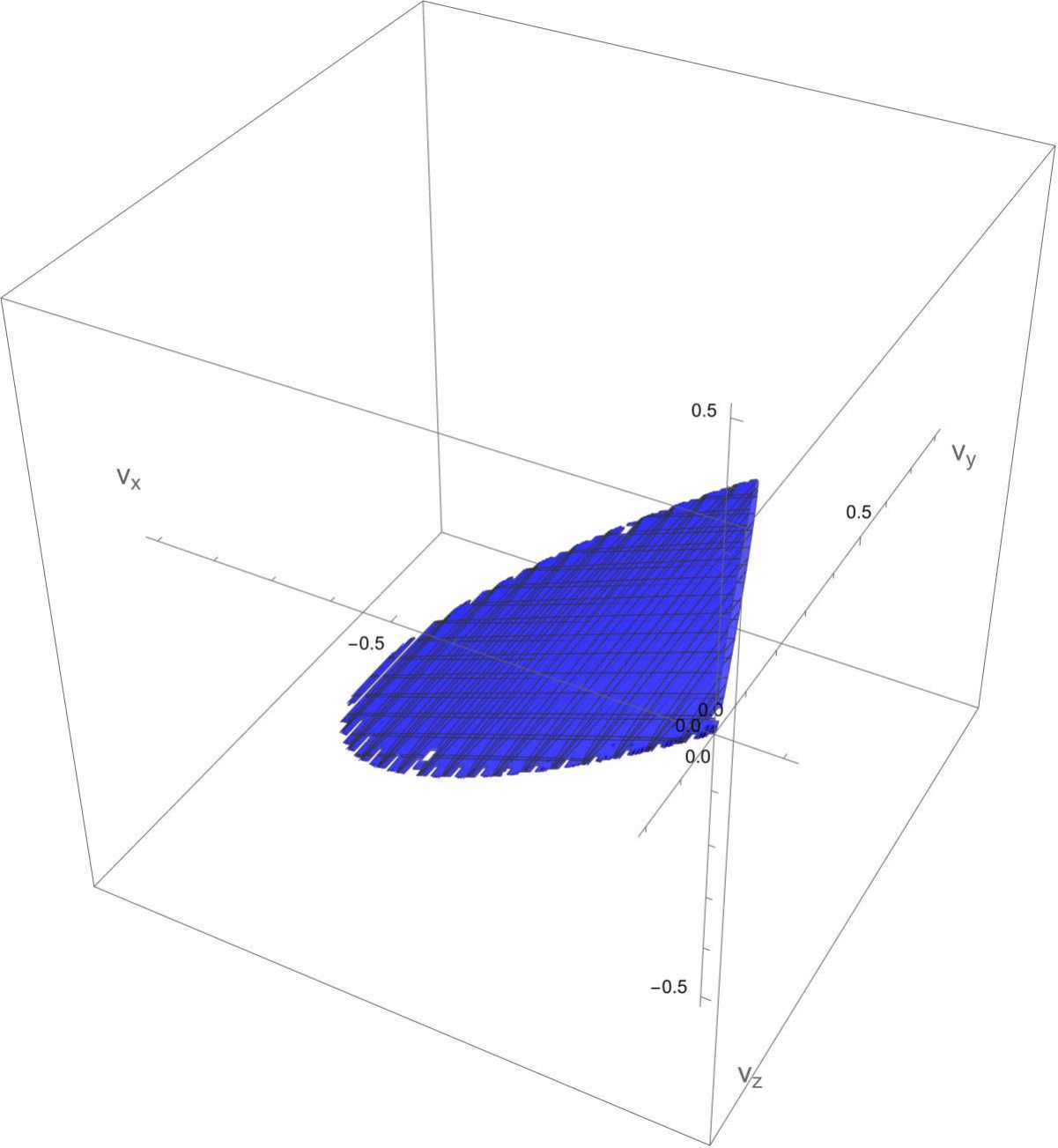}
    \caption{$p_\gamma = 1$, $p_\omega = 100$ and $p_{\gamma\omega} = 0$.}
    \label{fig:indicatrix-depnon-4}
\end{subfigure}

    \caption{
    The indicatrices for the depolarizing dynamics with non-unitary amplitude-damping drift at base point \(x=(0.5,0.3,0.2)\) with $\tc = 1$ and $\Delta = 1$ as surfaces shown. The four plots show how the indicatrix changes with respect to the control penalty \(p_\gamma\), \(p_{\gamma\omega}\), and \(p_\omega\).
    }
    \label{fig:indicatrices-depnon}
\end{figure}

For compactness, define
\begin{equation}
N_\gamma(v)
:=
\left(y^2-y+\frac{x^2}{2}\right)v_z
-\frac{z}{2}(xv_x+yv_y),
\end{equation}
and
\begin{equation}
N_\omega(v)
:=
\left(\frac{y}{2}-1\right)(zv_x-xv_z).
\end{equation}
Using \(D(v)=\Delta A\lambda_t(v)\), the controls can be written as
\begin{equation}
\gamma
=
\frac{N_\gamma(v)}{2A\lambda_t(v)},
\qquad
\omega
=
\frac{N_\omega(v)}{A\lambda_t(v)}.
\end{equation}
Therefore the control part of the cost becomes
\begin{equation}
G(v)
=
\frac{1}{A^2}
\left[
\frac{p_\gamma}{4}N_\gamma(v)^2
+
\frac{p_{\gamma\omega}}{2}N_\gamma(v)N_\omega(v)
+
p_\omega N_\omega(v)^2
\right].
\label{eq:nonunitary-drift-G}
\end{equation}
The cost equation \(L=C_t+p_\gamma\gamma^2+p_{\gamma\omega}\gamma\omega+p_\omega\omega^2\) then gives the indicatrix condition
\begin{equation}
\frac{1}{\lambda_t(v)}
=
\tc+\frac{G(v )}{\lambda_t(v)^2},
\label{eq:nonunitary-drift-indicatrix}
\end{equation}
Equivalently,
\begin{equation}
\tc\lambda_t(v)^2-\lambda_t(v)+G(v)=0.
\end{equation}
The indicatrix is as shown in figure \ref{fig:indicatrices-depnon}.
Unlike the unitary drift, which only adds a background rotational motion, this drift damps the transverse components while driving the state toward the preferred amplitude-damping fixed point. Thus, the displacement of the indicatrix is a bias toward the relaxation direction selected by the non-unitary drift. The weights deform the indicatrix as before: large $p_\gamma$ suppresses dissipation, large $p_\omega$ suppresses rotation, and $p_{\gamma\omega}$ skews the surface.

It is useful to make the quadratic form explicit. Define
\begin{equation}
a:=y-\frac{y^2}{2},
\qquad
B:=\frac{x^2}{2}+y^2-y,
\qquad
C:=\frac{y}{2}-1.
\end{equation}
Then
\begin{equation}
N_\gamma(v)=Bv_z-\frac{z}{2}(xv_x+yv_y),
\qquad
N_\omega(v)=C(zv_x-xv_z),
\end{equation}
and
\begin{equation}
G(v)
=
\frac{1}{z^2a^2}
\left[
\frac{p_\gamma}{4}N_\gamma(v)^2
+
\frac{p_{\gamma\omega}}{2}N_\gamma(v)N_\omega(v)
+
p_\omega N_\omega(v)^2
\right].
\end{equation}
The non-vanishing components of \(\widetilde{\Gamma}_{ij}\) are
\begin{equation}
\begin{aligned}
\widetilde{\Gamma}_{xx}
&=
\frac{1}{a^2}
\left(
\frac{p_\gamma x^2}{8}
-\frac{p_{\gamma\omega}xC}{2}
+2p_\omega C^2
\right),
\quad
\widetilde{\Gamma}_{xy}
=
\frac{1}{a^2}
\left(
\frac{p_\gamma xy}{8}
-\frac{p_{\gamma\omega}yC}{4}
\right),
\\[4pt]
\widetilde{\Gamma}_{yy}
&=
\frac{p_\gamma y^2}{8a^2},
\quad
\widetilde{\Gamma}_{xz}
=
\frac{1}{za^2}
\left(
-\frac{p_\gamma xB}{4}
+\frac{p_{\gamma\omega}C}{2}\left(\frac{x^2}{2}+B\right)
-2p_\omega C^2x
\right),
\\[4pt]
\widetilde{\Gamma}_{yz}
&=
\frac{1}{za^2}
\left(
-\frac{p_\gamma yB}{4}
+\frac{p_{\gamma\omega}Cxy}{4}
\right),
\quad
\widetilde{\Gamma}_{zz}
=
\frac{1}{z^2a^2}
\left(
\frac{p_\gamma B^2}{2}
-p_{\gamma\omega}BCx
+2p_\omega C^2x^2
\right).
\end{aligned}
\label{eq:nonunitary-drift-metric-components}
\end{equation}

Thus the non-unitary drift again gives a conic sub-Finsler structure of the form
\begin{equation}
F(v)
=
\tc\,\lambda_t(v)
+
\frac{\frac{p_\gamma}{4}N_\gamma(v)^2
+
\frac{p_{\gamma\omega}}{2}N_\gamma(v)N_\omega(v)
+
p_\omega N_\omega(v)^2}{z^2 a^2\lambda_t(v)},
\end{equation}
with
\begin{equation}
\lambda_t(v)
=
\frac{
z(xv_x+yv_y)-(x^2+y^2)v_z
}
{
\Delta z\left(y-\frac{y^2}{2}\right)
}.
\end{equation}

There are two cases where we encounter singular $\lambda_t$ in this example: $y = 0$ and $z = 0$. As in the previous example, the Finsler scaling function can also be found using \eqref{eq: genFins}:
\begin{eqnarray}
    F_{y=0}(v) &=& \Delta \frac{p_{\gamma} xv_x + p_{\gamma \omega} (2v_x-xv_y) - 8 p_\omega v_y}{4x^2}\nonumber\\
    &&+ 2 \sqrt{\frac{p_\gamma v_x^2 - 2 p_{\gamma \omega} v_x v_y + 4 p_\omega v_y^2}{4x^2} \left( \tc + \Delta^2 \frac{p_{\gamma} x^2 + 4 p_{\gamma\omega} x + 16 p_{\omega}}{16x^2} \right)},
    \label{eq:depol-ampdampdrift-singular-finsler-y}
\end{eqnarray}
and
\begin{eqnarray}
    F_{z=0}(v)
    &=&
    \frac{\Delta}{2r^4}
    \left[
        p_\gamma BS
        -p_{\gamma\omega}(BT-XS)
        -4p_\omega XT
    \right]  \nonumber\\
    &+&\sqrt{
        \frac{
            p_\gamma S^2
            -2p_{\gamma\omega}ST
            +4p_\omega T^2
        }{r^4}
        \left[
            \tc+\Delta^2
            \frac{
                p_\gamma B^2
                +2p_{\gamma\omega}BX
                +4p_\omega X^2
            }{4r^4}
        \right]
    },
\end{eqnarray}
where
\begin{equation}
    r^2=x^2+y^2,
    \qquad
    S=xv_x+yv_y,
    \qquad
    T=-yv_x+xv_y,
\end{equation}
and
\begin{equation}
    B=\frac{x^2}{2}+y^2-y,
    \qquad
    X=x\left(1-\frac{y}{2}\right).
\end{equation}

\begin{figure}[htbp]
\centering

\begin{subfigure}[t]{0.48\linewidth}
    \centering
    \includegraphics[width=\linewidth]{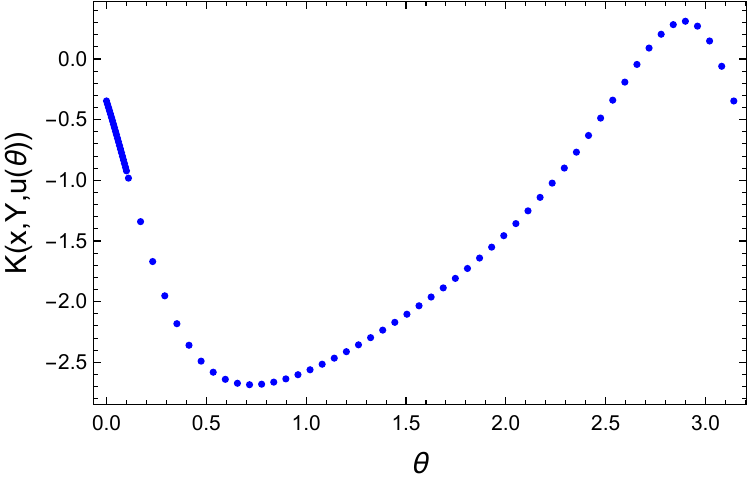}
    \caption{\(p_\gamma=1\), \(p_{\gamma\omega}=0\), \(p_\omega=1\).}
    \label{fig:nonunitary-drift-1-0-1}
\end{subfigure}
\hfill
\begin{subfigure}[t]{0.48\linewidth}
    \centering
    \includegraphics[width=\linewidth]{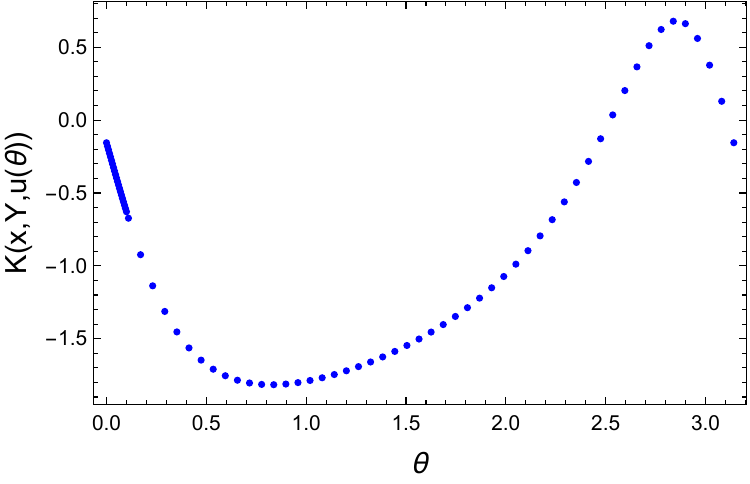}
    \caption{\(p_\gamma=1\), \(p_{\gamma\omega}=0.5\), \(p_\omega=1\).}
    \label{fig:nonunitary-drift-1-05-1}
\end{subfigure}

\vspace{0.4cm}

\begin{subfigure}[t]{0.48\linewidth}
    \centering
    \includegraphics[width=\linewidth]{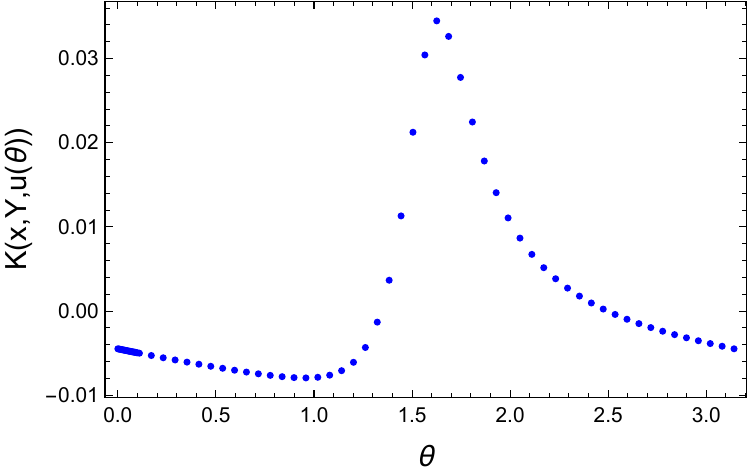}
    \caption{\(p_\gamma=100\), \(p_{\gamma\omega}=0\), \(p_\omega=1\).}
    \label{fig:nonunitary-drift-100-0-1}
\end{subfigure}
\hfill
\begin{subfigure}[t]{0.48\linewidth}
    \centering
    \includegraphics[width=\linewidth]{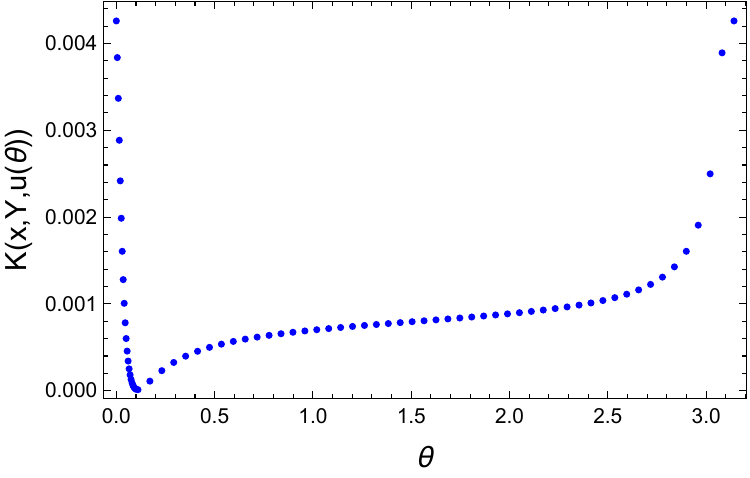}
    \caption{\(p_\gamma=1\), \(p_{\gamma\omega}=0\), \(p_\omega=100\).}
    \label{fig:nonunitary-drift-1-0-100}
\end{subfigure}

\caption{
Flag curvature for the depolarizing dynamics with non-unitary drift, evaluated at fixed base point \(x=(0.5,0.3,0.2)\) and fixed admissible flagpole \(Y\). The four panels show how \(K(x;Y,U(\theta))\) changes as the transverse flag direction \(U(\theta)\) is rotated in the physical control plane, for different choices of the quadratic control penalty \(p_\gamma\), \(p_{\gamma\omega}\), and \(p_\omega\). In all cases we take \(\Delta=1\) and \(\tc=1\).
}
\label{fig:nonunitary-drift-flag-curvature}
\end{figure}

\subsection{Example 5: damped quantum harmonic oscillator}

Consider the case for a damped harmonic oscillator with Lindbladian operators as follows~\cite{Rivas:2010lbs}:
\begin{equation}
    L_1 = \sqrt{\gamma (n + 1)} ~\hat{a}, \qquad L_2 = \sqrt{\gamma n} ~\hat{a}^\dagger,
\end{equation}
which give the following Master equation:
\begin{equation}
    \dot{\rho} = -\frac{i}{\hbar} \left[ H,\rho \right] + \gamma (n + 1) \left( \hat{a} \rho \hat{a}^\dagger - \frac{1}{2} \left\{ \hat{a}^\dagger \hat{a}, \rho \right\} \right) + \gamma n \left( \hat{a}^\dagger \rho \hat{a} - \frac{1}{2} \left\{ \hat{a} \hat{a}^\dagger, \rho \right\} \right).
\end{equation}

This example describes a harmonic oscillator coupled to a thermal bath. The operator $L_1$ corresponds to the loss of one excitation from the oscillator into the environment, while $L_2$ corresponds to the absorption of one excitation from the bath. The parameter $\gamma$ sets the damping rate, and $n$ is the mean occupation number of the thermal environment. In the zero-temperature limit $n=0$, only the decay channel remains, reducing the dynamics to ordinary amplitude damping.

Unlike the qubit examples considered above, the harmonic oscillator has an infinite-dimensional Hilbert space. However, for Gaussian states and quadratic Hamiltonians, the dynamics is closed at the level of the first and second moments of the ladder operators. Therefore, instead of tracking the full density matrix, we can describe the dissipative evolution through the
variances of $\langle \hat a \rangle, \langle \hat a^\dagger \rangle$ and their covariance. This gives a finite-dimensional description suitable for the same optimal-control and geometric analysis used in the previous examples.

This representation is useful because it separates the rotating degrees of freedom from the thermal fixed point more transparently. We define
\begin{eqnarray}
    V_{aa} &=& \langle\hat a^2\rangle - \langle\hat a\rangle^2,\nonumber\\
    V_{a^\dagger a^\dagger} &=& \langle\hat a^{\dagger 2}\rangle - \langle\hat a^\dagger\rangle^2,\\
    C_{aa^\dagger} &=& \frac{1}{2} \langle \{ \hat a, \hat a^\dagger \}\rangle - \langle\hat a\rangle \langle\hat a^\dagger\rangle.\nonumber
\end{eqnarray}
This gives the following dynamics:
\begin{eqnarray}
    \dot V_{aa} &=& -2i\omega V_{aa} - \gamma V_{aa},\nonumber\\
    \dot V_{a^\dagger a^\dagger} &=& 2i\omega V_{a^\dagger a^\dagger} - \gamma V_{a^\dagger a^\dagger},\\
    \dot C_{a a\dagger} &=& -\gamma C_{a a^\dagger} + \gamma \left( n+\frac{1}{2} \right).\nonumber
\end{eqnarray}
If we introduce the following transformation of variables,
\begin{equation}
    x = \frac{1}{2} \left( V_{aa} + V_{a^\dagger a^\dagger} \right),\qquad
    y = \frac{1}{2i} \left( V_{aa} - V_{a^\dagger a^\dagger} \right),\qquad
    z = C_{aa^\dagger} - \left( n + \frac12 \right),
\end{equation}
the previous dynamics takes the same form as a simple depolarizing dynamics with a single unitary rotation:
\begin{equation}
    \dot x = -\gamma x + 2\omega y,\qquad
    \dot y = -2\omega x - \gamma y,\qquad
    \dot z = -\gamma z.
\end{equation}
Thus, after shifting the covariance variable $C_{aa^\dagger}$ by its thermal steady-state
value, the damped harmonic oscillator reduces to a three-dimensional linear system with an isotropic contraction rate $\gamma$ and a rotation in the $(x,y)$ plane generated by $\omega$. In this form, the dynamics is formally identical to the depolarizing dynamics with a single unitary rotation studied earlier, with the thermal state mapped to the origin $(x,y,z)=(0,0,0)$.

\section{Discussion} \label{sec:Discussion}

In this paper, we studied a control-theoretic notion of circuit complexity for open quantum systems, defined by adapting Nielsen's approach to systems governed by a Lindbladian evolution. In this case it is more natural to work directly in the space of (mixed) states, rather than defining first an operator complexity and then a state complexity by minimization over the possible operators. 

We showed that this construction naturally leads to a Finsler or sub-Finsler geometry on the space of admissible state-space trajectories. The geometry depends non-trivially on the parameters that enter the cost functional, namely the weights for the controls (the penalty factors) and the terminal cost. The geometry has the property that its geodesics are solutions of the associated optimal control problem, so that many properties of the solutions are encoded in a single function, the Finslerian function, on a sub-bundle of the tangent bundle on the space of states.  

We derived explicit expressions for the Finslerian function, and illustrated this framework in explicit single-qubit examples. For depolarizing dynamics without drift, the reduced geometry becomes quadratic and hence sub-Riemannian on the admissible distribution. In the presence of drift, the admissibility condition is affine and the resulting geometry is genuinely Finslerian. We then used the associated flag curvature to characterize the local structure of the induced control geometry, interpreting it as a measure of directional sensitivity of nearby optimal trajectories.
We showed that the flag curvature can change sign depending on the parameters. In the usual unitary case, \cite{Brown:2016wib} interpreted the negative sectional curvature as arising generically from the structure of the algebra of generators. Schematically, with an algebra $[A,B]=C$, negative curvature appears if $C$ is a ``hard'' direction, with a sufficiently higher cost than $A,B$. Since in the non-unitary case we do not have a group structure, an analogous explanation is not readily apparent. It would be interesting to find a similar heuristic explanation also for the semigroup dynamics, in order to understand better the properties of non-unitary complexity in more general systems. 

As was pointed out by Lloyd and Viola \cite{Lloyd:2000dwz}, there is another very important aspect in which open-system dynamics differs from closed-systems. If one has a universal set of generators, such that their iterated commutators generate any possible Hamiltonian, then it is possible to reproduce, with arbitrary precision, any arbitrary  unitary time-evolution with a circuit that uses only those generators (one just uses the Trotter formula repeatedly). However this is not true for an arbitrary non-unitary evolution described by a quantum semigroup dynamics. One could try to get around this problem by using purification, or the Stinespring dilation, to lift the system to a unitary dynamics. However one then needs to use operators that act also on the environment, and not only on the system. Alternatively, one can generate an arbitrary dynamics by performing measurements and using feedback loop such that the subsequent operations depend on the outcome of the measurements. The dynamics then becomes non-Markovian and one has to generalize the Lindblad equation to a non-Markovian version. We leave for future work the important problem of extending our geometric formulation to the non-Markovian case. 

It would be interesting to compare our approach to open-system complexity with other approaches in the literature. Several papers \cite{Caceres:2019pgf,Bhattacharyya:2021fii, Bhattacharyya:2022rhm, Bhattacharya:2022wlp, Bhattacharyya:2024duw} have considered complexity for mixed states defined by minimizing the complexity of the corresponding purifications. This procedure is however very impractical, as the ancilla space used for the purification can \emph {a priori} have an arbitrary dimension.
 In  \cite{Ruan:2020vze} a different strategy was used: they did not resort to purification but used instead the Bures distance in the space of mixed states, which is the analogue of the Fubini-Study metric for pure states. As one can recover the Fubini-Study metric from Nielsen's complexity with appropriately chosen penalty factors, it is perhaps possible to do the same for the Bures metric and the non-unitary penalties. Of course, the relation could arise only when the geometry is Riemannian and not sub-Finslerian. Another different but complementary viewpoint was developed in \cite{Lumia:2026zwx}, where the object of study is not the reduced Lindblad evolution itself, but the ensemble of stochastic quantum trajectories generated by a chosen unraveling and analyzed through a data-driven geometric construction. Since that notion is sensitive to dynamical features such as chaos and integrability, it would be very interesting to understand more precisely how it is related to the control-geometric notion studied here, and whether the two approaches capture different facets of complexity in dissipative quantum dynamics.

In closed quantum systems, different notions of complexity, in particular circuit complexity and Krylov complexity, have been related in several works \cite{PhysRevLett.132.160402,Craps:2025kub}, and related multiseed constructions inspired by Nielsen complexity were proposed in \cite{PhysRevLett.134.050402}. Krylov complexity has also been studied in open systems \cite{Bhattacharya:2022gbz}. It would therefore be interesting to understand whether generalizations of similar relations between circuit complexity and Krylov complexity exist in open-system dynamics, and how dissipation, decoherence, and non-unitary evolution modify this connection.

One motivation behind our study is related to representation theory and conformal field theory. In many constructions that have appeared in the literature \cite{Caputa:2018kdj,Chagnet:2021uvi,Baiguera:2025uss}, the allowed gates are taken from the generators of the symmetry algebra, so that the complexity geometry effectively probes only states lying within a single representation, or equivalently within the conformal family of a chosen primary operator. In this sense, the motion in state space remains confined to a rather special orbit. 
A non-unitary extension of this setup can overcome this limitation and allow transitions between different primary families. Such a possibility would considerably enlarge the accessible state space and may lead to a richer notion of complexity, one that is sensitive not only to motion within a given symmetry representation but also to the structure relating distinct sectors of the theory\footnote{A different proposal for a version of non-unitary complexity with related motivations has also been considered recently in \cite{Demulder:2026bey}. See also \cite{Tang:2026axr,Bai:2026tbj} for studies of exact Lindbladian dynamics in WZW and related conformal field theories, and \cite{Bai:2026avl} for a recent analysis of non-unitary complex-time evolution in two-dimensional CFTs, all from perspectives different from ours.}. An intermediate step towards addressing the full-fledged CFT problem would be to consider a system of coupled spins and study the complexity as a function of the decomposable representation of the total spin.



As discussed in the introduction, one important motivation for the study of complexity is its role as a probe of the structure of spacetime in the holographic correspondence. In particular, the long-time growth of complexity can be associated to the presence of a black hole and it is one of the few probes of the black-hole interior that are in principle accessible to an outside observer. It would be very interesting to develop a more precise dictionary between the geometric properties of the black-hole interior and the behavior of complexity.\footnote{See \cite{Jorstad:2023kmq,Caceres:2024edr,Policastro:2025odh} for some explorations of the holographic complexity proposals in relation to the black-hole singularity.} The strategy of coupling the black hole in Anti-de Sitter space to an external bath, thus allowing it to evaporate, has proven fruitful in elucidating the information-processing properties of black holes and the release of information in the evaporation process \cite{Almheiri:2019psf,Almheiri:2019hni}. The framework developed in this paper should be of help in extending this understanding to the question of the evolution of complexity. In particular, we expect that there could be an interesting interplay between the tendency of complexity to increase, and the contracting properties of the dissipative evolution. A similar interplay in the relaxation properties has been investigated in solvable models, like SYK perturbed with a Lindbladian interaction  
\cite{Sa:2021tdr, Kulkarni:2021gtt}. It would also be interesting to understand how the evaporation is affected by changing the channels available for the non-unitary evolution.

Another potentially interesting direction is to apply the present framework to de Sitter space. Although the global de Sitter universe is not itself an open system, a local observer is restricted to a causal patch bounded by a cosmological horizon. The observable sector is then naturally described by a reduced density matrix obtained by tracing over the degrees of freedom beyond the horizon \cite{Alicki:2023rfv}. The resulting effective evolution is therefore expected to be non-unitary. This suggests that our complexity geometry for open systems may provide a natural notion of complexity for observables in a static patch of de Sitter space. In particular, one may ask whether the reduced dynamics admits an effective Lindblad description and, if so, whether the associated Finsler or sub-Finsler geometry captures the directional hardness of preparing local states in the presence of horizon-induced dissipation. The corresponding flag curvature could then provide a measure of how strongly the optimal state-preparation paths depend on the de Sitter temperature and on the approach to the late-time stationary state.

Finally, a very important question is to explore to what extent the geometric approach can be applied to the study of numerical optimal-control algorithms. In this context, we should notice that the presence of dissipation is not always a nuisance, but it can also be used as a resource. 
 For example, in cat-code systems, two-photon dissipation can be used to stabilize quantum states and improve control~\cite{Marquet:2023fpq,Marquet:2024ero}. Our method could describe the cost of using both coherent controls and dissipative processes to prepare or protect these states. Similarly, in ion-trap computers, one could choose a cost function based not only on physical effort, but also on task-specific requirements such as Secret Independence, which is a  measure of uniformity of noises in a quantum operation~\cite{Gustiani:2024ctx}. In our theoretical study we assumed that the non-unitary operations can be assigned an arbitrary cost, and naturally, the extent to which this is true has to be assessed on a case-by-case basis in an experimentally realizable setup.

\begin{acknowledgments}
It is a pleasure to acknowledge discussions with Aranya Bhattacharya, Prasanna Venkatesh B, Vaso Dominador Jr., Leonardus Brahmantyo Putra, Cica Gustiani and Benjamin Huard. We would like to thank Shira Chapman and Yaron Oz for their useful feedback on the draft. KG is supported by the Royal Society under grant RF\textbackslash{}ERE\textbackslash{}231142.
\end{acknowledgments}

\appendix

\section{Appendix}

\subsection{Computing the Complexity for Depolarizing Dynamics with Unitary Drift}

    In this appendix we give some details on the numerical computation of the optimal trajectories presented in section \ref{sec:depol_drift}. The main point is to introduce a new set of variables that simplify the PMP equations and the numerical evaluation of the complexity. 
    
    
    
    We recall that the control equations are given by \eqref{eq:depol-drift-eom}, and the objective is to minimize the cost function \eqref{eq:cost-function}.
   
    We start by defining the PMP Hamiltonian to optimize:
    \begin{eqnarray}
        H &=& p_x \left( -2\gamma x - \omega y + \Delta z \right) + p_y \left( \omega x - 2\gamma y \right) + p_z \left( -\Delta x - 2\gamma z \right)\nonumber\\
        &&- \left( p_\gamma \gamma^2 + p_{\gamma \omega} \gamma \omega + p_\omega \omega^2 + C_t \right). 
    \end{eqnarray}
    It is convenient to write the Hamiltonian equations for the coordinates and conjugate momenta in vector notation:
\begin{eqnarray}
    \begin{split}
        \dot {\bf x} & = -2 \gamma {\bf x} + \omega \, {\bf e_z} \wedge {\bf x} + \Delta  {\bf e_y}\wedge {\bf x} \, \\
        \dot {\bf p} & = 2 \gamma {\bf p} + \omega \, {\bf e_z} \wedge {\bf p} + \Delta {\bf e_y} \wedge {\bf p} \,,
    \end{split}
\end{eqnarray}
where ${\bf e_y},{\bf e_z}$ are the unit vectors in the respective directions. 
From these we can derive the equations of motion for the angular momenta ${\bf J} = {\bf x}\wedge {\bf p}$: 
\begin{equation}  \label{eq:app-eom-angmom}
    \dot {\bf J} = \omega \, {\bf e_z}\wedge {\bf J} + \Delta {\bf e_y} \wedge {\bf J}
\,. 
\end{equation}
It is easy to see that the two following quantities are constants of motion: 
\begin{equation}
    s = {\bf x} \cdot {\bf p} \,, \quad \ell^2 = {\bf J}^2 \,.
\end{equation}

We apply the stationarity condition by optimising the PMP Hamiltonian with respect to the controls $u=(\gamma, \omega)$:
    \begin{equation}
        \frac{\partial H}{\partial u} = 0.
    \end{equation}
    This gives the optimal controls: 
    \begin{eqnarray}
        \gamma^* &=& \frac{-4p_\omega (xp_x+yp_y+zp_z) - p_{\gamma\omega} (xp_y - yp_x)}{4p_\gamma p_\omega - p_{\gamma\omega}^2} = \frac{-4 p_\omega s - p_{\gamma \omega} J_z}{4p_\gamma p_\omega - p_{\gamma\omega}^2},\nonumber\\
        \omega^* &=& \frac{2p_\gamma (xp_y - yp_x) + 2p_{\gamma\omega} (xp_x+yp_y+zp_z)}{4p_\gamma p_\omega - p_{\gamma\omega}^2}=\frac{2p_\gamma J_z + 2 p_{\gamma \omega} s}{4p_\gamma p_\omega - p_{\gamma\omega}^2}.\nonumber
    \end{eqnarray}
   Integrating the equations  \eqref{eq:app-eom-angmom} we can express $J_y$ in terms of $J_z$: 
    \begin{equation}
        J_y = C - \frac{1}{\Delta} \frac{p_\gamma J_z^2 + 2 p_{\gamma \omega} s J_z}{4p_\gamma p_\omega - p_{\gamma\omega}^2},
    \end{equation}
    where $C$ is an integration constant, while $J_x$ can be expressed as $J_x = \pm \sqrt{\ell^2 - J_y^2 - J_z^2}$. Finally, from the integration of the equation of motion for $J_z$, we have
    \begin{equation}
        \Delta t = \pm \int~\frac{dJ_z}{\Delta \sqrt{\ell^2 - J_y^2(J_z) - J_z^2}}.
    \end{equation}
    This gives an elliptic function. The rest of the computation is done numerically with \texttt{Mathematica} using the variables defined above.

\bibliographystyle{JHEP}
\bibliography{mybib_two}

@article{Baiguera:2025dkc,
  author        = {Baiguera, Stefano and Balasubramanian, Vijay and Caputa, Pawel and Chapman, Shira and Haferkamp, Jonas and Heller, Michal P. and Halpern, Nicole Yunger},
  title         = {{Quantum complexity in gravity, quantum field theory, and quantum information science}},
  journal       = {Phys. Rept.},
  volume        = {1159},
  pages         = {1--77},
  year          = {2026},
  doi           = {10.1016/j.physrep.2025.11.001},
  eprint        = {2503.10753},
  archivePrefix = {arXiv},
  primaryClass  = {hep-th}
}

@article{Nielsen:2006cea,
    author = "Nielsen, Michael A. and Dowling, Mark R. and Gu, Mile and Doherty, Andrew C.",
    title = "{Quantum Computation as Geometry}",
    eprint = "quant-ph/0603161",
    archivePrefix = "arXiv",
    doi = "10.1126/science.1121541",
    journal = "Science",
    volume = "311",
    number = "5764",
    pages = "1133--1135",
    year = "2006"
}

@article{Dowling:2006tnk,
  author        = {Dowling, Mark R. and Nielsen, Michael A.},
  title         = {{The geometry of quantum computation}},
  journal       = {Quant. Inf. Comput.},
  volume        = {8},
  number        = {10},
  pages         = {0861--0899},
  year          = {2008},
  doi           = {10.26421/QIC8.10-1},
  eprint        = {quant-ph/0701004},
  archivePrefix = {arXiv},
  primaryClass  = {quant-ph}
}

@article{Lloyd:2000dwz,
    author = "Lloyd, Seth and Viola, Lorenza",
    title = "{Control of open quantum systems dynamics}",
    eprint = "quant-ph/0008101",
    archivePrefix = "arXiv",
    month = "8",
    year = "2000"
}

@article{Carlini:2007fqn,
    author = "Carlini, A. and Hosoya, A. and Koike, T. and Okudaira, Y.",
    title = "{Quantum Brachistochrone for Mixed States}",
    eprint = "quant-ph/0703047",
    archivePrefix = "arXiv",
    doi = "10.1088/1751-8113/41/4/045303",
    journal = "J. Phys. A",
    volume = "41",
    pages = "045303",
    year = "2008"
}

@article{Caceres:2019pgf,
    author = "Caceres, Elena and Chapman, Shira and Couch, Josiah D. and Hernandez, Juan P. and Myers, Robert C. and Ruan, Shan-Ming",
    title = "{Complexity of Mixed States in QFT and Holography}",
    eprint = "1909.10557",
    archivePrefix = "arXiv",
    primaryClass = "hep-th",
    doi = "10.1007/JHEP03(2020)012",
    journal = "JHEP",
    volume = "03",
    pages = "012",
    year = "2020"
}

@article{Chapman:2021jbh,
  author        = {Chapman, Shira and Policastro, Giuseppe},
  title         = {{Quantum computational complexity from quantum information to black holes and back}},
  journal       = {Eur. Phys. J. C},
  volume        = {82},
  number        = {2},
  pages         = {128},
  year          = {2022},
  doi           = {10.1140/epjc/s10052-022-10037-1},
  eprint        = {2110.14672},
  archivePrefix = {arXiv},
  primaryClass  = {hep-th}
}

@article{Susskind:2014rva,
  author        = {Susskind, Leonard},
  title         = {{Computational Complexity and Black Hole Horizons}},
  journal       = {Fortsch. Phys.},
  volume        = {64},
  pages         = {24--43},
  year          = {2016},
  doi           = {10.1002/prop.201500092},
  eprint        = {1403.5695},
  archivePrefix = {arXiv},
  primaryClass  = {hep-th}
}

@article{Susskind:2014moa,
  author        = {Susskind, Leonard},
  title         = {{Entanglement is not enough}},
  journal       = {Fortsch. Phys.},
  volume        = {64},
  pages         = {49--71},
  year          = {2016},
  doi           = {10.1002/prop.201500095},
  eprint        = {1411.0690},
  archivePrefix = {arXiv},
  primaryClass  = {hep-th}
}

@article{Brown:2015bva,
  author        = {Brown, Adam R. and Roberts, Daniel A. and Susskind, Leonard and Swingle, Brian and Zhao, Ying},
  title         = {{Holographic Complexity Equals Bulk Action?}},
  journal       = {Phys. Rev. Lett.},
  volume        = {116},
  number        = {19},
  pages         = {191301},
  year          = {2016},
  doi           = {10.1103/PhysRevLett.116.191301},
  eprint        = {1509.07876},
  archivePrefix = {arXiv},
  primaryClass  = {hep-th}
}

@article{Brown:2016wib,
  author        = {Brown, Adam R. and Susskind, Leonard and Zhao, Ying},
  title         = {{Quantum Complexity and Negative Curvature}},
  journal       = {Phys. Rev. D},
  volume        = {95},
  number        = {4},
  pages         = {045010},
  year          = {2017},
  doi           = {10.1103/PhysRevD.95.045010},
  eprint        = {1608.02612},
  archivePrefix = {arXiv},
  primaryClass  = {hep-th}
}

@article{Brown:2019whu,
  author        = {Brown, Adam R. and Susskind, Leonard},
  title         = {{Complexity geometry of a single qubit}},
  journal       = {Phys. Rev. D},
  volume        = {100},
  number        = {4},
  pages         = {046020},
  year          = {2019},
  doi           = {10.1103/PhysRevD.100.046020},
  eprint        = {1903.12621},
  archivePrefix = {arXiv},
  primaryClass  = {hep-th}
}

@article{Mori:2023qbd,
  author        = {Mori, Takashi},
  title         = {{Liouvillian-gap analysis of open quantum many-body systems in the weak dissipation limit}},
  journal       = {Phys. Rev. B},
  volume        = {109},
  number        = {6},
  pages         = {064311},
  year          = {2024},
  doi           = {10.1103/PhysRevB.109.064311},
  eprint        = {2311.10304},
  archivePrefix = {arXiv},
  primaryClass  = {cond-mat.stat-mech}
}

@article{Flory:2020eot,
  author        = {Flory, Mario and Heller, Michal P.},
  title         = {{Geometry of Complexity in Conformal Field Theory}},
  journal       = {Phys. Rev. Res.},
  volume        = {2},
  number        = {4},
  pages         = {043438},
  year          = {2020},
  doi           = {10.1103/PhysRevResearch.2.043438},
  eprint        = {2005.02415},
  archivePrefix = {arXiv},
  primaryClass  = {hep-th}
}

@article{Chagnet:2021uvi,
  author        = {Chagnet, Nathana{\"e}l and Chapman, Shira and de Boer, Jan and Zukowski, Claire},
  title         = {{Complexity for Conformal Field Theories in General Dimensions}},
  journal       = {Phys. Rev. Lett.},
  volume        = {128},
  number        = {5},
  pages         = {051601},
  year          = {2022},
  doi           = {10.1103/PhysRevLett.128.051601},
  eprint        = {2103.06920},
  archivePrefix = {arXiv},
  primaryClass  = {hep-th}
}

@article{Demulder:2026bey,
    author = "Demulder, Saskia",
    title = "{Non-invertible circuit complexity from fusion operations}",
    eprint = "2601.09535",
    archivePrefix = "arXiv",
    primaryClass = "hep-th",
    month = "1",
    year = "2026"
}

@article{Jefferson:2017sdb,
  author        = {Jefferson, Robert and Myers, Robert C.},
  title         = {{Circuit complexity in quantum field theory}},
  journal       = {JHEP},
  volume        = {10},
  pages         = {107},
  year          = {2017},
  doi           = {10.1007/JHEP10(2017)107},
  eprint        = {1707.08570},
  archivePrefix = {arXiv},
  primaryClass  = {hep-th}
}

@article{Gorini:1975nb,
  author  = {Gorini, Vittorio and Kossakowski, Andrzej and Sudarshan, E. C. G.},
  title   = {{Completely Positive Dynamical Semigroups of N Level Systems}},
  journal = {J. Math. Phys.},
  volume  = {17},
  pages   = {821},
  year    = {1976},
  doi     = {10.1063/1.522979}
}

@article{Lindblad:1975ef,
  author  = {Lindblad, G{\"o}ran},
  title   = {{On the Generators of Quantum Dynamical Semigroups}},
  journal = {Commun. Math. Phys.},
  volume  = {48},
  pages   = {119},
  year    = {1976},
  doi     = {10.1007/BF01608499}
}

@article{Chruscinski:2017,
  author  = {Chru{\'s}ci{\'n}ski, Dariusz and Pascazio, Saverio},
  title   = {{A Brief History of the GKLS Equation}},
  journal = {Open Syst. Inf. Dyn.},
  volume  = {24},
  number  = {3},
  pages   = {1740001},
  year    = {2017},
  doi     = {10.1142/S1230161217400017}
}

@book{Rivas:2012ugu,
  author    = {Rivas, {\'A}ngel and Huelga, Susana F.},
  title     = {{Open Quantum Systems}},
  publisher = {Springer},
  year      = {2012},
  doi       = {10.1007/978-3-642-23354-8}
}

@book{Breuer:2002pc,
  author    = {Breuer, Heinz-Peter and Petruccione, Francesco},
  title     = {{The Theory of Open Quantum Systems}},
  publisher = {Oxford University Press},
  year      = {2002}
}

@article{Lidar:2005gzv,
  author        = {Lidar, Daniel A. and Shabani, Alireza and Alicki, Robert},
  title         = {{Conditions for strictly purity-decreasing quantum Markovian dynamics}},
  journal       = {Chem. Phys.},
  volume        = {322},
  pages         = {82--86},
  year          = {2006},
  doi           = {10.1016/j.chemphys.2005.06.038},
  eprint        = {quant-ph/0411119},
  archivePrefix = {arXiv},
  primaryClass  = {quant-ph}
}

@book{Bao:2000,
  author    = {Bao, David and Chern, Shiing-Shen and Shen, Zhongmin},
  title     = {{An Introduction to Riemann-Finsler Geometry}},
  publisher = {Springer},
  address   = {New York},
  year      = {2000}
}

@article{Lopez:2000,
  author  = {L{\'o}pez, Carlos and Mart{\'i}nez, Eduardo},
  title   = {{Sub-Finslerian Metric Associated to an Optimal Control System}},
  journal = {SIAM J. Control Optim.},
  volume  = {39},
  number  = {3},
  pages   = {798--811},
  year    = {2000},
  doi     = {10.1137/S0363012999357562}
}

@article{Boscain:2021jlj,
  author        = {Boscain, Ugo and Sigalotti, Mario and Sugny, Dominique},
  title         = {{Introduction to the Pontryagin Maximum Principle for Quantum Optimal Control}},
  journal       = {PRX Quantum},
  volume        = {2},
  number        = {3},
  pages         = {030203},
  year          = {2021},
  doi           = {10.1103/PRXQuantum.2.030203},
  eprint        = {2010.09368},
  archivePrefix = {arXiv},
  primaryClass  = {quant-ph}
}

@book{PMP,
  author    = {Pontryagin, L. S. and Boltyanskii, V. G. and Gamkrelidze, R. V. and Mishchenko, E. F.},
  title     = {{The Mathematical Theory of Optimal Processes}},
  publisher = {John Wiley and Sons},
  address   = {New York},
  year      = {1962}
}

@book{Liberzon:2012,
  author    = {Liberzon, Daniel},
  title     = {{Calculus of Variations and Optimal Control Theory}},
  publisher = {Princeton University Press},
  year      = {2012}
}

@article{Rivas:2010lbs,
  author        = {Rivas, {\'A}ngel and Plato, A. Douglas K. and Huelga, Susana F. and Plenio, Martin B.},
  title         = {{Markovian master equations: a critical study}},
  journal       = {New J. Phys.},
  volume        = {12},
  number        = {11},
  pages         = {113032},
  year          = {2010},
  doi           = {10.1088/1367-2630/12/11/113032},
  eprint        = {1006.4666},
  archivePrefix = {arXiv},
  primaryClass  = {quant-ph}
}

@article{Sa:2021tdr,
  author        = {S{\'a}, Lucas and Ribeiro, Pedro and Prosen, Toma{\v z}},
  title         = {{Lindbladian dissipation of strongly-correlated quantum matter}},
  journal       = {Phys. Rev. Res.},
  volume        = {4},
  number        = {2},
  pages         = {L022068},
  year          = {2022},
  doi           = {10.1103/PhysRevResearch.4.L022068},
  eprint        = {2112.12109},
  archivePrefix = {arXiv},
  primaryClass  = {cond-mat.stat-mech}
}

@article{Kulkarni:2021gtt,
  author        = {Kulkarni, Abhishek and Numasawa, Tokiro and Ryu, Shinsei},
  title         = {{Lindbladian dynamics of the Sachdev-Ye-Kitaev model}},
  journal       = {Phys. Rev. B},
  volume        = {106},
  number        = {7},
  pages         = {075138},
  year          = {2022},
  doi           = {10.1103/PhysRevB.106.075138},
  eprint        = {2112.13489},
  archivePrefix = {arXiv},
  primaryClass  = {cond-mat.stat-mech}
}

@article{Zyczkowski:2005jvo,
  author        = {{\.Z}yczkowski, Karol and Alicki, Robert and Emerson, Joseph},
  title         = {{Scalable noise estimation with random unitary operators}},
  journal       = {J. Opt. B},
  volume        = {7},
  number        = {10},
  pages         = {S347},
  year          = {2005},
  doi           = {10.1088/1464-4266/7/10/021},
  eprint        = {quant-ph/0503243},
  archivePrefix = {arXiv},
  primaryClass  = {quant-ph}
}

@article{Dankert:2009yux,
  author        = {Dankert, Christoph and Cleve, Richard and Emerson, Joseph and Livine, Etera},
  title         = {{Exact and approximate unitary 2-designs and their application to fidelity estimation}},
  journal       = {Phys. Rev. A},
  volume        = {80},
  number        = {1},
  pages         = {012304},
  year          = {2009},
  doi           = {10.1103/PhysRevA.80.012304},
  eprint        = {quant-ph/0606161},
  archivePrefix = {arXiv},
  primaryClass  = {quant-ph}
}

@article{Magesan:2012mfg,
  author        = {Magesan, Easwar and Gambetta, Jay M. and Emerson, Joseph},
  title         = {{Characterizing quantum gates via randomized benchmarking}},
  journal       = {Phys. Rev. A},
  volume        = {85},
  number        = {4},
  pages         = {042311},
  year          = {2012},
  doi           = {10.1103/PhysRevA.85.042311},
  eprint        = {1109.6887},
  archivePrefix = {arXiv},
  primaryClass  = {quant-ph}
}

@article{Wallman:2015uzh,
  author        = {Wallman, Joel J. and Emerson, Joseph},
  title         = {{Noise tailoring for scalable quantum computation via randomized compiling}},
  journal       = {Phys. Rev. A},
  volume        = {94},
  number        = {5},
  pages         = {052325},
  year          = {2016},
  doi           = {10.1103/PhysRevA.94.052325},
  eprint        = {1512.01098},
  archivePrefix = {arXiv},
  primaryClass  = {quant-ph}
}

@article{Hashim:2020cop,
  author        = {Hashim, Akel and others},
  title         = {{Randomized compiling for scalable quantum computing on a noisy superconducting quantum processor}},
  journal       = {Phys. Rev. X},
  volume        = {11},
  number        = {4},
  pages         = {041039},
  year          = {2021},
  doi           = {10.1103/PhysRevX.11.041039},
  eprint        = {2010.00215},
  archivePrefix = {arXiv},
  primaryClass  = {quant-ph}
}

@article{Karpinski:2008ehv,
  author        = {Karpinski, Michal and Radzewicz, Czeslaw and Banaszek, Konrad},
  title         = {{Fiber-optic realization of anisotropic depolarizing quantum channels}},
  journal       = {J. Opt. Soc. Am. B},
  volume        = {25},
  pages         = {668},
  year          = {2008},
  doi           = {10.1364/JOSAB.25.000668},
  eprint        = {0707.3728},
  archivePrefix = {arXiv},
  primaryClass  = {quant-ph}
}

@article{Shaham:2010iju,
  author        = {Shaham, A. and Eisenberg, H. S.},
  title         = {{Realizing controllable noise in photonic quantum information channels}},
  journal       = {Phys. Rev. A},
  volume        = {83},
  pages         = {022303},
  year          = {2011},
  doi           = {10.1103/PhysRevA.83.022303},
  eprint        = {1006.5795},
  archivePrefix = {arXiv},
  primaryClass  = {quant-ph}
}

@article{Jeong:2012jbn,
  author        = {Jeong, Yoon-Chan and Lee, Joon-Cheol and Kim, Yoon-Ho},
  title         = {{Experimental implementation of a fully controllable depolarizing quantum operation}},
  journal       = {Phys. Rev. A},
  volume        = {87},
  pages         = {014301},
  year          = {2013},
  doi           = {10.1103/PhysRevA.87.014301},
  eprint        = {1204.0850},
  archivePrefix = {arXiv},
  primaryClass  = {quant-ph}
}

@article{Denis:2022ckr,
  author        = {Denis, J{\'e}r{\^o}me and Martin, John},
  title         = {{Extreme depolarization for any spin}},
  journal       = {Phys. Rev. Res.},
  volume        = {4},
  number        = {1},
  pages         = {013178},
  year          = {2022},
  doi           = {10.1103/PhysRevResearch.4.013178},
  eprint        = {2106.11680},
  archivePrefix = {arXiv},
  primaryClass  = {quant-ph}
}

@article{Krantz:2019jkw,
  author        = {Krantz, Philip and Kjaergaard, Morten and Yan, Fei and Orlando, Terry P. and Gustavsson, Simon and Oliver, William D.},
  title         = {{A quantum engineer's guide to superconducting qubits}},
  journal       = {Appl. Phys. Rev.},
  volume        = {6},
  number        = {2},
  pages         = {021318},
  year          = {2019},
  doi           = {10.1063/1.5089550},
  eprint        = {1904.06560},
  archivePrefix = {arXiv},
  primaryClass  = {quant-ph}
}

@article{Kubica:2022fqg,
  author        = {Kubica, Aleksander and Haim, Arbel and Vaknin, Yotam and Levine, Harry and Brand{\~a}o, Fernando and Retzker, Alex},
  title         = {{Erasure Qubits: Overcoming the $T_1$ Limit in Superconducting Circuits}},
  journal       = {Phys. Rev. X},
  volume        = {13},
  number        = {4},
  pages         = {041022},
  year          = {2023},
  doi           = {10.1103/PhysRevX.13.041022},
  eprint        = {2208.05461},
  archivePrefix = {arXiv},
  primaryClass  = {quant-ph}
}

@article{Fisher:2011hvf,
  author        = {Fisher, Kent and Prevedel, Robert and Kaltenbaek, Rainer and Resch, Kevin J.},
  title         = {{Optimal linear optical implementation of a single-qubit damping channel}},
  journal       = {New J. Phys.},
  volume        = {14},
  pages         = {033016},
  year          = {2012},
  doi           = {10.1088/1367-2630/14/3/033016},
  eprint        = {1109.2070},
  archivePrefix = {arXiv},
  primaryClass  = {quant-ph}
}

@article{Albert:2018reb,
  author        = {Albert, Victor V. and others},
  title         = {{Performance and structure of single-mode bosonic codes}},
  journal       = {Phys. Rev. A},
  volume        = {97},
  number        = {3},
  pages         = {032346},
  year          = {2018},
  doi           = {10.1103/PhysRevA.97.032346},
  eprint        = {1708.05010},
  archivePrefix = {arXiv},
  primaryClass  = {quant-ph}
}

@article{PhysRevB.80.155307,
  author  = {Stobbe, S{\o}ren and Johansen, Jeppe and Kristensen, Philip Tr{\o}st and Hvam, J{\o}rn M. and Lodahl, Peter},
  title   = {{Frequency dependence of the radiative decay rate of excitons in self-assembled quantum dots: Experiment and theory}},
  journal = {Phys. Rev. B},
  volume  = {80},
  number  = {15},
  pages   = {155307},
  year    = {2009},
  doi     = {10.1103/PhysRevB.80.155307}
}

@article{DOHERTY20131,
  author  = {Doherty, Marcus W. and Manson, Neil B. and Delaney, Paul and Jelezko, Fedor and Wrachtrup, J{\"o}rg and Hollenberg, Lloyd C. L.},
  title   = {{The nitrogen-vacancy colour centre in diamond}},
  journal = {Phys. Rept.},
  volume  = {528},
  number  = {1},
  pages   = {1--45},
  year    = {2013},
  doi     = {10.1016/j.physrep.2013.02.001}
}

@article{Goldman:2015ohd,
  author        = {Goldman, M. L. and Sipahigil, A. and Doherty, M. W. and Yao, N. Y. and Bennett, S. D. and Markham, M. and Twitchen, D. J. and Manson, N. B. and Kubanek, A. and Lukin, M. D.},
  title         = {{Phonon-Induced Population Dynamics and Intersystem Crossing in Nitrogen-Vacancy Centers}},
  journal       = {Phys. Rev. Lett.},
  volume        = {114},
  number        = {14},
  pages         = {145502},
  year          = {2015},
  doi           = {10.1103/PhysRevLett.114.145502},
  eprint        = {1406.4065},
  archivePrefix = {arXiv},
  primaryClass  = {quant-ph}
}

@article{Ruan:2020vze,
    author = "Ruan, Shan-Ming",
    title = "{Purification Complexity without Purifications}",
    eprint = "2006.01088",
    archivePrefix = "arXiv",
    primaryClass = "hep-th",
    doi = "10.1007/JHEP01(2021)092",
    journal = "JHEP",
    volume = "01",
    pages = "092",
    year = "2021"
}

@article{Marquet:2023fpq,
  author        = {Marquet, Antoine and others},
  title         = {{Autoparametric Resonance Extending the Bit-Flip Time of a Cat Qubit up to 0.3 s}},
  journal       = {Phys. Rev. X},
  volume        = {14},
  number        = {2},
  pages         = {021019},
  year          = {2024},
  doi           = {10.1103/PhysRevX.14.021019},
  eprint        = {2307.06761},
  archivePrefix = {arXiv},
  primaryClass  = {quant-ph}
}

@article{Marquet:2024ero,
  author        = {Marquet, Antoine and others},
  title         = {{Harnessing two-photon dissipation for enhanced quantum measurement and control}},
  journal       = {Phys. Rev. Applied},
  volume        = {22},
  number        = {3},
  pages         = {034053},
  year          = {2024},
  doi           = {10.1103/PhysRevApplied.22.034053},
  eprint        = {2403.07744},
  archivePrefix = {arXiv},
  primaryClass  = {quant-ph}
}

@article{Gustiani:2024ctx,
  author        = {Gustiani, Cica and Leichtle, Dominik and Miller, Jonathan and Grassie, Ross and Mills, Daniel and Kashefi, Elham},
  title         = {{On-Chip Verified Quantum Computation with an Ion-Trap Quantum Processing Unit}},
  journal       = {Phys. Rev. Lett.},
  volume        = {135},
  number        = {16},
  pages         = {160801},
  year          = {2025},
  doi           = {10.1103/jpms-v3kw},
  eprint        = {2410.24133},
  archivePrefix = {arXiv},
  primaryClass  = {quant-ph}
}

@article{Tang:2026axr,
    author = "Tang, Qicheng and Barad, Ruhanshi and Wen, Xueda",
    title = "{Exact operator dynamics in Lindbladian Wess-Zumino-Witten conformal field theories}",
    eprint = "2606.19465",
    archivePrefix = "arXiv",
    primaryClass = "cond-mat.stat-mech",
    month = "6",
    year = "2026"
}

@article{Stinespring1955,
  author  = {Stinespring, W. Forrest},
  title   = {Positive Functions on {C}*-Algebras},
  journal = {Proceedings of the American Mathematical Society},
  volume  = {6},
  number  = {2},
  pages   = {211--216},
  year    = {1955},
  doi     = {10.1090/S0002-9939-1955-0069403-4}
}

@book{NielsenChuang2010,
  author    = {Nielsen, Michael A. and Chuang, Isaac L.},
  title     = {Quantum Computation and Quantum Information},
  edition   = {10th Anniversary},
  publisher = {Cambridge University Press},
  address   = {Cambridge},
  year      = {2010},
  isbn      = {9781107002173}
}

@book{Watrous2018,
  author    = {Watrous, John},
  title     = {The Theory of Quantum Information},
  publisher = {Cambridge University Press},
  address   = {Cambridge},
  year      = {2018},
  doi       = {10.1017/9781316848142},
  isbn      = {9781107180567}
}

@article{Caputa:2018kdj,
    author = "Caputa, Pawel and Magan, Javier M.",
    title = "{Quantum Computation as Gravity}",
    eprint = "1807.04422",
    archivePrefix = "arXiv",
    primaryClass = "hep-th",
    reportNumber = "YITP-18-75",
    doi = "10.1103/PhysRevLett.122.231302",
    journal = "Phys. Rev. Lett.",
    volume = "122",
    number = "23",
    pages = "231302",
    year = "2019"
}

@article{Baiguera:2025uss,
    author = "Baiguera, Stefano and Chagnet, Nicolas and Chapman, Shira and Shoval, Osher",
    title = "{CFT complexity and penalty factors}",
    eprint = "2507.22118",
    archivePrefix = "arXiv",
    primaryClass = "hep-th",
    doi = "10.1007/JHEP02(2026)247",
    journal = "JHEP",
    volume = "02",
    pages = "247",
    year = "2026"
}

@article{Jorstad:2023kmq,
    author = "J{\o}rstad, Eivind and Myers, Robert C. and Ruan, Shan-Ming",
    title = "{Complexity=anything: singularity probes}",
    eprint = "2304.05453",
    archivePrefix = "arXiv",
    primaryClass = "hep-th",
    reportNumber = "YITP-23-41",
    doi = "10.1007/JHEP07(2023)223",
    journal = "JHEP",
    volume = "07",
    pages = "223",
    year = "2023"
}

@article{Policastro:2025odh,
    author = "Policastro, Giuseppe and Wittum, Simon",
    title = "{Probing the singularity of scalar-haired black holes with holographic complexity}",
    eprint = "2512.07403",
    archivePrefix = "arXiv",
    primaryClass = "hep-th",
    doi = "10.1007/JHEP05(2026)116",
    journal = "JHEP",
    volume = "05",
    pages = "116",
    year = "2026"
}

@article{Caceres:2024edr,
    author = "C{\'a}ceres, Elena and Murcia, {\'A}ngel J. and Patra, Ayan K. and Pedraza, Juan F.",
    title = "{Kasner eons with matter: holographic excursions to the black hole singularity}",
    eprint = "2408.14535",
    archivePrefix = "arXiv",
    primaryClass = "hep-th",
    reportNumber = "WI-27-2024, IFT-UAM/CSIC-24-123",
    doi = "10.1007/JHEP12(2024)077",
    journal = "JHEP",
    volume = "12",
    pages = "077",
    year = "2024"
}

@article{Almheiri:2019psf,
    author = "Almheiri, Ahmed and Engelhardt, Netta and Marolf, Donald and Maxfield, Henry",
    title = "{The entropy of bulk quantum fields and the entanglement wedge of an evaporating black hole}",
    eprint = "1905.08762",
    archivePrefix = "arXiv",
    primaryClass = "hep-th",
    doi = "10.1007/JHEP12(2019)063",
    journal = "JHEP",
    volume = "12",
    pages = "063",
    year = "2019"
}

@article{Almheiri:2019hni,
    author = "Almheiri, Ahmed and Mahajan, Raghu and Maldacena, Juan and Zhao, Ying",
    title = "{The Page curve of Hawking radiation from semiclassical geometry}",
    eprint = "1908.10996",
    archivePrefix = "arXiv",
    primaryClass = "hep-th",
    doi = "10.1007/JHEP03(2020)149",
    journal = "JHEP",
    volume = "03",
    pages = "149",
    year = "2020"
}

@article{Alicki:2023rfv,
    author = "Alicki, Robert and Barenboim, Gabriela and Jenkins, Alejandro",
    title = "{Quantum thermodynamics of de Sitter space}",
    eprint = "2307.04800",
    archivePrefix = "arXiv",
    primaryClass = "gr-qc",
    reportNumber = "IFIC/23-31",
    doi = "10.1103/PhysRevD.108.123530",
    journal = "Phys. Rev. D",
    volume = "108",
    number = "12",
    pages = "123530",
    year = "2023"
}

@article{Bhattacharyya:2024duw,
    author = "Bhattacharyya, Arpan and Brahma, Suddhasattwa and Haque, S. Shajidul and Lund, Jacob S. and Paul, Arpon",
    title = "{The early universe as an open quantum system: complexity and decoherence}",
    eprint = "2401.12134",
    archivePrefix = "arXiv",
    primaryClass = "hep-th",
    doi = "10.1007/JHEP05(2024)058",
    journal = "JHEP",
    volume = "05",
    pages = "058",
    year = "2024"
}

@article{Bhattacharyya:2022rhm,
    author = "Bhattacharyya, Arpan and Hanif, Tanvir and Haque, S. Shajidul and Paul, Arpon",
    title = "{Decoherence, entanglement negativity, and circuit complexity for an open quantum system}",
    eprint = "2210.09268",
    archivePrefix = "arXiv",
    primaryClass = "hep-th",
    doi = "10.1103/PhysRevD.107.106007",
    journal = "Phys. Rev. D",
    volume = "107",
    number = "10",
    pages = "106007",
    year = "2023"
}

@article{Bhattacharya:2022wlp,
    author = "Bhattacharya, Aranya and Bhattacharyya, Arpan and Maulik, Sabyasachi",
    title = "{Pseudocomplexity of purification for free scalar field theories}",
    eprint = "2209.00049",
    archivePrefix = "arXiv",
    primaryClass = "hep-th",
    doi = "10.1103/PhysRevD.106.086010",
    journal = "Phys. Rev. D",
    volume = "106",
    number = "8",
    pages = "086010",
    year = "2022"
}

@article{Bhattacharyya:2021fii,
    author = "Bhattacharyya, Arpan and Hanif, Tanvir and Haque, S. Shajidul and Rahman, Md. Khaledur",
    title = "{Complexity for an open quantum system}",
    eprint = "2112.03955",
    archivePrefix = "arXiv",
    primaryClass = "hep-th",
    doi = "10.1103/PhysRevD.105.046011",
    journal = "Phys. Rev. D",
    volume = "105",
    number = "4",
    pages = "046011",
    year = "2022"
}

@article{Bhattacharya:2022gbz,
    author = "Bhattacharya, Aranya and Nandy, Pratik and Nath, Pingal Pratyush and Sahu, Himanshu",
    title = "{Operator growth and Krylov construction in dissipative open quantum systems}",
    eprint = "2207.05347",
    archivePrefix = "arXiv",
    primaryClass = "quant-ph",
    doi = "10.1007/JHEP12(2022)081",
    journal = "JHEP",
    volume = "12",
    pages = "081",
    year = "2022"
}

@article{Craps:2025kub,
    author = "Craps, Ben and Pascuzzi, Gabriele and Pedraza, Juan F. and Qu, Le-Chen and Ruan, Shan-Ming",
    title = "{Explicit Connections Between Krylov and Nielsen Complexity}",
    eprint = "2511.15799",
    archivePrefix = "arXiv",
    primaryClass = "hep-th",
    reportNumber = "IFT-UAM/CSIC-25-123",
    month = "11",
    year = "2025"
}

@article{Lumia:2026zwx,
    author = "Lumia, Luca and Tirrito, Emanuele and Collura, Mario and Essler, Fabian H. L. and Fazio, Rosario",
    title = "{Complexity of Quantum Trajectories}",
    eprint = "2602.00232",
    archivePrefix = "arXiv",
    primaryClass = "quant-ph",
    month = "1",
    year = "2026"
}

@article{PhysRevLett.132.160402,
  title = {A Relation between Krylov and Nielsen Complexity},
  author = {Craps, Ben and Evnin, Oleg and Pascuzzi, Gabriele},
  journal = {Phys. Rev. Lett.},
  volume = {132},
  issue = {16},
  pages = {160402},
  numpages = {6},
  year = {2024},
  month = {Apr},
  publisher = {American Physical Society},
  doi = {10.1103/PhysRevLett.132.160402},
  url = {https://link.aps.org/doi/10.1103/PhysRevLett.132.160402}
}

@article{PhysRevLett.134.050402,
  title = {Multiseed Krylov Complexity},
  author = {Craps, Ben and Evnin, Oleg and Pascuzzi, Gabriele},
  journal = {Phys. Rev. Lett.},
  volume = {134},
  issue = {5},
  pages = {050402},
  numpages = {7},
  year = {2025},
  month = {Feb},
  publisher = {American Physical Society},
  doi = {10.1103/PhysRevLett.134.050402},
  url = {https://link.aps.org/doi/10.1103/PhysRevLett.134.050402}
}

@article{PhysRevA.110.042601,
  title = {Optimized continuous dynamical decoupling via differential geometry and machine learning},
  author = {Morazotti, Nicolas Andr\'e da Costa and da Silva, Adonai Hil\'ario and Audi, Gabriel and Fanchini, Felipe Fernandes and Napolitano, Reginaldo de Jesus},
  journal = {Phys. Rev. A},
  volume = {110},
  issue = {4},
  pages = {042601},
  numpages = {14},
  year = {2024},
  month = {Oct},
  publisher = {American Physical Society},
  doi = {10.1103/PhysRevA.110.042601},
  url = {https://link.aps.org/doi/10.1103/PhysRevA.110.042601}
}

@article{Bai:2026avl,
    author = "Bai, Chen and Mao, Weibo and Nozaki, Masahiro and Tan, Mao Tian and Wen, Xueda",
    title = "{Relaxation Process During Complex Time Evolution In Two-Dimensional Integrable and Chaotic CFTs}",
    eprint = "2601.09290",
    archivePrefix = "arXiv",
    primaryClass = "hep-th",
    reportNumber = "RIKEN-iTHEMS-Report-26",
    month = "1",
    year = "2026"
}

@article{Bai:2026tbj,
    author = "Bai, Chen",
    title = "{Exact Lindbladian Dynamics from Conformal Embeddings and Topological Defects in Conformal Field Theory}",
    eprint = "2607.08827",
    archivePrefix = "arXiv",
    primaryClass = "cond-mat.stat-mech",
    month = "7",
    year = "2026"
}
\end{document}